\documentstyle[amssymb,12pt]{article}

\newlength{\minitwocolumn}
\font\teneufm=eufm10
\font\seveneufm=eufm7
\font\fiveeufm=eufm5
\newfam\eufmfam
\textfont\eufmfam=\teneufm
\scriptfont\eufmfam=\seveneufm
\scriptscriptfont\eufmfam=\fiveeufm

\makeatletter
\@addtoreset{equation}{section}
\makeatother

\newtheorem{thm}{Theorem}[section]
\newtheorem{prop}[thm]{Proposition}

\newtheorem{conj}[thm]{Conjecture}

\newtheorem{df}{Definition}[section]

\title{\bf The Intergals of Motion for\\
the Deformed Virasoro Algebra}
\begin{document}
\maketitle
\begin{center}
{\it
Dedicated to 
Professor Masaki Kashiwara
on the occasion on the 60th birthday}
\\~\\~\\
{ Boris FEIGIN$~^{\alpha}$,
Takeo KOJIMA$~^{\beta}$,
Jun'ichi SHIRAISHI$~^{\gamma}$,
Hidekazu WATANABE$~^{\gamma}$}
\\~\\~\\
{\it
$~^\alpha$
L.D.Landau Institute for Theoretical Sciences, \\
Chernogolovka, Moscow 142432, RUSSIA
\\
$~^\beta$
Department of Mathematics,
College of Science and Technology,
Nihon University,\\
Surugadai, Chiyoda-ku, Tokyo 101-0062, 
JAPAN\\
$~^\gamma$
Graduate School of Mathematical Science,
University of Tokyo, \\
Komaba, Meguro-ku, Tokyo, 153-8914,
JAPAN
}
\end{center}
~\\
\begin{abstract}
We explicitly construct two classes of
infinitly many commutative operators in terms of
the deformed Virasoro algebra.
We call one of them 
local integrals and the other nonlocal one,
since they can be regarded as elliptic deformations of 
the local and nonlocal integrals of motion 
obtained by V.Bazhanov, S.Lukyanov and
Al.Zamolodchikov \cite{BLZ}.
\end{abstract}

\newpage

\section{Introduction}

The purpose of this paper is to construct
two classes of infinitly many commutative operators,
in terms of the deformed Virasoro algebra.
Let us recall some facts about soliton equation.
The classical sine-Gordon equation
\begin{eqnarray}
\partial_{\tau} \partial_t \phi(t,\tau)=
e^{\phi(t,\tau)}-e^{-\phi(t,\tau)},
\end{eqnarray}
is one of the simplest example of the Toda field theory.
One of the authors and E.Frenkel \cite{FF}
considered the so-called local integrals of motion $I^{(cl)}$ 
for the classical sine-Gordon theory
\begin{eqnarray}
\{I^{(cl)},H^{(cl)}\}_{P.B.}=0,
\end{eqnarray}
where $H^{(cl)}=\frac{1}{2}\int (e^{\phi(t)}+e^{-\phi(t)})dt$
is the Hamiltonian.
They showed the existence of infinitly many
commutative integrals of motion by a cohomological argumemnt,
and showed that they can be regarded as 
the conservation laws for the KdV equation
\begin{eqnarray}
\partial_\tau W(\tau,t)=\partial_t^3 W(\tau,t)+3
W(\tau,t)\partial_t W(\tau,t).
\end{eqnarray}
They gave similar construction
for the Toda field theory 
associated with the root system of finite and affine type.

In \cite{FF} they constructed
the quantum deformtion of the local integrals of motion.
In other words they showed the existence of quantum deformation
of the conservation laws of the KdV equation.
After quantization
Gel'fand-Dickij bracket $\{,\}_{P.B.}$ for the second 
Hamiltonian structure of the KdV, 
gives rise to the Virasoro algebra
\begin{eqnarray}
[L_m,L_n]=(m-n)L_{m+n}+\frac{c_{CFT}}{12}\delta_{m+n,0}.
\end{eqnarray}
V.Bazhanov, S.Lukyanov, Al.Zamolodchikov \cite{BLZ}
constructed field theoretical analogue of the commuting
transfer matrix ${\bf T}(z)$, acting on 
the irreducible highest weight 
module of the Virasoro algebra.
They constructed this commuting transfer matrix ${\bf T}(z)$
as the trace of the monodromy matrix associated with
the quantum affine symmetry $U_q(\widehat{sl_2})$,
and commutatin relation $[{\bf T}(z),{\bf T}(w)]=0$
is a direct consequence of the Yang-Baxter relation.
The coefficients of the asymptotic expansion
of the operator ${\rm log}~{\bf T}(z)$ at $z \to \infty$,
produce the local integrals of motion
for the Virasoro algebra, which recover
the conservation laws of the KdV equation
in the classical limit $c_{CFT} \to \infty$.
They call the coefficients of the Taylor expansion
of the operator ${\bf T}(z)$ at $z=0$,
the nonlocal integrals of motion
for the Virasoro algebra.
They have explicit integral
representation in terms of the screening currents.

The purpose of this paper is 
to construct the elliptic version
of the integrals of motion given by 
V.Bazhanov, S.Lukyanov, Al.Zamolodchikov 
\cite{BLZ}.
Their construction is based on
the free field realization of the Borel subalgebra
${\cal B}_\pm$ of $U_q(\widehat{sl_2})$ 
in terms of the screening currents,
which cannot be extended to the full 
symmetry $U_q(\widehat{sl_2})$.
By using this realization
they construct the monodromy matrix as the image of
the universal $R$-matrix $\bar{\cal R} \in {\cal B}_+
\otimes {\cal B}_-$, and derive the transfer matrix
${\bf T}(z)$ as the trace of this monodromy matrix.
Because the universl $R$-matrix ${\cal \widetilde{R}}$
of the elliptic quantum group
does not exist in the tensor of the Borel subalgebras 
$ {\cal B}_+ \otimes {\cal B}_-$,
it is impossible to construct the elliptic
deformation of the transfer matris ${\bf T}(z)$
as the same manner as \cite{BLZ}.
Hence our method of construction should be
completely different from 
those of \cite{BLZ}.
Instead of considering the transfer mtrix ${\bf T}(z)$,
we directly give the integral representations of 
the integrals of motion 
${\cal I}_n, {\cal G}_n$ for the deformed Virasoro algebra.
The commutativity of the integrals of motion
are not understood as a direct consequence
of the Yang-Baxter equation.
They are understood as a consequence of
the commutative subalgebra of the Feigin-Odesskii algebra
\cite{FO}.

We would like to mention about two important degenerating limits
of the deformed Virasoro algebra.
One is the CFT-limit\cite{BLZ} 
and the other is the classical limit\cite{FR}.
In the CFT-limit V.Bazhanov, S.Lukyanov and Al.Zamolodchikov
\cite{BLZ} constructed infinitly many integrals of motion
for the Virasoro algebra, as we mentioned above.
In the classical limit,
the deformed Virasoro algebra
degenerates to the Poisson-Virasoro algebra introduced
by E.Frenkel and N.Reshetikhin \cite{FR}.
In the classical limit,
E.Frenkel \cite{F} constructed infinitly many integrals of
motion for the Poisson-Virasoro algebra.

The organization of this paper is as follows.
In Section 2, we review
the deformed Virasoro algebra,
including free field realization, screening currents
\cite{SKAO, FF}.
In Section 3, we
give explicit formulae for the local 
integrals of motion ${\cal I}_n$,
and show the commutation relation $[{\cal I}_m,{\cal I}_n]=0$ 
and Dynkin-automorphis invariance $\eta({\cal I}_n)={\cal I}_n$.
In Section 4, 
we give explicit formulae
for the nonlocal integrals of motion
${\cal G}_n$,
and show the commutation relation $[{\cal G}_m,{\cal G}_n]=0$,
$[{\cal I}_m,{\cal G}_n]=0$
and 
the Dynkin-automorphis invariance $\eta({\cal G}_n)={\cal G}_n$.
In Section 5, we study specialization to $s=2$.
We discuss about relation to the Poisson-Virasoro algebra
\cite{FR, F} and compare our results with those of CFT by
\cite{BLZ}.
In Appendix we summarize the normal ordering of 
the basic operators.

\section{The Deformed Virasoro Algebra}

In this section we review 
the deformed Virasoro algebra and
its screening currents \cite{SKAO, FF}.
We prepare the notations to be used in this paper.
Throughout this paper, we fix generic three parameters 
$0<x<1$, $r \in {\mathbb C}$ and $s \in {\mathbb C}$.
Let us set $z=x^{2u}$.
Let us set $r^*=r-1$.
The symbol $[u]_r$ for ${\rm Re}(r)>0$ stands for 
the Jacobi theta function
\begin{eqnarray}
~[u]_r&=&x^{\frac{u^2}{r}-u}\frac{\Theta_{x^{2r}}(x^{2u})}{(x^{2r};x^{2r})^3},
~~
\Theta_q(z)=
(z;q)_\infty (qz^{-1};q)_\infty (q;q)_\infty,
\end{eqnarray}
where we have used the standard notation
\begin{eqnarray}
(z;q)_\infty=\prod_{j=0}^\infty (1-q^jz).
\end{eqnarray}
We set the parametrizations $\tau, \tau^*$
\begin{eqnarray}
x=e^{-\pi \sqrt{-1}/r\tau}=e^{-\pi \sqrt{-1}/r^* \tau^*}.
\end{eqnarray}
The theta function $[u]_r$ enjoys 
the quasi-periodicity property
\begin{eqnarray}
[u+r]_r=-[u]_r,~~~[u+r\tau]_r=-e^{-\pi \sqrt{-1}\tau-\frac{2\pi \sqrt{-1}}{r}
u}[u]_r.
\end{eqnarray}
%In what follows we use three theta functions
%$[u]_r, [u]_{-r^*}, [u]_s$ which are associated with three periods
%$r, r^*=r-1, s$.
The symbol $[a]$ stands for
\begin{eqnarray} 
[a]=\frac{x^a-x^{-a}}{x-x^{-1}}.
\end{eqnarray}

\subsection{Free Field Realization}

Let $\beta_m^1,\beta_m^2$ 
be the oscillators $(m \in {\mathbb Z}_{\neq 0})$
with the commutation relations
\begin{eqnarray}
~[\beta_n^i,\beta_m^i]&=&n\frac{[r^*n]}{[rn]}
\frac{[(s-1)n]}{[sn]}\delta_{n+m,0}~(i=1,2),\\
~[\beta_n^i,\beta_m^j]&=&-n\frac{[r^*n]}{[rn]}
\frac{[n]}{[sn]}x^{sn~{\rm sgn}(i-j)}\delta_{n+m,0}~(1\leqq i\neq j \leqq 2).
\end{eqnarray}
Let $P, Q$ be the zero mode operators 
with the commutation realtions
\begin{eqnarray}
~[P,iQ]&=&2.
\end{eqnarray}
Let us map $\eta$ on this algebra by
\begin{eqnarray}
\eta(\beta_m^1)=x^{-sm}\beta_m^2,~~
\eta(\beta_m^2)=x^{sm}\beta_m^1,~~\eta(P)=-P,~~\eta(Q)=-Q.
\end{eqnarray}
It preserves the commutation relation,
$[\beta_n^i,\beta_m^j]=
[\eta(\beta_n^i),\eta(\beta_m^j)],
~[P,iQ]=[\eta(P),\eta(iQ)]$,
$\eta^2=id$.
In what follows we call the map $\eta$ Dynkin-automorphism.

We deal with the bosonic Fock space 
${\cal F}_{l,k} (l,k \in {\mathbb Z})$
generated by $\beta_{-m}^{i} (m>0,i=1,2)$
over the vacuum vectors $|l,k\rangle$ :
\begin{eqnarray}
\beta_m^j|l,k\rangle&=&0,  (m>0, j=1,2),\\
P|l,k\rangle&=&(l\sqrt{\frac{r}{r^*}}-k\sqrt{\frac{r^*}{r}})|l,k\rangle,\\
|l,k\rangle&=&e^{(l\sqrt{\frac{r}{r^*}}-k\sqrt{\frac{r^*}{r}})
\frac{i}{2}Q}|0,0\rangle.
\end{eqnarray}
We use the abberiviation $\hat{\pi}=\sqrt{rr^*}P$.
In what follows we work on the Fock space ${\cal F}_{l,k}$
with fixed values $l,k \in {\mathbb Z}$.

\subsection{The Deformed
Virasoro Algebra}

Let us set the fundamental operators $\Lambda_j(z)~(j=1,2)$
associated with vertex of Dynkin-diagram of the affine symmetry 
$A_1^{(1)}$
\begin{eqnarray}
\Lambda_1(z)&=&
x^{-\hat{\pi}}:\exp\left(
\sum_{m\neq 0}\frac{1}{m}(x^{rm}-x^{-rm})\beta_m^1 z^{-m}\right):,
\\
\Lambda_2(z)&=&
x^{\hat{\pi}}:\exp\left(
\sum_{m\neq 0}\frac{1}{m}(x^{rm}-x^{-rm})\beta_m^2 z^{-m}\right):.
\end{eqnarray}
%The generating operator $T_j(z)~(j=1,2)$ for three parameter 
%analogues of Virasoro algebra, are given by
%the elliptic analogue of the Miura-transformation.
%\begin{eqnarray}
%:(x^{-2D_z}-\Lambda_1(x^{-1}z))(x^{-2D_z}-\Lambda_2(xz)):=
%T_0(z)x^{-4D_z}-T_1(x^{-1}z)x^{-2D_z}+T_2(z).
%\end{eqnarray}
%Here we used $D_z=z\frac{d}{dz}$ such that $x^{D_z}f(z)x^{-D_z}=f%(xz)$.
%We set $T_0(z)=id$. We have
Here the symbol
$:*:$ stands for 
usual normal ordering of bosons,
{\it i.e.}, $\beta_m^i$ with $m>0$
should be moved to the right.
Let us set the operators $T_1(z), T_2(z)$ by
\begin{eqnarray}
T_1(z)=\Lambda_1(z)+\Lambda_2(z),~~
T_2(z)=:\Lambda_1(x^{-1}z)\Lambda_2(xz):.
\end{eqnarray}
%\begin{eqnarray}
%&&f_{11}(z_2/z_1)\Lambda_1(z_1)\Lambda_1(z_2)=
%f_{11}(z_1/z_2)\Lambda_1(z_2)\Lambda_1(z_1),\\
%&&f_{11}(z_2/z_1)\Lambda_2(z_1)\Lambda_2(z_2)=
%f_{11}(z_1/z_2)\Lambda_2(z_2)\Lambda_2(z_1),\\
%&&f_{11}(z_2/z_1)\Lambda_1(z_1)\Lambda_2(z_2)
%-
%f_{11}(z_1/z_2)\Lambda_2(z_2)\Lambda_1(z_1)\nonumber\\
%&=&\frac{(1-x^{2r-2})(1-x^{-2r})}{
%(1-x^{-2})}:\Lambda_1(z_1)\Lambda_2(z_2):
%(\delta(z_2/z_1)-\delta(x^{-2}z_2/z_1)).
%\end{eqnarray}
\begin{prop}~~~\label{def:3dVir}
The operators $T_j(z)~(j=1,2)$ satisfy
\begin{eqnarray}
&&f_{11}(z_2/z_1)T_1(z_1)T_1(z_2)-
f_{11}(z_1/z_2)T_1(z_2)T_1(z_1)\nonumber\\
&=&c
(\delta(x^2z_2/z_1)T_2(x^{-1}z_1)-\delta(x^2z_1/z_2)T_2(x^{-1}z_2)),
\label{def:dVir1}\\
&&f_{12}(z_2/z_1)T_1(z_1)T_2(z_2)=
f_{21}(z_1/z_2)T_2(z_2)T_1(z_1),\label{def:dVir2}\\
&&f_{22}(z_2/z_1)T_2(z_1)T_2(z_2)=
f_{22}(z_1/z_2)T_2(z_2)T_2(z_1).\label{def:dVir3}
\end{eqnarray}
Here the structure functions $f_{ij}(z)$ and 
the constant $c$ are given by
\begin{eqnarray}
f_{11}(z)&=&\exp\left(
\sum_{m=1}^\infty \frac{1}{m}
(1-x^{2rm})(1-x^{-2r^* m})\frac{(1-x^{2m(s-1)})}{(1-x^{2sm})}z^m
\right),\\
f_{12}(z)&=&\exp\left(
\sum_{m=1}^\infty \frac{1}{m}
(1-x^{2rm})(1-x^{-2r^*m})\frac{(1-x^{2m(s-2)})}{(1-x^{2sm})}(xz)^m
\right)=f_{21}(z),\\
f_{22}(z)&=&\exp\left(
\sum_{m=1}^\infty \frac{1}{m}
(1-x^{2rm})(1-x^{-2r^*m})\frac{(1+x^{2m})(1-x^{2m(s-2)})}{(1-x^{2sm})}z^m
\right),
\end{eqnarray}
\begin{eqnarray}
c=-\frac{(1-x^{-2r^*})(1-x^{2r})}{(1-x^2)}.
\end{eqnarray}
\end{prop}
Note that
the structure functions $f_{11}(z), f_{12}(z)=f_{21}(z)$ and 
$f_{22}(z)$ enjoy the fusion properties
\begin{eqnarray}
f_{11}(z)&=&
\frac{1}{(1-z)}\frac{ 
(x^{2s-2}z;x^{2s})_\infty
(x^{2r}z;x^{2s})_\infty 
(x^{-2r^*}z;x^{2s})_\infty
}{
(x^2z;x^{2s})_\infty
(x^{2r^*+2s}z;x^{2s})_\infty
(x^{2s-2r}z;x^{2s})_\infty
},\\
f_{12}(z)&=&
\frac{f_{11}(xz)f_{11}(x^{-1}z)}{\Delta(z)}=f_{21}(z),~
f_{22}(z)=f_{12}(x^{-1}z)f_{21}(xz),
\end{eqnarray}
where we have set
\begin{eqnarray}
\Delta(z)&=&
\frac{(1-x^{r+r^*}z)(1-x^{-r-r^*}z)}{(1-xz)(1-x^{-1}z)}.
\end{eqnarray}
The Dynkin automorphism $\eta$ acts on $\Lambda_j(z)$ as
\begin{eqnarray}
\eta(\Lambda_1(z))=\Lambda_2(x^s z),~~~
\eta(\Lambda_2(z))=\Lambda_1(x^{-s} z).
\end{eqnarray}

\begin{df}~~~
The deformed Virsoro algebra
is defined by the generators
$\widehat{T}_m^{(1)}$,
$\widehat{T}_m^{(2)},~(m \in {\mathbb Z})$
with the defining relations
(\ref{def:dVir1}), (\ref{def:dVir2}) 
and (\ref{def:dVir3}).
Here we should understand 
$\widehat{T}_m^{(1)}$,
$\widehat{T}_m^{(2)}$ as the Fourier coefficients
of the operators
$\widehat{T}_1(z)=\sum_{m\in {\mathbb Z}}
\widehat{T}_m^{(1)}z^{-m}$,
$\widehat{T}_2(z)=\sum_{m\in {\mathbb Z}}
\widehat{T}_m^{(2)}z^{-m}$.
\end{df}

\subsection{Screening Current}

We introduce the screening currents $F_j(z), E_j(z)~(j=0,1)$ as
\begin{eqnarray}
F_1(z)&=&e^{i \sqrt{\frac{r^*}{r}}Q}
z^{\frac{1}{r}\hat{\pi}
+\frac{r^*}{r}}:\exp\left(\sum_{m\neq 0}\frac{1}{m}(\beta_m^1-\beta_m^2)
z^{-m}\right):,\\
F_0(z)&=&e^{-i \sqrt{\frac{r^*}{r}}Q}
z^{-\frac{1}{r}\hat{\pi}
+\frac{r^*}{r}}:\exp\left(\sum_{m\neq 0}\frac{1}{m}
(-x^{sm}\beta_m^1+x^{-sm}\beta_m^2)
z^{-m}\right):,\\
E_1(z)&=&e^{-i\sqrt{\frac{r}{r^*}}Q}
z^{-\frac{1}{r^*}\hat{\pi}
+\frac{r}{r^*}}:\exp\left(-
\sum_{m\neq 0}\frac{1}{m}\frac{[rm]}{[r^*m]}(\beta_m^1-\beta_m^2)
z^{-m}\right):,\\
E_0(z)&=&e^{i\sqrt{\frac{r}{r^*}}Q}z^{\frac{1}{r^*}\hat{\pi}
+\frac{r}{r^*}}:\exp\left(-\sum_{m\neq 0}\frac{1}{m}
\frac{[rm]}{[r^* m]}
(-x^{sm}\beta_m^1+x^{-sm}\beta_m^2)
z^{-m}\right):.
\end{eqnarray}
We summarize convenient commutation relations
\begin{eqnarray}
~[\beta_m^1-\beta_m^2,\beta_n^1-\beta_n^2]&=&
m\frac{[r^*m]}{[rm]}(x^m+x^{-m})\delta_{m+n,0},\\
~[\beta_m^1-\beta_m^2,
-x^{sn}\beta_n^1+x^{-sn}\beta_n^2]&=&
m\frac{[r^*m]}{[rm]}(x^{(s-1)m}+x^{(1-s)m})\delta_{m+n,0},\\
~[-x^{sm}\beta_m^1+x^{-sm}\beta_m^2,
-x^{sn}\beta_n^1+x^{-sn}\beta_n^2]&=&
m\frac{[r^*m]}{[rm]}(x^m+x^{-m})\delta_{m+n,0}.
\end{eqnarray}
The Dynkin-automorphism $\eta$ maps 
the screening currents as
\begin{eqnarray}
\eta(F_1(z))=F_0(z),~~\eta(F_0(z))=F_1(z),~~
\eta(E_1(z))=E_0(z),~~\eta(E_0(z))=E_1(z).
\end{eqnarray}
The adjoint actions of the operator
$T_1(z)=\Lambda_1(z)+\Lambda_2(z)$ 
on the screening currents $F_j(z), E_j(z)$
can be regarded as
the differences of currents ${\cal A}(z), {\cal B}(z)$.
\begin{prop}~~~
The adjoint actions of $\Lambda_j(z)$ 
on the screenings $F_j(z), E_j(z)$ are 
\begin{eqnarray}
~[\Lambda_1(z_1),F_1(z_2)]&=&(x^{-r^*}-x^{r^*})
\delta(x^rz_1/z_2){\cal A}(x^{-r}z_2),\\
~[\Lambda_2(z_1),F_1(z_2)]&=&(x^{r^*}-x^{-r^*})
\delta(x^{-r}z_1/z_2){\cal A}(x^rz_2),\\
~[\Lambda_1(z_1),F_0(z_2)]&=&
(x^{r^*}-x^{-r^*})
\delta(x^{-r+s}z_1/z_2)
\eta({\cal A}(x^{r}z_2)),\\
~[\Lambda_2(z_1),F_0(z_2)]&=&
(x^{-r^*}-x^{r^*})
\delta(x^{r-s}z_1/z_2)
\eta({\cal A}(x^{-r}z_2)),
\end{eqnarray}
\begin{eqnarray}
~[\Lambda_1(z_1),E_1(z_2)]&=&(x^{r}-x^{-r})
\delta(x^{-r^*}z_1/z_2){\cal B}(x^{r^*}z_2),\\
~[\Lambda_2(z_1),E_1(z_2)]&=&(x^{-r}-x^{r})\delta(x^{r^*}z_1/z_2)
{\cal B}(x^{-r^*}z_2),\\
~[\Lambda_1(z_1),E_0(z_2)]&=&
(x^{-r}-x^{r})
\delta(x^{r^*+s}z_1/z_2)
\eta({\cal B}(x^{-r^*}z_2)),\\
~[\Lambda_2(z_1),E_0(z_2)]&=&
(x^{r}-x^{-r})
\delta(x^{-r^*-s}z_1/z_2)
\eta({\cal B}(x^{r^*}z_2)).
\end{eqnarray}
Here we have set
\begin{eqnarray}
{\cal A}(z)&=&
e^{i\sqrt{\frac{r^*}{r}}Q}z^{\frac{1}{r}\hat{\pi}+\frac{r^*}{r}}
:\exp\left(\sum_{m \neq 0}\frac{1}{m}(x^{rm}\beta_m^1-x^{-rm}\beta_m^2)z^{-m}
\right):,\\
{\cal B}(z)&=&
e^{-i\sqrt{\frac{r}{r^*}}Q}z^{-\frac{1}{r^*}\hat{\pi}+\frac{r}{r^*}}
:\exp\left(\sum_{m \neq 0}\frac{1}{m}\frac{[rm]}{[r^*m]}(-x^{-r^*m}\beta_m^{1}
+x^{r^*m}\beta_m^2)z^{-m}\right):.
\end{eqnarray}
\end{prop}
%In what follows, we use the abberiviation $r^*=r-1$.
\begin{prop}~~~
For regime ${\rm Re}(r)>0$, 
the commutation relations of the currents $F_j(z)$
are given as
\begin{eqnarray}
\frac{[u_1-u_2]_{r}}{[u_1-u_2-1]_{r}}F_j(z_1)F_j(z_2)&=&
\frac{[u_2-u_1]_r}{[u_2-u_1-1]_r}F_j(z_2)F_j(z_1),~(j=1,2)\\
\frac{[u_1-u_2+\frac{s}{2}-1]_r}{[u_1-u_2+\frac{s}{2}]_r}
F_0(z_1)F_1(z_2)&=&
\frac{[u_2-u_1+\frac{s}{2}-1]_r}{[u_2-u_1+\frac{s}{2}]_r}
F_1(z_2)F_0(z_1).
\end{eqnarray}
For regime ${\rm Re}(r)<0$, 
we have
\begin{eqnarray}
\frac{[u_1-u_2]_{-r}}{[u_1-u_2+1]_{-r}}F_j(z_1)F_j(z_2)&=&
\frac{[u_2-u_1]_{-r}}{[u_2-u_1+1]_{-r}}F_j(z_2)F_j(z_1),~(j=1,2)\\
\frac{[u_1-u_2-\frac{s}{2}+1]_{-r}}{[u_1-u_2-\frac{s}{2}]_{-r}}
F_0(z_1)F_1(z_2)&=&
\frac{[u_2-u_1-\frac{s}{2}+1]_{-r}}{
[u_2-u_1-\frac{s}{2}]_{-r}}
F_1(z_2)F_0(z_1).
\end{eqnarray}
For regime ${\rm Re}(r^*)>0$, 
we have
\begin{eqnarray}
\frac{[u_1-u_2]_{r^*}}{[u_1-u_2+1]_{r^*}}E_j(z_1)E_j(z_2)&=&
\frac{[u_2-u_1]_{r^*}}{[u_2-u_1+1]_{r^*}}E_j(z_2)E_j(z_1),~(j=1,2)\\
\frac{[u_1-u_2-\frac{s}{2}+1]_{r^*}}{
[u_1-u_2-\frac{s}{2}]_{r^*}}
E_0(z_1)E_1(z_2)&=&
\frac{[u_2-u_1-\frac{s}{2}+1]_{r^*}}{[u_2-u_1-\frac{s}{2}]_{r^*}}
E_1(z_2)E_0(z_1).
\end{eqnarray}
For regime ${\rm Re}(r^*)<0$, 
we have
\begin{eqnarray}
\frac{[u_1-u_2]_{-r^*}}{[u_1-u_2-1]_{-r^*}}E_j(z_1)E_j(z_2)&=&
\frac{[u_2-u_1]_{-r^*}}{[u_2-u_1-1]_{-r^*}}E_j(z_2)E_j(z_1),~(j=1,2)\\
\frac{[u_1-u_2+\frac{s}{2}-1]_{-r^*}}{
[u_1-u_2+\frac{s}{2}]_{-r^*}}
E_0(z_1)E_1(z_2)&=&
\frac{[u_2-u_1+\frac{s}{2}-1]_{-r^*}}{[u_2-u_1+\frac{s}{2}]_{-r^*}}
E_1(z_2)E_0(z_1).
\end{eqnarray}
We have
\begin{eqnarray}
&&~[E_1(z_1),F_1(z_2)]\nonumber\\
&=&\frac{1}{x-x^{-1}}
(\delta(xz_2/z_1)H(x^{r}z_2)
-\delta(xz_1/z_2)H(x^{-r}z_2)),\\
&&~[E_0(z_1),F_0(z_2)]\nonumber\\
&=&
\frac{1}{x-x^{-1}}
(\delta(xz_2/z_1)\eta(H(x^{r}z_2))-
\delta(xz_1/z_2)\eta(H(x^{-r}z_2))).
\end{eqnarray}
Here we have set
\begin{eqnarray}
H(z)&=&e^{-\frac{1}{\sqrt{rr^*}}iQ}
z^{-\frac{1}{\sqrt{rr^*}}P+\frac{1}{rr^*}}
:\exp\left(
-\sum_{m\neq 0}\frac{1}{m}\frac{[m]}{[r^*m]}z^{-m}
(\beta_m^1-\beta_m^2)
\right):.
\end{eqnarray}
We have
\begin{eqnarray}
&&E_1(z_1)F_0(z_2)=F_0(z_2)E_1(z_1),~~
E_0(z_1)F_1(z_2)=F_1(z_2)E_0(z_1).
\end{eqnarray}
\end{prop}

%\begin{eqnarray}
%&=& con. K_1(z)K_2(z)^{-1},\\
%K_1(z)&=&
%e^{-\frac{1}{2\sqrt{rr^*}}iQ}
%z^{-\frac{1}{2\sqrt{rr^*}}P}
%:\exp\left(-\sum_{m\neq 0}\frac{1}{m}\frac{[m]}{[r^*m]}
%\beta_m^1 z^{-m}\right):
%,\\
%K_2(z)&=&
%e^{\frac{1}{2\sqrt{rr^*}}iQ}
%z^{\frac{1}{2\sqrt{rr^*}}P}
%:\exp\left(-\sum_{m\neq 0}\frac{1}{m}\frac{[m]}{[r^*m]}
%\beta_m^2 z^{-m}\right):.
%\end{eqnarray}

\subsection{
Comparsion with another definition}

At first glance,
our definition of the deformed Virasoro is
different from those in \cite{SKAO}.
In this subsection we compare two definitions,
and show they are essentially the same thing.
We show that
operators $T_1(z), T_2(z)$
are realized as the tensor product
of the deformed Virasoro algebra
${\cal V}ir_{q,t},~(q=x^{2r},t=x^{2r-2})$ 
and a proper operator ${\cal Z}(z)$.
Let us set the auxiliary bosons
$B_m^{1}, B_m^{2}$ by
\begin{eqnarray}
B_m^{1}&=&
\beta_m^{1}-\beta_m^{2},\\
B_m^{2}&=&
x^m \beta_m^{1}+x^{-m}\beta_m^{2}.
\end{eqnarray}
We have
\begin{eqnarray}
&&[B_n^1,B_m^1]=n\frac{[r^*n]}{[rn]}(x^n+x^{-n})\delta_{m+n,0},~~[B_n^1,B_m^2]=0,
\\
&&[B_n^2,B_m^2]=n\frac{[r^*n][(s-2)n]}{[rn][sn]}(x^n+x^{-n})\delta_{m+n,0}~~
(m,n\in {\mathbb Z}_{\neq 0}).
\end{eqnarray}
Then we have the following decomposition
\begin{eqnarray}
\Lambda_1(z)=\Lambda_1^{DV}(z){\cal Z}(z),~~
\Lambda_2(z)=\Lambda_2^{DV}(z){\cal Z}(z),
\end{eqnarray}
where we have set
\begin{eqnarray}
\Lambda_1^{DV}(z)&=&x^{-\hat{\pi}}:\exp\left(
\sum_{m\neq 0}\frac{1}{m}\frac{x^{rm}-x^{-rm}}{x^m+x^{-m}}
B_m^{1}(xz)^{-m}\right):,\\
\Lambda_2^{DV}(z)&=&x^{-\hat{\pi}}:\exp\left(
-\sum_{m\neq 0}\frac{1}{m}\frac{x^{rm}-x^{-rm}}{x^m+x^{-m}}
B_m^{1}(x^{-1}z)^{-m}
\right):,\\
{\cal Z}(z)&=&:\exp\left(\sum_{m\neq 0}\frac{1}{m}
\frac{x^{rm}-x^{-rm}}{x^m+x^{-m}}B_m^2 z^{-m}\right):.
\end{eqnarray}
Here we have 
\begin{eqnarray}
T_1(z)=T^{DV}(z){\cal Z}(z),~~~T_2(z)=:{\cal Z}(x^{-1}z){\cal Z}(xz):,
\end{eqnarray}
where
\begin{eqnarray}
T^{DV}(z)=\Lambda_1^{DV}(z)+\Lambda_2^{DV}(z).
\end{eqnarray}
\begin{prop}~
The operator $T^{DV}(z)$ is the generting function of 
the deformed Virasoro algebra 
${\cal V}ir_{q,t}$ defined in \cite{SKAO} with
$q=x^{2r}, t=x^{2r-2}$
\begin{eqnarray}
&&f^{DV}(z_2/z_1)T^{DV}(z_1)T^{DV}(z_2)-
f^{DV}(z_1/z_2)T^{DV}(z_2)T^{DV}(z_1)\nonumber\\
&&=c(\delta(x^2z_2/z_1)-\delta(x^{2}z_1/z_2)),
\end{eqnarray}
where
\begin{eqnarray}
f^{DV}(z)&=&\exp\left(
\sum_{m=1}^\infty
\frac{1}{m}\frac{(1-x^{2rm})(1-x^{-2r^*m})}{(1+x^{2m})}z^{-m}
\right),\\
c&=&-\frac{(1-x^{-2r^*})(1-x^{2r})}{(1-x^2)}.
\end{eqnarray}
We have
\begin{eqnarray}
T^{DV}(z_1){\cal Z}(z_2)&=&{\cal Z}(z_2)T^{DV}(z_1),\\
f^{{\cal Z}}(z_2/z_1){\cal Z}(z_1){\cal Z}(z_2)&=&
f^{{\cal Z}}(z_1/z_2){\cal Z}(z_2){\cal Z}(z_1),
\end{eqnarray}
where
\begin{eqnarray}
f^{{\cal Z}}(z)&=&\exp\left(
-\sum_{m=1}^\infty
\frac{1}{m}\frac{(1-x^{2rm})(1-x^{-2r^*m})(x^{2m}-x^{2(s-1)m})}
{(1-x^{2sm})(1+x^{2m})}z^{-m}
\right).
\end{eqnarray}
\end{prop}
Therefore three parameter deformed Virasoro algebra
$T_1(z), T_2(z)$ is realized as an extension of
two parmeter deformed Virasoro algebra $T^{DV}(z)$.
Note that upon the specializtion $s=2$ we have
\begin{eqnarray}
~[B_n^2,B_m^2]=0,~~[B_n^1,B_m^2]=0,~~f^{{\cal Z}}(z)=1.
\end{eqnarray}
Hence we can regard $B_m^2=0$ and $T_1(z)=T^{DV}(z), T_2(z)=1$.

\section{Local Integrals of Motion ${\cal I}_n$}

In this section 
we construct the local integrals of motion ${\cal I}_n$ and
${\cal I}_n^*$.
We study the generic case :
$0<x<1$, $r \in {\mathbb C}$ and ${\rm Re}(s)>0$.

\subsection{Local Integral of Motion}

Let us set the function $h(u)$ and $h^*(u)$ by
\begin{eqnarray}
h(u)&=&\frac{[u]_s [u+r]_s}{
[u+1]_s [u+r^*]_s},~~~
h^*(u)=\frac{[u]_s [u-r^*]_s}{
[u+1]_s [u-r]_s},
\end{eqnarray}
where we have set $z=x^{2u}$. We have 
$h^*(u)=h(u)|_{r \to -r^*}$.

\begin{df}~~We define ${\cal I}_n$ for regime ${\rm Re}(s)>2$ and 
${\rm Re}(r^*)<0$ by
\begin{eqnarray}
{\cal I}_n=\int \cdots \int_{C} 
\prod_{j=1}^n \frac{dz_j}{2\pi\sqrt{-1}z_j}
\prod_{1\leqq j<k \leqq n}h(u_k-u_j)
T_1(z_1)\cdots T_1(z_n)~~~(n=1,2,\cdots).
\label{def:LocalIM1}
\end{eqnarray}
Here, the contour $C$ encircles $z_j=0$ in such a way that
$z_j=x^{-2+2sl}z_k, x^{-2r^*+2sl}z_k ~(l=0,1,2,\cdots)$ is inside and 
$z_j=x^{2-2sl}z_k, x^{2r^*-2sl}z_k ~(l=0,1,2,\cdots)$ 
is outside for $1\leqq j<k \leqq n$.
We define ${\cal I}_n^*$ for
regime ${\rm Re}(s)>2$ and 
${\rm Re}(r)>0$ by
\begin{eqnarray}
{\cal I}_n^*=\oint \cdots \oint_{C^*} 
\prod_{j=1}^n \frac{dz_j}{2\pi\sqrt{-1}z_j}
\prod_{1\leqq j<k \leqq n}h^*(u_k-u_j)
T_1(z_1)\cdots T_1(z_n)~~~(n=1,2,\cdots).
\end{eqnarray}
The contour $C^*$ encircles $z_j=0$ in such a way that
$z_j=x^{-2+2sl}z_k, x^{2r+2sl}z_k~(l=1,2,\cdots)$ is inside and 
$z_j=x^{2-2sl}z_k, x^{-2r-2sl}z_k~(l=1,2,\cdots)$ is outside 
for $1\leqq j<k \leqq n$.

We call ${\cal I}_n$ and ${\cal I}_n^*$ 
the local integrals of motion
for the deformed Virasoro algebra.
The definitions of ${\cal I}_n, {\cal I}_n^*$ for
generic ${\rm Re}(s)>0$ and $r \in {\mathbb C}$ should be
understood as analytic continuation.
\end{df}
We have the involution ${\cal I}_m^*={\cal I}_m|_{r\to -r^*}$
and ${\cal I}_m={\cal I}_m|_{r^* \to -r}$.
The followings are some of
{\bf Main Results} of this paper.

\begin{thm}~~~\label{thm:Local-Com}
The local integrals of motion 
${\cal I}_n$ and ${\cal I}_n^*$
commute with each other
\begin{eqnarray}
~[{\cal I}_n,{\cal I}_m]=[{\cal I}_n^*,{\cal I}_m^*]=0
~~~(m,n=1,2,\cdots ).
\end{eqnarray}
\end{thm}

\begin{thm}~~~\label{thm:Local-Dynkin}
The local integrals of motion ${\cal I}_n$ and
${\cal I}_n^*$ are invariant under the action
of the Dynkin-automorphism
\begin{eqnarray}
\eta({\cal I}_n)={\cal I}_n,~~\eta({\cal I}_n^*)={\cal I}_n^*~~(n=1,2,\cdots).
\end{eqnarray}
\end{thm}

\begin{conj}~~\label{conj:Local}
\begin{eqnarray}
~[{\cal I}_m, {\cal I}_n^*]=0
~~~(m,n=1,2,\cdots).
\end{eqnarray} 
\end{conj}

\subsection{Another Formulae}

In this subsection we prepare another formulae of 
the local integrals of motion ${\cal I}_n$.
Because the integral contour of the definition of
the local integrals of motion ${\cal I}_n$
is not annulus. {\it i.e.} $|x^{-2}z_k|<|z_j|<|x^2z_k|$,
the defining relations of the deformed Virasoro
(\ref{def:dVir1}), (\ref{def:dVir2}), 
(\ref{def:dVir3})
should be used carefully. 
Hence, in order to show
the commutation relations 
$[{\cal I}_m,{\cal I}_n]=0$,
it is better for us to deform the integral representations
of the local integrals of motion ${\cal I}_n$
to another formulae, in which 
the defining relations of the deformed Virasoro
(\ref{def:dVir1}), (\ref{def:dVir2}), 
(\ref{def:dVir3})
can be used safely.

Let us set the auxiliary function $s(z)$ and $s^*(z)$ by
$h(u)=s(z)f_{11}(z),~h^*(u)=s^*(z)f_{11}(z)$,
where $h(u)$ and $f_{11}(z)$ are given in the previous section.
We have explicitly
\begin{eqnarray}
s(z)&=&x^{-2r^*}\frac{(z;x^{2s})_\infty 
(x^{2s-2r}z;x^{2s})_\infty }
{(x^{2s-2}z;x^{2s})_\infty 
(x^{-2r^*}z;x^{2s})_\infty }\times
\frac{(1/z;x^{2s})_\infty 
(x^{2s-2r}/z;x^{2s})_\infty }
{(x^{2s-2}/z;x^{2s})_\infty 
(x^{-2r^*}/z;x^{2s})_\infty },
\end{eqnarray}
and
\begin{eqnarray}
s^*(z)&=&x^{2r}
\frac{(z;x^{2s})_\infty 
(x^{2s+2r^*}z;x^{2s})_\infty }
{(x^{2s-2}z;x^{2s})_\infty 
(x^{2r}z;x^{2s})_\infty }\times
\frac{(1/z;x^{2s})_\infty 
(x^{2s+2r^*}/z;x^{2s})_\infty }
{(x^{2s-2}/z;x^{2s})_\infty 
(x^{2r}/z;x^{2s})_\infty }.
\end{eqnarray}
Let us introduce a ``weak sense'' equality
\begin{df}~~
We say the operators ${\cal P}(z_1,z_2,\cdots, z_n)$
and ${\cal Q}(z_1,z_2,\cdots,z_n)$ are equal in 
the "weak sense" if
\begin{eqnarray}
\prod_{1\leqq i<j \leqq n}s(z_j/z_i){\cal P}(z_1,z_2,\cdots,z_n)=
\prod_{1\leqq i<j \leqq n}s(z_j/z_i){\cal Q}(z_1,z_2,\cdots,z_n).
\end{eqnarray}
We write
${\cal P}(z_1,z_2,\cdots,z_n)\sim {\cal Q}(z_1,z_2,\cdots,z_n)$, 
showing the weak equality.
\end{df}
For example 
$\delta(z_1/z_2)\sim 0$
and
$\frac{1}{z_1-z_2}\delta(z_1/z_2)\sim 0$.

Let us set
the auxiliary functions $g_{11}(z), g_{12}(z)$ and $g_{22}(z)$ 
by fusion procedure
\begin{eqnarray}
g_{12}(z)=
g_{11}(xz)g_{11}(x^{-1}z)
=g_{21}(z),~~
g_{22}(z)=g_{21}(x^{-1}z)g_{12}(xz),
\end{eqnarray}
where $g_{11}(z)=f_{11}(z)$ 
is defined in Proposition \ref{def:3dVir}.

\begin{prop}~~~~\label{prop:dVir-Delta}
The following equalities hold in the weak sense
\begin{eqnarray}
&&\delta(x^2z_3/z_2)g_{12}(x^{-1}z_2/z_1)T_1(z_1)T_2(x^{-1}z_2)
\sim
\delta(x^2z_3/z_2)g_{21}(xz_1/z_2)T_2(x^{-1}z_2)T_1(z_1),\nonumber\\
\\
&&\prod_{j=3,4
}\delta(x^2z_j/z_1)g_{22}(z_2/z_1)T_2(z_1)T_2(z_2)
\sim
\prod_{j=3,4}
\delta(x^2z_j/z_1)g_{22}(z_1/z_2)T_2(z_2)T_2(z_1).
\end{eqnarray} 
\end{prop}
{\it Proof.}~~~~
By defining relation of the deformation of the Virasoro algegra,
we have
\begin{eqnarray}
&&g_{12}(x^{-1}z_2/z_1)T_1(z_1)T_2(x^{-1}z_2)
-g_{21}(xz_1/z_2)T_2(x^{-1}z_2)T_1(z_1)\nonumber\\
&=&f_{12}(x^{-1}z_2/z_1)(\delta(z_2/z_1)-\delta(x^{-2}z_2/z_1))
T_1(z_1)T_2(x^{-1}z_2),
\end{eqnarray}
where we have used delta-function relation
\begin{eqnarray}
\Delta(z)-\Delta(z^{-1})=c(\delta(xz)-\delta(x^{-1}z)),~~~
\Delta(z)=\frac{(1-x^{2r-1}z)(1-x^{-2r+1}z)}{(1-xz)(1-x^{-1}z)}.
\end{eqnarray}
Because $
\prod_{1\leqq i<j \leqq 3}s(z_j/z_i)\delta(x^2z_3/z_2)
f_{12}(x^{-1}z_2/z_1)(\delta(z_2/z_1)-\delta(x^{-2}z_2/z_1))
T_1(z_1)T_2(x^{-1}z_2)=0$, we conclude the first equation
of proposition.
The second equation is obtained 
in the same manner.
~~Q.E.D.\\\\
Let us introduce $S_n$-invariant in the ``weak sense''.
\begin{df}~
We call 
the operator ${\cal P}(z_1,z_2,\cdots,z_n)$ is $S_n$-invariant 
in the "weak sense"
if 
\begin{eqnarray}
{\cal P}(z_1,z_2,\cdots,z_n)\sim{\cal P}(z_{\sigma(1)},z_{\sigma(2)},
\cdots,z_{\sigma(n)}),~~(\sigma \in S_n).
\end{eqnarray} 
\end{df}
{\bf Example}~~~The operator ${\cal O}_2(z_1,z_2)=
g_{11}(z_2/z_1)T_1(z_1)T_1(z_2)-
c\delta(x^2z_2/z_1)T_2(x^{-1}z_1)$
is $S_2$-invariant.\\\\
In what follows we use the notation of the ordered product
\begin{eqnarray}
\prod_{\longrightarrow \atop{l \in L}}T_1(z_l)=T_1(z_{l_1})T_1(z_{l_2})
\cdots T_1(z_{l_n}),~~~(L=\{l_1,\cdots,l_n|l_1<l_2<\cdots<l_n\}).
\end{eqnarray}
Let us set the auxiliarry operator 
${\cal O}_n(z_1,z_2,\cdots,z_n)$ by
\begin{eqnarray}
{\cal O}_n(z_1,z_2,\cdots,z_n)
&=&
\sum_{\alpha=0}^{[\frac{n}{2}]}(-1)^\alpha c^\alpha
%\frac{1}{\alpha!}
\sum_{A_1,A_2,\cdots,A_\alpha %\subset \{1,2,\cdots,n\}
%\atop{J_t \cap J_{t'}=\phi ~(t \neq t'), |J_t|=2 ~(t=1,\cdots,s)
%\atop{J=\{j_1, \cdots, j_s|j_t={\rm Min}(J_t)\},
%\atop{K=\{k_1, \cdots, k_s|k_t={\rm Max}(J_t)\}}}}
}
\prod_{1\leqq j \leqq \alpha}
\delta\left(x^2\frac{z_{Max(A_j)}}{z_{Min(A_j)}}\right)
\nonumber\\
&\times&
\prod_{\longrightarrow
\atop{1\leqq j \leqq n
\atop{j \notin A}}
}T_1(z_j)
\prod_{\longrightarrow
\atop{1\leqq j \leqq n
\atop{j \in A_{Min}}}}T_2(x^{-1}z_j)
\label{def:Sn-InvLocal}
\\
&\times&
\prod_{1\leqq j<k \leqq n
\atop{j,k \notin A}}g_{11}(z_k/z_j)
\prod_{1\leqq j<k \leqq n
\atop{j,k \in A_{Min}}}g_{22}(z_k/z_j)
\prod_{1\leqq j, k \leqq n
\atop{k\in A_{Min}~j\notin A}}g_{12}(x^{-1} z_k/z_j)\nonumber,
\end{eqnarray}
where the summation
$\sum_{A_1,\cdots,A_\alpha}$ 
is taken over the set
$A_1,A_2,\cdots,A_\alpha
 \subset
\{1,2,\cdots,n\}$
such that 
$A_j \cap A_k=\phi~(1\leqq j\neq k \leqq \alpha)$,
$|A_j|=2~(1\leqq j \leqq \alpha)$,
and $Min(A_j)<Min(A_k)~(1\leqq j<k \leqq \alpha)$.
We have set
$A_{Min}=\{Min(A_1),\cdots, Min(A_\alpha)\}$,
$A_{Max}=\{Max(A_1),\cdots, Max(A_\alpha)\}$, 
and $A=A_1 \cup A_2 \cup \cdots \cup A_\alpha$.\\
{\bf Example}~~We summarize the current ${\cal O}_n$ 
very explicitly for $n=1,2,3,4$.
\begin{eqnarray}
{\cal O}_1(z_1)&=&T_1(z_1),\\
{\cal O}_2(z_1,z_2)&=&g_{11}(z_2/z_1)T_1(z_1)T_1(z_2)-
c\delta(x^2z_2/z_1)T_2(x^{-1}z_1),\\
{\cal O}_3(z_1,z_2,z_3)&=&g_{11}(z_2/z_1)g_{11}(z_3/z_1)g_{11}(z_3/z_2)
T_1(z_1)T_1(z_2)T_1(z_3)\nonumber\\
&-&cg_{12}(x^{-1}z_2/z_1)T_1(z_1)T_2(x^{-1}z_2)\delta(x^2z_3/z_2)\nonumber\\
&-&cg_{12}(x^{-1}z_1/z_2)T_1(z_2)T_2(x^{-1}z_1)\delta(x^2z_3/z_1)\nonumber\\
&-&cg_{12}(x^{-1}z_1/z_3)T_1(z_3)T_2(x^{-1}z_1)\delta(x^2z_2/z_1),
\end{eqnarray}
\begin{eqnarray}
{\cal O}_4(z_1,z_2,z_3,z_4)&=&
\prod_{1\leqq j<k \leqq 4}g_{11}(z_k/z_j)
T_1(z_1)T_1(z_2)T_1(z_3)T_1(z_4)\nonumber\\
&-&cg_{11}(z_2/z_1)g_{12}(x^{-1}z_3/z_1)g_{12}(x^{-1}z_3/z_2)
T_1(z_1)T_1(z_2)T_2(x^{-1}z_3)\delta(x^2z_4/z_3)\nonumber\\
&-&cg_{11}(z_3/z_1)g_{12}(x^{-1}z_2/z_1)g_{12}(x^{-1}z_2/z_3)
T_1(z_1)T_1(z_3)T_2(x^{-1}z_2)\delta(x^2z_4/z_2)\nonumber\\
&-&cg_{11}(z_4/z_1)g_{12}(x^{-1}z_2/z_1)g_{12}(x^{-1}z_2/z_4)
T_1(z_1)T_1(z_4)T_2(x^{-1}z_2)\delta(x^2z_3/z_2)\nonumber\\
&-&cg_{11}(z_3/z_2)g_{12}(x^{-1}z_1/z_2)g_{12}(x^{-1}z_1/z_3)
T_1(z_2)T_1(z_3)T_2(x^{-1}z_1)\delta(x^2z_4/z_1)\nonumber\\
&-&cg_{11}(z_4/z_2)g_{12}(x^{-1}z_1/z_2)g_{12}(x^{-1}z_1/z_4)
T_1(z_2)T_1(z_4)T_2(x^{-1}z_1)\delta(x^2z_3/z_1)\nonumber\\
&-&cg_{11}(z_4/z_3)g_{12}(x^{-1}z_1/z_3)g_{12}(x^{-1}z_1/z_4)
T_1(z_3)T_1(z_4)T_2(x^{-1}z_1)\delta(x^2z_2/z_1)\nonumber\\
&+&c^2 g_{22}(z_3/z_1)\delta(x^2z_2/z_1)\delta(x^2z_4/z_3)
T_2(x^{-1}z_1)T_2(x^{-1}z_3)\nonumber\\
&+&c^2 g_{22}(z_2/z_1)\delta(x^2z_3/z_1)\delta(x^2z_4/z_2)
T_2(x^{-1}z_1)T_2(x^{-1}z_2)\nonumber\\
&+&c^2 g_{22}(z_2/z_1)\delta(x^2z_4/z_1)\delta(x^2z_3/z_2)
T_2(x^{-1}z_1)T_2(x^{-1}z_2).
\end{eqnarray}

\begin{prop}~~~\label{prop:Sn-InvLocal}
The operator ${\cal O}_n$ defined in 
(\ref{def:Sn-InvLocal}) is $S_n$-invariant
in the weak sense.
\begin{eqnarray}
{\cal O}_n(z_1,z_2,\cdots,z_n)
\sim{\cal O}_n(z_{\sigma(1)},z_{\sigma(2)},
\cdots,z_{\sigma(n)})~~~
(\sigma \in S_n).
\end{eqnarray}
\end{prop}

Before showing the complete proof,
we consider some simple examples as warming-up exercise.
When $n=2$ case, ${\cal O}_2(z_1,z_2)={\cal O}_2(z_2,z_1)$
is exactly the same as the defining relation of the deformed Virasoro algebra
Proposition \ref{def:3dVir}.
When $n=3$ case, it is enough to show $S_3$-invariance for two generators
$\sigma=(1,2), (2,3)$.
Let us study the case $\sigma=(1,2)$ case.
We have
\begin{eqnarray}
&&{\cal O}_3(z_1,z_2,z_3)-{\cal O}_3(z_2,z_1,z_3)\nonumber\\
&=&g_{11}(z_3/z_1)g_{11}(z_3/z_2)
(g_{11}(z_2/z_1)T_1(z_1)T_1(z_2)-g_{11}(z_1/z_2)T_1(z_2)T_1(z_1))
T_1(z_3)\\
&-&c T_1(z_3)(g_{12}(x^{-1}z_1/z_3)T_2(x^{-1}z_1)\delta(x^2z_2/z_1)-
g_{12}(x^{-1}z_2/z_3)T_2(x^{-1}z_2)\delta(x^2z_1/z_2)).\nonumber
\end{eqnarray}
Changing the ordering of $T_1(z_1)T_1(z_2), T_1(z_2)T_1(z_1)$ 
and $T_1(z_3)$ 
in the first term, we have the following in the "weak sense"
\begin{eqnarray}
&&T_1(z_3)g_{11}(z_1/z_3)g_{11}(z_2/z_3)
(g_{11}(z_2/z_1)T_1(z_1)T_1(z_2)-g_{11}(z_1/z_2)T_1(z_2)T_1(z_1))\\
&-&c T_1(z_3)(g_{12}(x^{-1}z_1/z_3)T_2(x^{-1}z_1)\delta(x^2z_2/z_1)-
g_{12}(x^{-1}z_2/z_3)T_2(x^{-1}z_2)\delta(x^2z_1/z_2)).\nonumber
\end{eqnarray}
Substituting the defining relation of the deformed 
Virasoro algebra (\ref{def:dVir1}), 
(\ref{def:dVir2}), (\ref{def:dVir3}) 
in Proposition \ref{def:3dVir},
we have
\begin{eqnarray}
&&c T_1(z_3)g_{11}(z_1/z_3)g_{11}(z_2/z_3)
(T_2(x^{-1}z_1)\delta(x^2z_2/z_1)-T_2(x^{-1}z_2)\delta(x^2z_1/z_2)
)\\
&-&c T_1(z_3)(g_{12}(x^{-1}z_1/z_3)T_2(x^{-1}z_1)\delta(x^2z_2/z_1)-
g_{12}(x^{-1}z_2/z_3)T_2(x^{-1}z_2)\delta(x^2z_1/z_2))=0.\nonumber
\end{eqnarray}
Here we have used the fusion relation $g_{12}(z)=g_{11}(x^{-1}z)g_{11}(xz)$.
The case $\sigma=(2,3)$ is similar.

~\\
{\it Proof.}~~~We show $S_n$-invariance in the weak sense for
general $n$ case.
It is enough to prove theorem for generators
$\sigma=(i,i+1)~(1\leqq i \leqq n-1)$.
At first we prepare convenient formula.
Moving  $T_1(z_i)$ and $T_1(z_{i+1})$
to the right, by using Proposition \ref{def:3dVir}, we have
\begin{eqnarray}
&&(T_1(z_i)T_1(z_{i+1})g_{11}(z_{i+1}/z_i)-
T_1(z_{i+1})T_1(z_i)g_{11}(z_i/z_{i+1}))\nonumber
\\
&\times&
T_1(z_{i+2})\cdots T_1(z_n)
\prod_{j=i+2}^n g_{11}(z_j/z_i)g_{11}(z_j/z_{i+1})\sim
c T_1(z_{i+2})\cdots T_1(z_n)\\
&\times&
(\delta(x^2z_{i+1}/z_i)
T_2(x^{-1}z_i)\prod_{j=i+2}^n g_{21}(x^{-1}z_i/z_j)-
\delta(x^2z_i/z_{i+1})
T_2(x^{-1}z_{i+1})\prod_{j=i+2}^n g_{21}(x^{-1}z_{i+1}/z_j)).\nonumber
\end{eqnarray}
Let us study
${\cal O}_n(z_1,\cdots,z_i,z_{i+1},\cdots,z_n)-
{\cal O}_n(z_1,\cdots,z_{i+1},z_i,\cdots,z_n)$.
Collecting $T_1(z_i), T_1(z_{i+1})$ and
$T_2(x^{-1}z_i), T_2(x^{-1}z_{i+1})$ in the center,
by using Proposition
\ref{def:3dVir} and Proposiotion \ref{prop:dVir-Delta}, we have
\begin{eqnarray}
&&{\cal O}_n(z_1,\cdots,z_i,z_{i+1},\cdots,z_n)-
{\cal O}_n(z_1,\cdots,z_{i+1},z_i,\cdots,z_n)\nonumber\\
&\sim&
\sum_{\alpha=0}^{[\frac{n}{2}]}(-1)^\alpha c^\alpha
%\frac{1}{\alpha!}
\sum_{A_1,A_2,\cdots,A_{\alpha-1} 
\subset \{1,2,\cdots,n\}-\{i,i+1\}
%\atop{J_t \cap J_{t'}=\phi ~(t \neq t'), |J_t|=2 ~(t=1,\cdots,s)
%\atop{J=\{j_1, \cdots, j_s|j_t={\rm Min}(J_t)\},
%\atop{K=\{k_1, \cdots, k_s|k_t={\rm Max}(J_t)\}}}}
}
\prod_{1\leqq j \leqq \alpha }\delta(x^2z_{Max(A_j)}/z_{Min(A_j)})
\nonumber\\
&\times&
\prod_{\longrightarrow
\atop{1\leqq j \leqq n
\atop{j \neq i,i+1
\atop{j \notin A}}
}}T_1(z_j)
(g_{11}(z_{i+1}/z_i)T_1(z_i)T_1(z_{i+1})
-g_{11}(z_{i}/z_{i+1})T_1(z_{i+1})T_1(z_i)
)
\prod_{\longrightarrow
\atop{1\leqq k \leqq n
\atop{k \in A_{Min}}}}T_2(x^{-1}z_k)
\nonumber\\
&\times&
\prod_{1\leqq j<k \leqq n
\atop{j,k\neq i,i+1
\atop{j,k \notin A}}}
g_{11}(z_k/z_j)
\prod_{1\leqq j \leqq n
\atop{j\neq i,i+1
\atop{j \notin A}}}
g_{11}(z_i/z_j)g_{11}(z_{i+1}/z_j)
\prod_{1\leqq j,k \leqq n
\atop{k\in A_{Min}
\atop{j\notin A}}}g_{12}(x^{-1} z_k/z_j)
\prod_{1\leqq j<k \leqq n
\atop{j,k \in A_{Min}}}g_{22}(z_k/z_j)\nonumber\\
%%%%%%%%%%%%%%%%%%%%%%%%%%%%%%%%%%%%%%%%%%%%%%%%%%%%%%%
&+&
\sum_{\beta=1}^{[\frac{n}{2}]}(-1)^\beta c^\beta
%\frac{1}{(\beta-1)!}
\sum_{
B_1,B_2,\cdots,B_{\beta} \subset \{1,2,\cdots,n\}
\atop{{\rm s.t}~
B_\gamma=\{i,i+1\}~{\rm for~some}~\gamma
}
%\atop{J_t \cap J_{t'}=\phi ~(t \neq t'), |J_t|=2 ~(t=1,\cdots,s)
%\atop{J=\{j_1, \cdots, j_s|j_t={\rm Min}(J_t)\},
%\atop{K=\{k_1, \cdots, k_s|k_t={\rm Max}(J_t)\}}}}
}
\prod_{1\leqq j\leqq \beta-1}\delta(x^2z_{Max(A_j)}/
z_{Min(A_j)})
\prod_{\longrightarrow
\atop{1\leqq j \leqq n
\atop{j \notin B}}
}T_1(z_j)
\nonumber\\
&\times&
(\prod_{1\leqq j \leqq n
\atop{j \notin B}}
g_{12}(x^{-1}z_i/z_j)
\prod_{1\leqq k \leqq n
\atop{k \neq i
\atop{k \in B_{Min}}
}}
g_{22}(x^{-1}z_k/z_i)
T_2(x^{-1}z_i)\delta(x^2z_{i+1}/z_i)\nonumber\\
&-&
\prod_{1\leqq j \leqq n
\atop{j \notin B}}
g_{12}(x^{-1}z_{i+1}/z_j)
\prod_{1\leqq k \leqq n
\atop{k \neq i
\atop{k \in B_{Min}}
}}
g_{22}(x^{-1}z_k/z_{i+1})
T_2(x^{-1}z_{i+1})\delta(x^2z_i/z_{i+1})
)\nonumber\\
&\times&
\prod_{\longrightarrow
\atop{1\leqq k \leqq n
\atop{k \neq i
\atop{k \in B_{Min}}}}}T_2(x^{-1}z_k)
\prod_{1\leqq j<k \leqq n
\atop{j,k \notin B}}g_{11}(z_k/z_j)
\prod_{1\leqq j<k \leqq n
\atop{j,k \neq i
\atop{j,k \in B_{Min}}}}
g_{22}(z_k/z_j)
\prod_{1\leqq j, k \leqq n
\atop{
k \neq i
\atop{k\in B_{Min},~j\notin B}}}
g_{12}(x^{-1} z_k/z_j).\nonumber\\
\label{eqn:Sn-InvLocal}
\end{eqnarray}
In the second sum, the operator part
is deformed by the fusion relation of coefficients functions.
\begin{eqnarray}
&&\prod_{1\leqq j \leqq n
\atop{j \notin B}}
g_{12}(x^{-1}z_i/z_j)
\prod_{1\leqq k \leqq n
\atop{k \neq i
\atop{k \in B_{Min}}
}}
g_{22}(x^{-1}z_k/z_i)
T_2(x^{-1}z_i)\delta(x^2z_{i+1}/z_i)\nonumber\\
&-&\prod_{1\leqq j \leqq n
\atop{j \notin B}}
g_{12}(x^{-1}z_i/z_j)
\prod_{1\leqq k \leqq n
\atop{k \neq i
\atop{k \in B_{Min}}
}}
g_{22}(x^{-1}z_k/z_{i+1})
T_2(x^{-1}z_{i+1})\delta(x^2z_{i}/z_{i+1})\nonumber\\
&=&
\prod_{1\leqq j \leqq n
\atop{j \notin B}}
g_{11}(z_i/z_j)g_{11}(z_{i+1}/z_j)
\prod_{1\leqq k \leqq n
\atop{k \neq i
\atop{k \in B_{Min}}
}}
g_{12}(x^{-1}z_k/z_i)g_{12}(x^{-1}z_k/z_{i+1})\nonumber\\
&\times&
(T_2(x^{-1}z_i)\delta(x^2z_{i+1}/z_i)-T_2(x^{-1}z_{i+1})
\delta(x^2z_i/z_{i+1})
).
\end{eqnarray}
Substituting this into
the second sum in (\ref{eqn:Sn-InvLocal})
and changing the summation variable $\beta=\alpha+1$,
we conclude
${\cal O}_n(z_1,\cdots,z_i,z_{i+1},\cdots,z_n)\sim
{\cal O}_n(z_1,\cdots,z_{i+1},z_i,\cdots,z_n)$ is
direct consequence of the commutation relation of
$T_1(z)$ in Proposition \ref{def:3dVir}.~~{\it Q.E.D.}\\

Let us set the formal power series 
${\cal A}(z_1,z_2,\cdots,z_n)$ by
\begin{eqnarray}
{\cal A}(z_1,z_2,\cdots,z_n)=\sum_{k_1,\cdots,k_n \in {\mathbb Z}}
a_{k_1,\cdots,k_n}z_1^{k_1}z_2^{k_2}\cdots z_n^{k_n}.
\end{eqnarray}
We define the symbol $[\cdots]_{1,z_1\cdots z_n}$ by
\begin{eqnarray}
\left[
{\cal A}(z_1,z_2,\cdots,z_n)
\right]_{1,z_1\cdots z_n}=a_{0,0,\cdots ,0}.
\end{eqnarray}
Let us set
$D=\{(z_1,\cdots,z_n)\in {\mathbb C}^n|
\sum_{k_1,\cdots,k_n \in {\mathbb Z}}
|a_{k_1,\cdots,k_n}z_1^{k_1}z_2^{k_2}\cdots z_n^{k_n}|<+\infty
\}$.
When we assume closed curve $J$ is contained in $D$,
we have
\begin{eqnarray}
[{\cal A}(z_1,z_2,\cdots,z_n)]_{1,z_1 \cdots z_n}=\int \cdots \int_J 
\prod_{j=1}^n \frac{dz_j}{2\pi\sqrt{-1}z_j}
{\cal A}(z_1,z_2,\cdots,z_n).
\end{eqnarray}

Let us set the auxiliary functions, $s_{11}(z)=s(z),~
h_{11}(z)=h(u)$, and
\begin{eqnarray}
s_{12}(z)=s_{21}(z)=s_{11}(x^{-2}z)s_{11}(x^2z),~
s_{22}(z)=s_{11}(x^{-2}z)s_{11}(z)^2s_{11}(x^{2}z),\\
h_{12}(z)=h_{21}(z)=h_{11}(x^{-2}z)h_{11}(x^2z),~
h_{22}(z)=h_{11}(x^{-2}z)h_{11}(z)^2h_{11}(x^{2}z).
\end{eqnarray}

We will use the following formulae of
the local integrals of motion to
show the commutation relations,
$[{\cal I}_m,{\cal I}_n]=[{\cal I}_m^*,{\cal I}_n^*]=0$.

\begin{thm}~~~
For ${\rm Re}(s)>2$ and ${\rm Re}(r)<0$,
the local integrals of motion ${\cal I}_n$ are written as
\begin{eqnarray}
{\cal I}_n&=&\left[\prod_{1\leqq j<k \leqq n}s(z_k/z_j)
{\cal O}_n(z_1, z_2, \cdots,z_n)\right]_{1,z_1\cdots z_n}.
\end{eqnarray}
For ${\rm Re}(s)>2$ and ${\rm Re}(r^*)>0$,
the local integrals 
of motion ${\cal I}_n^*$ are written as
\begin{eqnarray}
{\cal I}_n^*&=&\left[\prod_{1\leqq j<k \leqq n}s^*(z_k/z_j)
{\cal O}_n(z_1, z_2, \cdots,z_n)\right]_{1,z_1 \cdots z_n}.
\end{eqnarray}
\end{thm}
{\it Proof.}~~~
We start from
\begin{eqnarray}
{\cal I}_n=
\int \cdots \int_C \prod_{j=1}^n
\frac{dz_j}{2\pi \sqrt{-1}z_j}
\prod_{1\leqq j < k\leqq N}h(u_k-u_j)
\prod_{\longrightarrow
\atop{1\leqq j \leqq n}}T_1(z_j),
\end{eqnarray}
where $C$ is given by
\begin{eqnarray}
C:|x^{-2}z_k|,|x^{2+2s}z_k|,|x^{-2r^*}z_k|
<|z_j|<
|x^{-2-2s}z_k|,|x^2z_k|,|x^{2r^*}z_k|~(1\leqq j<k \leqq n).
\end{eqnarray}
Let us take the residue at
$z_n=x^{-2}z_J~(1\leqq J \leqq n-1)$.
We have
\begin{eqnarray}
{\cal I}_n&=&\int \cdots \int_{\widehat{C}(1)}
\prod_{j=1}^n \frac{dz_j}{2\pi \sqrt{-1}z_j}
\prod_{1\leqq j<k\leqq n}h(u_k-u_j)
\prod_{\longrightarrow
\atop{1\leqq j \leqq n}}T_1(z_j)\nonumber
\\
&-&\sum_{J=1}^{n-1}
\left.\int \cdots \int_{\widehat{C}(J)}
\prod_{j=1}^{n-1} \frac{dz_j}{2\pi \sqrt{-1}z_j}
{\rm Res}_{z_n=x^{-2}z_J}
\frac{dz_n}{z_n}
\prod_{1\leqq j<k\leqq n}h(u_k-u_j)
\prod_{\longrightarrow
\atop{1\leqq j \leqq n}}T_1(z_j)~
\right|_{z_n=x^{-2}z_J}.
\end{eqnarray}
Here the contours $\widehat{C}(J)$,
$(1\leqq J\leqq n-1)$ are given by
\begin{eqnarray}
&&|x^{-2}z_k|<|z_n|<
|x^{2-2s}z_k|~~~(J\leqq k \leqq n-1)\nonumber\\
&&|x^{-2}z_k|<|z_n|<|x^{2}z_k|~~~(1\leqq k \leqq J-1)\nonumber\\
&&|x^{-2}z_k|<|z_j|<|x^{2}z_k|~~~(1\leqq j<k \leqq n-1).
\end{eqnarray}
The region $\{(z_k,z_n)\in {\mathbb C}^2|
|x^{-2}z_k|<|z_n|<|x^{2-2s}z_k|\}$ 
for $J\leqq k \leqq n$,
are annulus. 
Hence defining relations of the deformed Virasoro
can be used.
Hence we can use the following relation,
which is derived by defining relations of 
the deformed Virasoro algebra.
\begin{eqnarray}
&&{\rm Res}_{w_2=x^{-2}w_1}\frac{dw_2}{w_2}
\left(
\prod_{j=1}^M h(u_j-v_1)h(v_2-u_j) \right)h(v_2-v_1)
\cdot
T_1(w_1)
\left(\prod_{\longrightarrow
\atop{1\leqq j \leqq M}}T_1(z_j)\right)
T_1(w_2)\nonumber\\
&&=c \prod_{j=1}^M
h_{21}\left(xz_j/w_1\right)
\cdot
T_2(x^{-1}w_1)
\left(
\prod_{\longrightarrow
\atop{1\leqq j \leqq M}}T_1(z_j)
\right).
\end{eqnarray}
By the same arguments as above,
we can use the weak sense relation
\begin{eqnarray}
&&\delta(x^2z_3/z_2)g_{12}(x^{-1}z_2/z_1)T_1(z_1)T_2(x^{-1}z_2)
\sim
\delta(x^2z_3/z_2)g_{21}(xz_1/z_2)T_2(x^{-1}z_2)T_1(z_1).
\nonumber
\end{eqnarray}
Hence we have
\begin{eqnarray}
{\cal I}_n&=&\int \cdots \int_{\widehat{C}(1)}
\prod_{j=1}^n \frac{dz_j}{2\pi \sqrt{-1}z_j}
\prod_{1\leqq j<k\leqq n}h(u_k-u_j)
\prod_{\longrightarrow
\atop{1\leqq j \leqq n}}T_1(z_j)\nonumber
\\
&-c &\sum_{J=1}^{n-1}
\int \cdots \int_{{C}(J)}
\prod_{j=1}^{n-1} \frac{dz_j}{2\pi \sqrt{-1}z_j}
\prod_{1\leqq j<k\leqq n
\atop{j,k \neq J}}h_{11}(z_j/z_k)
\prod_{1\leqq k \leqq n-1
\atop{k \neq J}}h_{12}(x^{-1}z_J/z_k)
\nonumber\\
&\times&
\prod_{\longrightarrow
\atop{1\leqq j \leqq n-1
\atop{j \neq J}}}T_1(z_j)\cdot
T_2(x^{-1}z_J).
\end{eqnarray}
Here the contours $C(J)$ for $1\leqq J \leqq n-1$
are given by
\begin{eqnarray}
|x^{2s-2}z_k|<|z_J|<|x^{-2s+4}z_k|~~(1\leqq k \neq J \leqq n-1),
\nonumber
\\
|x^{-2}z_k|<|z_j|<|x^2z_k|~~(1\leqq j<k \leqq n-1; j,k \neq J).
\end{eqnarray}
Repeating the same arguments for $z_{n-1}, z_{n-2}, \cdots$,
we have this theorem.
~~
Q.E.D.

\subsection{Proof of $[{\cal I}_m,{\cal I}_n]=[{\cal I}_m^*,{\cal I}_n^*]=0$}

In this section we show the commutation relation
$[{\cal I}_m,{\cal I}_n]=[{\cal I}_m^*,{\cal I}_n^*]=0$.

\begin{prop}
\begin{eqnarray}
{\cal O}_n(z_1,\cdots,z_n){\cal O}_m(z_{n+1},\cdots,z_{n+m})
\sim
\prod_{1\leqq j \leqq n
\atop{n+1\leqq k \leqq n+m}}\frac{1}{f_{11}(z_k/z_j)}
{\cal O}_{n+m}(z_1,\cdots,z_{n+m}).
\end{eqnarray}
\end{prop}
{\it Proof}~~~
This is direct consequence of the following explicit formulae
\begin{eqnarray}
&&{\cal O}_{n+m}(z_1,\cdots,z_{n+m})\sim
\prod_{1 \leqq j\leqq n
\atop{n+1\leqq k \leqq n+m}}g_{11}(z_k/z_j)
{\cal O}_n(z_1,\cdots,z_n){\cal O}_m(z_{n+1},\cdots,z_{n+m})\nonumber\\
&+&
\sum_{\alpha=1}^{[\frac{n+m}{2}]}(-1)^\alpha c^\alpha \frac{1}{(\alpha !)^2}
\sum_{L\subset \{1,\cdots,n\}
\atop{|L|=\alpha}}
\sum_{R\subset \{n+1,\cdots,n+m\}
\atop{|R|=\alpha}}
\prod_{\longrightarrow 
\atop{1\leqq j\leqq n+m
\atop{j \notin L \cup R}}}T_1(z_j)
\prod_{\longrightarrow 
\atop{1\leqq j\leqq n+m
\atop{j \in L}}}T_2(x^{-1}z_j)
\prod_{1\leqq t \leqq \alpha}
\delta\left(x^2\frac{z_{R(t)}}{z_{L(t)}}\right)
\nonumber\\
&\times&
\prod_{1\leqq j<k\leqq n+m
\atop{
j,k \notin L \cup R}}g_{11}(z_k/z_j)
\prod_{1\leqq j<k \leqq n+m
\atop{j,k \in L}}g_{22}(z_k/z_j)
\prod_{1\leqq j,k \leqq n+m
\atop{k \in L, j \notin L \cup R}}
g_{12}(x^{-1}z_k/z_j).
\end{eqnarray}
Here we have set
the index $L(t)$ and $R(t)$ by
$L=\{L(1)<L(2)<\cdots<L(\alpha)\}$ and 
$R=\{R(1)<R(2)<\cdots<R(\alpha)\}$.~~
Q.E.D.
\\~\\
{\it Proof~of~Theorem \ref{thm:Local-Com}}~~~
At first we restrict ourself to the regime,
${\rm Re}(s)>2$ and ${\rm Re }(r)<0$, in order to use
the power series formulae of the local integrals of motion,
${\cal I}_n$.
In Proposition \ref{prop:Sn-InvLocal}
we showed
\begin{eqnarray}
\prod_{1\leqq j<k \leqq n}s(z_k/z_j)
{\cal O}_{n}(z_1,\cdots, z_{n})
=
\prod_{1\leqq j<k \leqq n}
s(z_{\sigma(k)}/z_{\sigma(j)})
{\cal O}_{n}
(z_{\sigma(1)},\cdots, z_{\sigma(n)})~~(\sigma \in S_{n}).
\end{eqnarray}
Hence we have
\begin{eqnarray}
&&{\cal I}_n \cdot {\cal I}_m\nonumber\\
&=&
\left[\prod_{1\leqq j<k \leqq n}s(z_k/z_j){\cal O}_n(z_1,\cdots,z_n)
\prod_{n+1\leqq j<k \leqq n+m}s(z_k/z_j)
{\cal O}_m(z_{n+1},\cdots,z_{n+m})\right]_{1,z_1\cdots z_{n+m}}\nonumber
\\&=&
\left[\frac{1}{(n+m)!}
\sum_{\sigma \in S_{n+m}}
\prod_{j=1}^n
\prod_{k=n+1}^{n+m}
\frac{1}{h(u_{\sigma(k)}-u_{\sigma(j)})}
\prod_{1\leqq j<k \leqq n+m}s(z_k/z_j)
{\cal O}_{n+m}(z_1,\cdots, z_{n+m})
\right]_{1,z_1 \cdots z_{n+m}}.\nonumber\\
\end{eqnarray}
Hence the commutation relation
${\cal I}_n \cdot {\cal I}_m={\cal I}_m\cdot
{\cal I}_n$ is reduced to the following theta identity.
\begin{eqnarray}
{\rm LHS}(n,m)={\rm RHS}(n,m)
\end{eqnarray}
where we have set
\begin{eqnarray}
{\rm LHS}(n,m)&=&\sum_{J \subset \{1,2,\cdots,n+m\}
\atop{|J|=n}}
\prod_{j \in J}
\prod_{k \notin J}
\frac{[u_k-u_j+1]_s[u_k-u_j+r^*]_s}{[u_{k}-u_{j}]_s[u_k-u_j+r]_s},
\\
{\rm RHS}(n,m)&=&\sum_{J^c \subset \{1,2,\cdots,n+m\}
\atop{|J^c|=m}}
\prod_{j \in J^c}
\prod_{k \notin J^c}
\frac{[u_k-u_j+1]_s[u_k-u_j+r^*]_s}{[u_{k}-u_{j}]_s[u_k-u_j+r]_s}.
\end{eqnarray}
where we have used $h(u)=\frac{[u]_s [u+r]_s}{[u+1]_s[u+r^*]_s}$.
LHS$(n,m)$ and RHS$(n,m)$ are an elliptic functions.
Therefore, from Liouville thorem, it is enough 
to check whether all the residues of LHS$(n,m)$ and RHS$(n,m)$ coincide or not.
Candidates of poles are
$u_\alpha=u_\beta~(\alpha \neq \beta)$ 
and $u_\alpha=u_\beta-r~(\alpha \neq \beta)$.
Let us consider $u_\alpha=u_\beta~(\alpha \neq \beta)$ 
\begin{eqnarray}
{\rm LHS}(n,m)&=&\left(
\frac{[u_\alpha-u_\beta+1]_s
[u_\alpha-u_\beta+r^*]_s}{[u_\alpha-u_\beta]_s
[u_\alpha-u_\beta-r]_s}+
\frac{[u_\beta-u_\alpha+1]_s[u_\beta-u_\alpha+r^*]_s}{
[u_\beta-u_\alpha]_s[u_\beta-u_\alpha-r]_s}\right)
\nonumber\\
&\times&\sum_{J\cup J^c\{1,\cdots,n+m\}-\{\alpha,\beta\}
\atop{|J|=n-1}}
\prod_{j \in J}
\prod_{k \notin J}
\frac{[u_k-u_j+1]_s[u_k-u_j+r^*]_s}{[u_{k}-u_{j}]_s[u_k-u_j+r]_s}
\end{eqnarray}
Hence we have ${\rm Res}_{u_\alpha=u_\beta}{\rm LHS}(n,m)=0$.
As the same manner
we have ${\rm Res}_{u_\alpha=u_\beta}{\rm RHS}(n,m)=0$.
Therefore $u_\alpha=u_{\beta}$ is not pole.
We only have to consider poles 
$u_\alpha=u_\beta~(\alpha \neq \beta)$.
We show the ${\rm LHS}(n,m)={\rm RHS}(n,m)$ by the induction
of the number $n+m$.
We assume $n>m\geqq 1$ without loosing generality.
At first we show the starting point $n>m=1$.
\begin{eqnarray}
\sum_{k=1}^{n+1}\prod_{j=1
\atop{j\neq k}}^{n+1}\frac{
[u_j-u_k+1]_s[u_j-u_k+r^*]_s}{[u_j-u_k]_s[u_j-u_k+r]_s}
=
\sum_{k=1}^{n+1}\prod_{j=1
\atop{j\neq k}}^{n+1}\frac{
[u_k-u_j+1]_s[u_k-u_j+r^*]_s}{[u_k-u_j]_s[u_k-u_j+r]_s}
\end{eqnarray}
Both LHS$(n,1)$ and RHS$(n,1)$ have simple poles
at $u_\alpha=u_\beta-r~(\alpha \neq \beta)$
modulo ${\mathbb Z}+{\mathbb Z}\tau$.
Beacuse both LHS$(n,1)$ and RHS$(n,1)$ are symmetric with respect with
$u_1,u_2,\cdots,u_{n+1}$,
it is enough to check the pole at $u_2=u_1-r$. We have
\begin{eqnarray}
&&{\rm Res}_{~u_2=u_1-r}
{\rm LHS}(n,1)={\rm Res}_{~u_2=u_1-r}
{\rm RHS}(n,1)\nonumber\\
&&=
{\rm Res}_{u=0}\frac{[-r^*]_s[-1]_s}{[-r]_s[u]_s}
\prod_{j=3}^{n+1}\frac{[u_j-u_1+1]_s[u_j-u_1+r^*]_s}
{[u_j-u_1]_s[u_j-u_1+r]_s}.
\end{eqnarray}
We have shown $n>m=1$ case.
We show general $n>m\geqq 1$ case.
We assume the equation ${\rm LHS}(n,1)={\rm RHS}(n,1)$
for some $(m,n)$.
Beacuse both LHS$(n+1,m+1)$ and RHS$(n+1,m+1)$ are symmetric with respect with
$u_1,u_2,\cdots,u_{n+m+2}$,
it is enough to check the pole at $u_2=u_1-r$.
Let us take the residue at $u_2=u_1-r$ for $(m+1,n+1)$.
\begin{eqnarray}
{\rm Res}_{~u_2=u_1-r}
&&\left(
\sum_{J \subset \{1,2,\cdots,n+m+2\}
\atop{|J|=n+1}}
\prod_{j \in J}
\prod_{k \notin J}
\frac{[u_k-u_j+1]_s[u_k-u_j+r^*]_s}{[u_{k}-u_{j}]_s[u_k-u_j+r]_s}
\right.
\nonumber\\
&&\left.
-\sum_{J^c \subset \{1,2,\cdots,n+m+2\}
\atop{|J^c|=m+1}}
\prod_{j \in J^c}
\prod_{k \notin J^c}
\frac{[u_k-u_j+1]_s[u_k-u_j+r^*]_s}{[u_{k}-u_{j}]_s[u_k-u_j+r]_s}.
\right)\nonumber\\
&=&
\prod_{j=3}^{n+m+2}
\frac{[u_j-u_1+1]_s[u_j-u_1+r^*]_s}{
[u_j-u_1]_s[u_j-u_1+r]_s}\\
&\times&\left(
\sum_{J \subset \{3,4,\cdots,n+m+2\}
\atop{|J|=n}}
\prod_{j \in J}
\prod_{k \notin J}
\frac{[u_k-u_j+1]_s[u_k-u_j+r^*]_s}{[u_{k}-u_{j}]_s[u_k-u_j+r]_s}
\right.
\nonumber\\
&&\left.
-\sum_{J^c \subset \{3,4,\cdots,n+m+2\}
\atop{|J^c|=m}}
\prod_{j \in J^c}
\prod_{k \notin J^c}
\frac{[u_k-u_j+1]_s[u_k-u_j+r^*]_s}{[u_{k}-u_{j}]_s[u_k-u_j+r]_s}.
\right)=0.\nonumber
\end{eqnarray}
We have used the assumption of induction 
${\rm LHS}(n,m)={\rm RHS}(n,m)$.
We have proved the equation $[{\cal I}_n,{\cal I}_m]=0$ for general 
$n>m \geqq 1$.
Generalization to generic parameter case : ${\rm Re}(s)>0$ and
$r \in {\mathbb C}$,
should be understood as analytic continuation.
The commutation relation
$[{\cal I}_n^*,{\cal I}_m^*]=0$ is shown by the same manner
as above.
~~Q.E.D.
%As the same manner, we have sufficient condition
%for $[{\cal I}_m,{\cal I}_n^*]=0$ as follows.
%\begin{eqnarray}
%&&\sum_{\sigma \in S_{n+m}}\prod_{i=1}^m \prod_{j=m+1}^{m+n}
%\frac{1}{h(u_{\sigma(j)}-u_{\sigma(i)})}
%\prod_{m+1\leqq i<j \leqq m+n}
%\frac{h^*(u_{\sigma(j)}-u_{\sigma(i)})}{h(
%u_{\sigma(j)}-u_{\sigma(i)})}\nonumber\\
%&=&
%\sum_{\sigma \in S_{n+m}}\prod_{i=1}^n \prod_{j=n+1}^{m+n}
%\frac{1}{h(u_{\sigma(j)}-u_{\sigma(i)})}
%\prod_{1 \leqq i<j \leqq n}
%\frac{h^*(u_{\sigma(j)}-u_{\sigma(i)})}{h(
%u_{\sigma(j)}-u_{\sigma(i)})}.
%\end{eqnarray}

\subsection{Proof of Dynkin Automorphism Invariance
$\eta({\cal I}_n)={\cal I}_n$}

In this section we show the Dynkin automorphism invariance
$\eta({\cal I}_n)={\cal I}_n$.
Before showing the complete proof, we consider
some simple examples as warming-up exercise.
We study the case :
${\rm Re}(s)>2$ and $1-{\rm Re}(s)<{\rm Re}(r)<0$.
Let us study $n=2$ case.
The following relation is convenient for our calculation.
\begin{eqnarray}
h_{11}(z)-h_{11}(x^{2s}z)&=&
c_h (\delta(x^2 z)-
\delta(x^{2r-2+2s}z)),\\
c_h&=&
-\frac{(x^2;x^{2s})_\infty
(x^{2r-2};x^{2s})_\infty
(x^{2s-2};x^{2s})_\infty
(x^{2s-2r+2};x^{2s})_\infty
}{
(x^{2r-4};x^{2s})_\infty
(x^{2s};x^{2s})_\infty
(x^{2s};x^{2s})_\infty
(x^{2s-2r+4};x^{2s})_\infty
}.\nonumber
\end{eqnarray}
Using the above relation, we have
\begin{eqnarray}
&&\eta(h_{11}(z_2/z_1)T_1(z_1)T_1(z_2))
=h_{11}(z_2/z_1)(\Lambda_1(x^{-s}z_1)\Lambda_1(x^{-s}z_2)
+\Lambda_2(x^{s}z_1)\Lambda_2(x^{s}z_2))\nonumber\\
&+&
h_{11}(x^{2s}z_2/z_1)\Lambda_1(x^{-s}z_1)\Lambda_2(x^sz_2)
+h_{11}(x^{-2s}z_2/z_1)\Lambda_2(x^{s}z_1)\Lambda_1(x^{-s}z_2)\nonumber\\
&+&c_h (\delta(x^2z_2/z_1)-\delta(x^{2r-2+2s}z_2/z_1))
\Lambda_1(x^{-s}z_1)\Lambda_2(x^sz_2)\nonumber\\
&-&c_h (\delta(x^{2-2s}z_2/z_1)-\delta(x^{2r-2}z_2/z_1))
\Lambda_2(x^{s}z_1)\Lambda_1(x^{-s}z_2).
\end{eqnarray}
The delta-function yields
\begin{eqnarray}
&&\delta(x^{2r-2+2s}z_2/z_1)
\Lambda_1(x^{-s}z_1)\Lambda_2(x^sz_2)=
\delta(x^{2r-2}z_2/z_1)
\Lambda_2(x^{s}z_1)\Lambda_1(x^{-s}z_2)=0,\\
&&c_h \delta(x^{2-2s}z_2/z_1)
\Lambda_2(x^{s}z_1)\Lambda_1(x^{-s}z_2)=
c s(x^{-2}) \delta(x^{2-2s}z_2/z_1)
T_2(x^{s-1}z_1),\\
&&c_h \delta(x^2z_2/z_1)\Lambda_1(x^{-s}z_1)\Lambda_2(x^sz_2)
=c s(x^{-2}) \delta(x^2z_2/z_1) \eta(T_2(x^{-1}z_1)).
\end{eqnarray}
Hence we have $\eta({\cal I}_2)={\cal I}_2$.
Let us study $n=3$ case.
In what follows we use the following abbreviation.
\begin{eqnarray}
h_{12}(z)=h_{11}(x^{-1}z)h_{11}(xz),
~~~h_{12}^{\eta}(z)=h_{11}(x^{s-1}z)h_{11}(x^{-s+1}z).
\end{eqnarray}
We have
\begin{eqnarray}
&&\eta\left(\left[\prod_{1\leqq j<k\leqq 3}
h_{11}(z_k/z_j)
\Lambda_2(z_1)\Lambda_2(z_2)\Lambda_1(z_3)
\right]_{1,z_1 z_2 z_3}\right)\nonumber\\
&&=
\left[\prod_{1\leqq j<k\leqq 3}
h_{11}(z_k/z_j)
\Lambda_1(z_1)\Lambda_1(z_2)\Lambda_2(z_3)
\right]_{1,z_1 z_2 z_3}\nonumber
\\
&&+\left[c s(x^{-2})h_{12}^\eta(x^{-1}z_1/z_2)
\delta(x^2z_3/z_1)\Lambda_1(z_2)\eta(T_2(x^{-1}z_1))
\right]_{1,z_1 z_2 z_3}\nonumber\\
&&+\left[c s(x^{-2})h_{12}^\eta(x^{-1}z_2/z_1)
\delta(x^2z_3/z_2)\Lambda_1(z_1)\eta(T_2(x^{-1}z_2))
\right]_{1,z_1 z_2 z_3}.
\end{eqnarray}
Moreover we have
\begin{eqnarray}
&&\eta\left(\left[\prod_{1\leqq j<k\leqq 3}
h_{11}(z_k/z_j)
T_1(z_1)T_1(z_2)T_1(z_3)
\right]_{1,z_1 z_2 z_3}\right)\nonumber\\
&&=
\left[\prod_{1\leqq j<k\leqq 3}h_{11}(z_k/z_j)
T_1(z_1)T_1(z_2)T_1(z_3)\right]_{1,z_1 z_2 z_3}\nonumber
\\
&&-\left[c s(x^{-2})h_{12}(x^{-1}z_2/z_1)
\delta(x^2z_3/z_2)T_1(z_1)T_2(x^{-1}z_2)
\right]_{1,z_1 z_2 z_3}\nonumber\\
&&-\left[c s(x^{-2})h_{12}(x^{-1}z_3/z_1)
\delta(x^2z_3/z_1)T_1(z_2)T_2(x^{-1}z_1)
\right]_{1,z_1 z_2 z_3}\nonumber\\
&&-\left[c s(x^{-2})h_{12}(x^{-1}z_1/z_3)
\delta(x^2z_2/z_1)T_1(z_3)T_2(x^{-1}z_1)
\right]_{1,z_1 z_2 z_3}\nonumber\\
&&+\left[c s(x^{-2})h_{12}^\eta(x^{-1}z_2/z_1)
\delta(x^2z_3/z_2)T_1(z_1)\eta(T_2(x^{-1}z_2))
\right]_{1,z_1 z_2 z_3}\nonumber\\
&&+\left[c s(x^{-2})h_{12}^\eta(x^{-1}z_3/z_1)
\delta(x^2z_3/z_1)T_1(z_2)\eta(T_2(x^{-1}z_1))
\right]_{1,z_1 z_2 z_3}\nonumber\\
&&+\left[c s(x^{-2})h_{12}^\eta(x^{-1}z_1/z_3)
\delta(x^2z_2/z_1)T_1(z_3)\eta(T_2(x^{-1}z_1))
\right]_{1,z_1 z_2 z_3}.
\end{eqnarray}
As the same manner, we have
\begin{eqnarray}
&&\eta([h_{12}(x^{-1}z_2/z_1)\delta(x^2z_3/z_2)
T_1(z_1)T_2(x^{-1}z_2)]_{1,z_1 z_2 z_3})
\nonumber\\
&=&
[h_{12}^\eta(x^{-1}z_2/z_1)\delta(x^2z_3/z_2)T_1(z_1)
\eta(T_2(x^{-1}z_2))]_{1,z_1 z_2 z_3}.
\end{eqnarray}
Summing up all the above, we have $\eta({\cal I}_3)={\cal I}_3$.
\\
\\
{\it Proof.}~~
Let us start general $n$ case.
For a while we study :
${\rm Re}(s)>2$ and $1-{\rm Re}(s)<{\rm Re}(r)<0$.
In what follows we use the following abberiviations.
\begin{eqnarray}
&&
h_{22}^\eta(z)=
h_{11}(x^{-s}z)
h_{11}(x^{-s+2}z)
h_{11}(x^{s-2}z)h_{11}(x^sz),\\
&&h_{22}^{\eta \eta}(z)
=h_{11}(x^{-2s+2}z)h_{11}(z)^2h_{11}(x^{2s-2}z).
\end{eqnarray}
We have
\begin{eqnarray}
&&\eta\left(\left[\prod_{1\leqq j<k \leqq n}
h_{11}(z_k/z_j)
T_1(z_1)T_1(z_2)\cdots T_1(z_n)\right]_{1,z_1 \cdots z_n}\right)\nonumber\\
&=&\left[\sum_{\alpha, \beta \geqq 0
\atop{0\leqq \alpha+\beta \leqq [\frac{n}{2}]}}
(-1)^\alpha (c s(x^{-2}))^{\alpha+\beta}
%\frac{1}{\alpha ! \beta !}
\sum_{A_1,\cdots,A_\alpha,B_1,\cdots,B_\beta}
\prod_{1\leqq j \leqq \alpha}
\delta\left(x^2\frac{z_{Max(A_j)}}{z_{Min(A_j)}}\right)
\prod_{1\leqq j \leqq \beta}
\delta\left(x^2\frac{z_{Max(A_k)}}{z_{Min(A_j)}}\right)\right.
\nonumber\\
&\times&
\prod_{\longrightarrow
\atop{j \notin A \cup B}
}T_1(z_j)
\prod_{\longrightarrow
\atop{j \in A_{Min}}
}T_2(x^{-1}z_j)
\prod_{\longrightarrow
\atop{j \in B_{Min}}
}\eta(T_2(x^{-1}z_j))\times\prod_{j<k
\atop{j,k \notin A\cup B}}
h_{11}(z_k/z_j)
\prod_{j\notin A\cup B
\atop{k \in A_{Min}}}
h_{12}(z_k/z_j)
\nonumber\\
&\times&\left.
\prod_{j\notin A\cup B
\atop{k \in B_{Min}}}
h_{12}^\eta(z_k/z_j)\prod_{j<k
\atop{j,k \in A_{Min}}}
h_{22}(z_k/z_j)
\prod_{j<k
\atop{j,k \in B_{Min}}}
h_{22}^{\eta \eta}(z_k/z_j)
\prod_{j<k
\atop{j\in A_{Min}
\atop{k \in B_{Min}}}}
h_{22}^\eta(z_k/z_j)\right]_{1,z_1 \cdots z_n}.\nonumber\\
\label{eqn:Dyn-Local1}
\end{eqnarray}
We explain the notation of the above formulae (\ref{eqn:Dyn-Local1}).
The summation
$\sum_{A_1,\cdots,A_\alpha,B_1,\cdots,B_\beta}$ 
is taken over the set
$A_1,A_2,\cdots,A_\alpha,
B_1,B_2,\cdots,B_\beta
 \subset
\{1,2,\cdots,n\}$
such that 
$A_j \cap A_k=\phi~(1\leqq j\neq k \leqq \alpha)$,
$B_j \cap B_k=\phi~(1\leqq j\neq k \leqq \beta)$,
$|A_j|=|B_k|=2~(1\leqq j \leqq \alpha, 1\leqq k \leqq \beta)$,
and $Min(A_j)<Min(A_k)~(1\leqq j<k \leqq \alpha)$,
$Min(B_j)<Min(B_k)~(1\leqq j<k \leqq \beta)$.
We have set
$A_{Min}=\{Min(A_1),\cdots, Min(A_\alpha)\}$,
$A_{Max}=\{Max(A_1),\cdots, Max(A_\alpha)\}$, 
$B_{Min}=\{Min(B_1),\cdots, Min(B_\beta)\}$, 
$B_{Min}=\{Min(B_1),\cdots, Min(B_\beta)\}$, 
and $A=A_1 \cup \cdots \cup A_\alpha$,
$B=B_1 \cup \cdots \cup B_\beta$.
We have
\begin{eqnarray}
&&\eta\left(\left[\prod_{1\leqq j<k \leqq n}
h_{11}(z_k/z_j)
\prod_{1\leqq j \leqq n
\atop{n+1 \leqq k \leqq n+m}}h_{12}(x^{-1}z_k/z_j)
\prod_{n+1\leqq j<k \leqq n+m}h_{22}(z_k/z_j)
\right.\right.\nonumber\\
&&\times \left.\left.
T_1(z_1)\cdots T_1(z_n)
T_2(x^{-1}z_{n+1})
\cdots T_2(x^{-1}z_{n+m})
\prod_{1\leqq j \leqq m}
\delta(x^2z_{n+m+j}/z_{n+j})
\right]_{1,z_1 \cdots z_{n+m}}\right)\nonumber\\
&=&\left[\sum_{\alpha, \beta \geqq 0
\atop{0\leqq \alpha+\beta \leqq [\frac{n}{2}]}}
(-1)^\alpha (c s(x^{-2}))^{\alpha+\beta}
%\frac{1}{\alpha ! \beta !}
\sum_{A_1,\cdots,A_\alpha,B_1,\cdots,B_\beta}
\prod_{1\leqq j \leqq \alpha}
\delta\left(x^2\frac{z_{Max(A_j)}}{z_{Min(A_j)}}\right)
\prod_{1\leqq j\leqq \beta}
\delta\left(x^2\frac{z_{Max(A_j)}}{z_{Min(A_j)}}\right)
\right.\nonumber\\
&\times&\prod_{1\leqq j \leqq m}
\delta\left(x^2\frac{z_{n+m+j}}{z_{n+j}}\right)
\times
\prod_{\longrightarrow
\atop{1\leqq j \leqq n
\atop{j \notin A \cup B}}
}T_1(z_j)
\prod_{\longrightarrow
\atop{j \in A_{Min}}
}T_2(x^{-1}z_j)
\prod_{\longrightarrow
\atop{j \in B_{Min} \cup \{n+1,\cdots,n+m\}}
}\eta(T_2(x^{-1}z_j))\nonumber\\
&\times&
\prod_{1\leqq j<k\leqq n
\atop{j,k \notin A\cup B}}
h_{11}(z_k/z_j)
\prod_{1\leqq j \leqq n
\atop{j\notin A\cup B
\atop{k \in A_{Min}}}}
h_{12}(x^{-1}z_k/z_j)
\prod_{1\leqq j \leqq n
\atop{j\notin A\cup B
\atop{k \in B_{Min}\cup\{n+1,\cdots,n+m\}}}}
h_{12}^\eta(x^{-1}z_k/z_j)\nonumber
\\
&\times&\left.
\prod_{j<k
\atop{j,k \in A_{Min}}}
h_{22}(z_k/z_j)
\prod_{j<k
\atop{j\in A_{Min}
\atop{k \in B_{Min}\cup\{n+1,\cdots,n+m\}}}}
h_{22}^{\eta}(z_k/z_j)
\prod_{j<k
\atop{j,k \in B_{Min}\cup\{n+1,\cdots,n+m\}}}
h_{22}^{\eta \eta}(z_k/z_j)
\right]_{1,z_1 \cdots z_{n+m}}.\nonumber\\
\label{eqn:Dyn-Local2}
\end{eqnarray}
Using the formulae (\ref{def:Sn-InvLocal}), 
(\ref{eqn:Dyn-Local1}) and (\ref{eqn:Dyn-Local2}), we have 
$\eta({\cal I}_n)={\cal I}_n$
for general $n$.
Cancellations of coefficients of
$\prod T_1 \prod T_2 \prod \eta(T_2)$
come from relation
$0=(1-1)^k=\sum_{l=0}^k (-1)^k~_kC_l$.
Generalization to generic parameter ${\rm Re}(s)>0$ and $r \in {\mathbb C}$
should be understood as analytic continuation.
~~~{Q.E.D.}

\section{Nonlocal Integrals of Motion ${\cal G}_n$}

In this section we explicitly construct 
the nonlocal integrals of motion ${\cal G}_n$ and ${\cal G}_n^*$.
In this section we study generic case :
$0<x<1$, ${\rm Re}(r) \neq 0$ and $s \in {\mathbb C}$ (
resp. $0<x<1$, ${\rm Re}(r^*) \neq 0$ and $s \in {\mathbb C}$).
Let us use the parametrization :
$z_j=x^{2u_j}, w_j=x^{2v_j}$.

\subsection{Nonlocal Integrals of Motion ${\cal G}_n$}

We explicitly construct 
the nonlocal integrals of motion ${\cal G}_n$ and ${\cal G}_n^*$,
and state the main results.

Let us set the theta function $\vartheta_{\alpha,r}(u)$ by
\begin{eqnarray}
\vartheta_{\alpha,r}(u)=[u-\hat{\pi}+\alpha]_r
[u-\alpha]_r+[u-\hat{\pi}-\alpha]_r[u+\alpha]_r.
\end{eqnarray}
This function $\vartheta_{\alpha,r}(u)$ satisfies
\begin{eqnarray}
\vartheta_{\alpha,r}(u+r\tau)=e^{-2\pi\sqrt{-1}\tau+
\frac{2\pi\sqrt{-1}}{r}(\hat{\pi}+2u)}\vartheta_{\alpha,r}(u),~~
\vartheta_{\alpha,r}(u)=\vartheta_{-\alpha,r}(u).
\end{eqnarray}

%%%%%%%%%%%%%%%%%%%%%%%%%%%%%%%%%%%%%%%%%%%%%%%%%%%%%%%
%%%%%%%%%%%%%%%%%%%%%%%%%%%%%%%%%%%%%%%%%%%%%%%%%%%%%%%
\begin{df}~\\
$\bullet$~We define ${\cal G}_n~(n=1,2,\cdots)$ 
for the regime ${\rm Re}(r)>0$ and 
$0<{\rm Re}(s)<2$ by
\begin{eqnarray}
{\cal G}_n&=&\int \cdots \int_{I}
\prod_{j=1}^n\frac{dz_j}{2\pi\sqrt{-1}z_j}
\prod_{j=1}^n\frac{dw_j}{2\pi\sqrt{-1}w_j}
\prod_{\longrightarrow
\atop{1\leqq j \leqq n}}F_1(z_j)
\prod_{\longrightarrow
\atop{1\leqq j \leqq n}}F_0(w_j)\nonumber\\
&\times&
\frac{\prod_{1\leqq i<j \leqq n}
[u_i-u_j]_r[u_j-u_i-1]_r[v_i-v_j]_r[v_j-v_i-1]_r
}{
\prod_{i,j=1}^n
[u_i-v_j+\frac{s}{2}]_r[v_j-u_i+\frac{s}{2}-1]_r}
\vartheta_{\alpha,r}\left(
\sum_{j=1}^n(u_j-v_j)\right).\nonumber\\
\end{eqnarray}
Here the contour $I$ encircles
$z_i, w_i=0$ in such a way that\\
(1) $z_j=x^{s+2lr}w_i, x^{2-s+2lr}w_i~(l=0,1,2,\cdots)$ is inside and 
$z_j=x^{-s-2lr}w_i, x^{s-2-2lr}w_i~(l=0,1,2,\cdots)$ is outside
for $i,j=1,2,\cdots,n$,\\
(2) $z_p=x^{2r^*+2lr}z_q~(l=0,1,2,\cdots)$ is inside 
$z_p=x^{-2r^*-2lr}z_q~(l=0,1,2,\cdots)$ is outside for $1\leqq p<q\leqq n$,\\
(3) $w_p=x^{2r^*+2lr}w_q~(l=0,1,2,\cdots)$ is inside 
$w_p=x^{-2r^*-2lr}w_q~(l=0,1,2,\cdots)$ is outside for $1\leqq p<q\leqq n$.\\\\
%%%%%%%%%%%%%%%%%%%%%%%%%%%%%%%%%%%%%%%%%%%%%%%%%%%%%%
%%%%%%%%%%%%%%%%%%%%%%%%%%%%%%%%%%%%%%%%%%%%%%%%%%%%%%
$\bullet$~We define ${\cal G}_n~(n=1,2,\cdots)$ 
for the regime ${\rm Re}(r)<0$ and 
$0<{\rm Re}(s)<2$ by
\begin{eqnarray}
{\cal G}_n&=&\int \cdots \int_{I}
\prod_{j=1}^n\frac{dz_j}{2\pi\sqrt{-1}z_j}
\prod_{j=1}^n\frac{dw_j}{2\pi\sqrt{-1}w_j}
\prod_{\longrightarrow
\atop{1\leqq j \leqq n}}F_1(z_j)
\prod_{\longrightarrow
\atop{1\leqq j \leqq n}}F_0(w_j)\nonumber\\
&\times&
\frac{\prod_{1\leqq i<j \leqq n}
[u_i-u_j]_{-r}[u_j-u_i+1]_{-r}[v_i-v_j]_{-r}[v_j-v_i+1]_{-r}
}{
\prod_{i,j=1}^n
[u_i-v_j-\frac{s}{2}]_{-r}[v_j-u_i-\frac{s}{2}+1]_{-r}}
\vartheta_{\alpha,-r}\left(
\sum_{j=1}^n(v_j-u_j)\right).\nonumber\\
\end{eqnarray}
Here the contour $I$ encircles
$z_i, w_i=0$ in such a way that\\
(1) $z_j=x^{s-2lr}w_i, x^{2-s-2lr}w_i~(l=0,1,2,\cdots)$ is inside and 
$z_j=x^{-s+2lr}w_i, x^{s-2+2lr}w_i~(l=0,1,2,\cdots)$ is outside
for $i,j=1,2,\cdots,n$,\\
(2) $z_p=x^{-2r^*-2lr}z_q~(l=0,1,2,\cdots)$ is inside 
$z_p=x^{2r^*+2lr}z_q~(l=0,1,2,\cdots)$ is outside for $1\leqq p<q\leqq n$,\\
(3) $w_p=x^{-2r^*-2lr}w_q~(l=0,1,2,\cdots)$ is inside 
$w_p=x^{2r^*+2lr}w_q~(l=0,1,2,\cdots)$ is outside for $1\leqq p<q\leqq n$.\\\\
%%%%%%%%%%%%%%%%%%%%%%%%%%%%%%%%%%%%%%%%%%%%%%
%%%%%%%%%%%%%%%%%%%%%%%%%%%%%%%%%%%%%%%%%%%%%%
$\bullet$~We define ${\cal G}_n^*~(n=1,2,\cdots)$ 
for the regime ${\rm Re}(r^*)>0$ and 
$0<{\rm Re}(s)<2$ by
\begin{eqnarray}
{\cal G}_n^*&=&\int \cdots \int_{I^*}
\prod_{j=1}^n\frac{dz_j}{2\pi\sqrt{-1}z_j}
\prod_{j=1}^n\frac{dw_j}{2\pi\sqrt{-1}w_j}
\prod_{\longrightarrow
\atop{1\leqq j \leqq n}}E_1(z_j)
\prod_{\longrightarrow
\atop{1\leqq j \leqq n}}E_0(w_j)\nonumber\\
&\times&
\frac{\prod_{1\leqq i<j \leqq n}
[u_i-u_j]_{r^*}[u_j-u_i+1]_{r^*}
[v_i-v_j]_{r^*}[v_j-v_i+1]_{r^*}
}{
\prod_{i,j=1}^n
[u_i-v_j-\frac{s}{2}]_{r^*}[v_j-u_i-\frac{s}{2}+1]_{r^*}}
\vartheta_{\alpha,r^*}\left(
\sum_{j=1}^n(v_j-u_j)\right).\nonumber\\
\end{eqnarray}
Here the contour $I^*$ encircles
$z_i, w_i=0$ in such a way that\\
(1) $z_j=x^{s+2lr^*}w_i, x^{2-s+2lr^*}w_i~(l=0,1,2,\cdots)$ is inside and 
$z_j=x^{-s-2lr^*}w_i, x^{s-2-2lr^*}w_i~(l=0,1,2,\cdots)$ is outside
for $i,j=1,2,\cdots,n$,\\
(2) $z_p=x^{2r+2lr^*}z_q~(l=0,1,2,\cdots)$ is inside 
$z_p=x^{-2r-2lr^*}z_q~(l=0,1,2,\cdots)$ is outside for $1\leqq p<q\leqq n$,\\
(3) $w_p=x^{2r+2lr^*}w_q~(l=0,1,2,\cdots)$ is inside 
$w_p=x^{-2r-2lr^*}w_q~(l=0,1,2,\cdots)$ is outside for $1\leqq p<q\leqq n$.\\\\
%%%%%%%%%%%%%%%%%%%%%%%%%%%%%%%%%%%%%%%%%%%%
%%%%%%%%%%%%%%%%%%%%%%%%%%%%%%%%%%%%%%%%%%%%
$\bullet$~We define ${\cal G}_n^*~(n=1,2,\cdots)$ 
for the regime ${\rm Re}(r^*)<0$ and 
$0<{\rm Re}(s)<2$ by
\begin{eqnarray}
{\cal G}_n^*&=&\int \cdots \int_{I^*}
\prod_{j=1}^n\frac{dz_j}{2\pi\sqrt{-1}z_j}
\prod_{j=1}^n\frac{dw_j}{2\pi\sqrt{-1}w_j}
\prod_{\longrightarrow
\atop{1\leqq j \leqq n}}E_1(z_j)
\prod_{\longrightarrow
\atop{1\leqq j \leqq n}}E_0(w_j)\nonumber\\
&\times&
\frac{\prod_{1\leqq i<j \leqq n}
[u_i-u_j]_{-r^*}[u_j-u_i-1]_{-r^*}
[v_i-v_j]_{-r^*}[v_j-v_i-1]_{-r^*}
}{
\prod_{i,j=1}^n
[u_i-v_j+\frac{s}{2}]_{-r^*}
[v_j-u_i+\frac{s}{2}-1]_{-r^*}}
\vartheta_{\alpha,-r^*}\left(
\sum_{j=1}^n(u_j-v_j)\right).\nonumber\\
\end{eqnarray}
Here the contour $I^*$ encircles
$z_i, w_i=0$ in such a way that\\
(1) $z_j=x^{s-2lr^*}w_i, x^{2-s-2lr^*}w_i~(l=0,1,2,\cdots)$ is inside and 
$z_j=x^{-s+2lr^*}w_i, x^{s-2+2lr^*}w_i~(l=0,1,2,\cdots)$ is outside
for $i,j=1,2,\cdots,n$,\\
(2) $z_p=x^{-2r-2lr^*}z_q~(l=0,1,2,\cdots)$ is inside 
$z_p=x^{2r+2lr^*}z_q~(l=0,1,2,\cdots)$ is outside for $1\leqq p<q\leqq n$,\\
(3) $w_p=x^{-2r-2lr^*}w_q~(l=0,1,2,\cdots)$ is inside 
$w_p=x^{2r+2lr^*}w_q~(l=0,1,2,\cdots)$ is outside for $1\leqq p<q\leqq n$.\\\\
We call ${\cal G}_n$ and ${\cal G}_n^*$ 
the nonlocal integrals of motion
for the deformed Virasoro algebra.
The definitions of ${\cal G}_n$
and ${\cal G}_n^*$ for generic $s \in {\mathbb C}$,
should be understood as analytic continuation.
\end{df}

~\\
{\bf Example}~~For ${\rm Re}(r)>0$ and $0<{\rm Re}(s)<2$ we have
\begin{eqnarray}
{\cal G}_1=\int \int_I \frac{dz_1}{2\pi \sqrt{-1}z_1}
\frac{dz_2}{2\pi \sqrt{-1}z_2}F_1(z_1)F_0(z_2)\frac{\vartheta_{\alpha,r}(u_1-u_2)}{[u_1-u_2+\frac{s}{2}]_r[u_1-u_2-\frac{s}{2}+1]_r}.
\end{eqnarray}
For ${\rm Re}(r)<0$ and $0<{\rm Re}(s)<2$ we have
\begin{eqnarray}
{\cal G}_1=\int \int_I \frac{dz_1}{2\pi \sqrt{-1}z_1}
\frac{dz_2}{2\pi \sqrt{-1}z_2}F_1(z_1)F_0(z_2)
\frac{\vartheta_{\alpha,-r}(u_2-u_1)}{
[u_1-u_2-\frac{s}{2}]_{-r}[u_1-u_2+\frac{s}{2}-1]_{-r}}.
\end{eqnarray}
For ${\rm Re}(r^*)>0$ and $0<{\rm Re}(s)<2$ we have
\begin{eqnarray}
{\cal G}_1^*=\int \int_I \frac{dz_1}{2\pi \sqrt{-1}z_1}
\frac{dz_2}{2\pi \sqrt{-1}z_2}E_1(z_1)E_0(z_2)
\frac{\vartheta_{\alpha,r^*}(u_2-u_1)}{
[u_1-u_2-\frac{s}{2}]_{r^*}[u_1-u_2+\frac{s}{2}-1]_{r^*}}.
\end{eqnarray}
For ${\rm Re}(r^*)<0$ and $0<{\rm Re}(s)<2$ we have
\begin{eqnarray}
{\cal G}_1^*=\int \int_I \frac{dz_1}{2\pi \sqrt{-1}z_1}
\frac{dz_2}{2\pi \sqrt{-1}z_2}E_1(z_1)E_0(z_2)
\frac{\vartheta_{\alpha,-r^*}(u_1-u_2)}
{[u_1-u_2+\frac{s}{2}]_{-r^*}[u_1-u_2-\frac{s}{2}+1]_{-r^*}}.
\end{eqnarray}
The contour $I$ and $I^*$ encircles $z_1=0$ in such a way
that $z_1=x^{s+2rl}z_2, x^{2-s+2rl}z_2~(l=0,1,2,\cdots)$
is inside and
$z_1=x^{-s-2rl}z_2, x^{s-2-2rl}z_2~(l=0,1,2,\cdots)$
is outside.

~\\
The followings are some of {\bf Main Results}
of our paper.

\begin{thm}~~~\label{thm:Nonlocal-Com1}
The nonlocal integrals of motion ${\cal G}_n$ 
(resp. ${\cal G}_n^*$) commute
with each other for generic ${\rm Re}(r)\neq 0$ and $s\neq 2$
(resp ${\rm Re}(r^*)\neq 0$ and $s\neq 2$)
\begin{eqnarray}
[{\cal G}_n,{\cal G}_m]=0,~~~[{\cal G}_n^*, {\cal G}_m^*]=0~~~
(n,m=1,2,\cdots).
\end{eqnarray}
\end{thm}

\begin{thm}~~~\label{thm:Nonlocal-Com2}
The nonlocal integrals of motion ${\cal G}_n$ 
and ${\cal G}_n^*$ 
commute with each other for generic 
$0<{\rm Re}(r)<1$ and $s\neq 2$
\begin{eqnarray}
[{\cal G}_n,{\cal G}_m^*]=0~~~(n,m=1,2,\cdots).
\end{eqnarray}
\end{thm}

\begin{thm}~~~The nonlocal integrals of motion ${\cal G}_n$ 
(resp. ${\cal G}_n^*$) are invariant under the action of 
Dynkin automorphism $\eta$,
for generic 
${\rm Re}(r)\neq 0$ and $s\neq 2$ (resp.
${\rm Re}(r^*)\neq 0$ and $s\neq 2$).
\begin{eqnarray}
\eta({\cal G}_n)={\cal G}_n,~~~
\eta({\cal G}_n^*)={\cal G}_n^*~~~
(n=1,2,\cdots).
\end{eqnarray}
\end{thm}

\begin{thm}~~~\label{thm:LocalNonLocal}
For generic parameter
${\rm Re}(r)\neq 0,~{\rm Re}(s)>0$, we have
\begin{eqnarray}
~[{\cal I}_n,{\cal G}_m]=0,~~
[{\cal I}_n^*,{\cal G}_m]=0~~
(n,m=1,2,\cdots ).
\end{eqnarray}
For generic parameter
${\rm Re}(r^*)\neq 0,~{\rm Re}(s)>0$, we have
\begin{eqnarray}
~[{\cal I}_n,{\cal G}_m^*]=0,~~
[{\cal I}_n^*,{\cal G}_m^*]=0,~~
(n,m=1,2,\cdots ).
\end{eqnarray}
\end{thm}

\subsection{Proof of $[{\cal G}_n, {\cal G}_m]=
[{\cal G}_n^*, {\cal G}_m^*]=0$}

In this section we show the commutation relations
$[{\cal G}_n, {\cal G}_m]=
[{\cal G}_n^*, {\cal G}_m^*]=0$.
At first we show the theta function identity,
which gives a generalization of the one in \cite{FO}.

\begin{thm}~~
The following theta function identity holds
\begin{eqnarray}
&&\sum_{K \cup K^c=\{1,\cdots,n+m\}
\atop{|K|=n,~K \cap K^c=\varnothing}}
\sum_{L \cup L^c=\{1,\cdots,n+m\}
\atop{|L|=n,~ L \cap L^c=\varnothing}}
\widehat{\vartheta}_{\beta,r}
\left(\sum_{j\in K^c}u_j-\sum_{j\in L^c}v_j\right)
\widehat{\vartheta}_{\alpha,r}
\left(\sum_{j\in K}u_j-\sum_{j \in L}v_j\right)
\nonumber
\\
&&\times\prod_{i\in K^c \atop{k\in K}}
\prod_{j \in L^c \atop{l\in L}}
\frac{[v_j-u_k+\frac{s}{2}]_r
[u_i-v_l+\frac{s}{2}]_r
[u_k-v_j+\frac{s}{2}-1]_r
[v_l-u_i+\frac{s}{2}-1]_r}{
[u_i-u_k]_r[v_j-v_l]_r[u_k-u_i-1]_r[v_l-v_j-1]_r}
\nonumber\\
&=&\sum_{K \cup K^c=\{1,\cdots,n+m\}
\atop{|K|=n,~K \cap K^c=\varnothing}}
\sum_{L \cup L^c=\{1,\cdots,n+m\}
\atop{|L|=n,~ L \cap L^c=\varnothing}}
\widehat{\vartheta}_{\alpha,r}
\left(\sum_{j\in K^c}u_j-\sum_{j\in L^c}v_j\right)
\widehat{\vartheta}_{\beta,r}
\left(\sum_{j\in K}u_j-\sum_{j \in L}v_j\right)\nonumber
\\
&&\times\prod_{i\in K^c \atop{k\in K}}
\prod_{j \in L^c \atop{l\in L}}
\frac{[v_l-u_i+\frac{s}{2}]_r
[u_k-v_j+\frac{s}{2}]_r
[u_i-v_l+\frac{s}{2}-1]_r
[v_j-u_k+\frac{s}{2}-1]_r}{
[u_k-u_i]_r[v_l-v_j]_r[u_i-u_k-1]_r[v_j-v_l-1]_r}.\nonumber\\
\label{eqn:Nonlocaltheta}
\end{eqnarray}
Here the theta functions
$\widehat{\vartheta}_{\alpha,r}(u)$ 
and $\widehat{\vartheta}_{\beta,r}(u)$ 
are characterized by
\begin{eqnarray}
\widehat{\vartheta}_{\alpha,r}(u+r)&=&
\widehat{\vartheta}_{\alpha,r}(u),\\
\widehat{\vartheta}_{\alpha,r}(u+r\tau)&=&
e^{-2\pi \sqrt{-1}\tau+\frac{2\pi\sqrt{-1}}{r}
(\hat{\pi}+2u)+\nu_\alpha}
\widehat{\vartheta}_{\alpha,r}(u),~~(\nu_\alpha \in {\mathbb C}),
\\
\widehat{\vartheta}_{\beta,r}(u+r)&=&
\widehat{\vartheta}_{\beta,r}(u),\\
\widehat{\vartheta}_{\beta,r}(u+r\tau)&=&
e^{-2\pi \sqrt{-1}\tau+\frac{2\pi\sqrt{-1}}{r}
(\hat{\pi}+2u)+\nu_\beta}
\widehat{\vartheta}_{\beta,r}(u),~~(\nu_\beta \in {\mathbb C}).
\end{eqnarray}
\end{thm}
%%%%%%%%%%%%%%%%%%%%%%%%%%%%%%%%%%%%%%%%%%%%%
%%%%%%%%%%%%%%%%%%%%%%%%%%%%%%%%%%%%%%%%%%%%%
{\it Proof}~~
In order to consider elliptic function, 
we divide the above theta identity
by $\widehat{\vartheta}_{r,\gamma}(\sum_{j=1}^{n+m}(u_j-v_j))$
with $\nu_\gamma \in {\mathbb C}$.
Let us set
\begin{eqnarray}
{\rm LHS}(n,m)&=&\sum_{K \cup K^c=\{1,\cdots,n+m\}
\atop{|K|=n,~K \cap K^c=\varnothing}}
\sum_{L \cup L^c=\{1,\cdots,n+m\}
\atop{|L|=n,~ L \cap L^c=\varnothing}}\nonumber\\
&\times&
\frac{\widehat{\vartheta}_{\alpha,r}
(\sum_{j\in K^c}u_j-\sum_{j\in L^c}v_j)
\widehat{\vartheta}_{\beta,r}(\sum_{j\in K}u_j-\sum_{j \in L}v_j)}
{\widehat{\vartheta}_{\gamma,r}(\sum_{j=1}^{n+m}(u_j-v_j))}
\\
&&\times\prod_{i\in K^c \atop{k\in K}}
\prod_{j \in L^c \atop{l\in L}}
\frac{[v_j-u_k+\frac{s}{2}]_r
[u_i-v_l+\frac{s}{2}]_r
[u_k-v_j+\frac{s}{2}-1]_r
[v_l-u_i+\frac{s}{2}-1]_r}{
[u_i-u_k]_r[v_j-v_l]_r[u_k-u_i-1]_r[v_l-v_j-1]_r}.
\nonumber
\end{eqnarray}
\begin{eqnarray}
{\rm RHS}(n,m)&=&\sum_{K \cup K^c=\{1,\cdots,n+m\}
\atop{|K|=n,~K \cap K^c=\varnothing}}
\sum_{L \cup L^c=\{1,\cdots,n+m\}
\atop{|L|=n,~ L \cap L^c=\varnothing}}\nonumber\\
&\times&
\frac{\widehat{\vartheta}_{\beta,r}
(\sum_{j\in K^c}u_j-\sum_{j\in L^c}v_j)
\widehat{\vartheta}_{\alpha,r}(\sum_{j\in K}u_j-\sum_{j \in L}v_j)}
{\widehat{\vartheta}_{\gamma,r}(\sum_{j=1}^{n+m}(u_j-v_j))}
\\
&&\times\prod_{i\in K^c \atop{k\in K}}
\prod_{j \in L^c \atop{l\in L}}
\frac{[v_l-u_i+\frac{s}{2}]_r
[u_k-v_j+\frac{s}{2}]_r
[u_i-v_l+\frac{s}{2}-1]_r
[v_j-u_k+\frac{s}{2}-1]_r}{
[u_k-u_i]_r[v_l-v_j]_r[u_i-u_k-1]_r[v_j-v_l-1]_r}.\nonumber
\end{eqnarray}
We will show ${\rm LHS}(n,m)={\rm RHS}(n,m)$
by induction.
Both LHS$(n,m)$ and RHS$(n,m)$ are elliptic functions.
Therefore, from Liouville theorem, it is enough to check whether
all residues of LHS$(n,m)$ and RHS$(n,m)$ coincide or not.
Candidates of poles are $u_\alpha=u_\beta$, $v_\alpha=v_\beta$,
$u_\alpha=u_\beta+1$, $v_\alpha=v_\beta+1$ and
$\widehat{\vartheta}_{\gamma,r}(\{u_\alpha\}|\{u_\beta\})=0$.
(Some of them are real pole and some of them are fake.)
Let us consider $u_\alpha=u_\beta~(\alpha\neq \beta)$.
Take the residue of the LHS$(n,m)$ at $u_\alpha=u_\beta$.
We have
\begin{eqnarray}
&&{\rm Res}_{u_\alpha=u_\beta}
\left(\frac{1}{[u_\alpha-u_\beta]_r[u_\beta-u_\alpha-1]_r}+
\frac{1}{[u_\beta-u_\alpha]_r[u_\alpha-u_\beta-1]_r}\right)
\nonumber\\
&\times&\sum_{L \cup L^c=\{1,\cdots,n+m\}
\atop{|L|=n,~ L \cap L^c=\varnothing}}
\prod_{j \in L^c \atop{l\in L}}
\frac{[v_l-u_\alpha+\frac{s}{2}]_r
[u_\alpha-v_j+\frac{s}{2}]_r
[u_\alpha-v_l+\frac{s}{2}-1]_r
[v_j-u_\alpha+\frac{s}{2}-1]_r}{
[v_l-v_j]_r[v_j-v_l-1]_r}
\nonumber\\
&\times&
\sum_{K \cup K^c=\{1,\cdots,n+m\}-\{\alpha,\beta\}
\atop{|K|=n-1,~K \cap K^c=\varnothing}}
\prod_{i\in K^c\cup\{\alpha\} \atop{k\in K\cup\{\alpha\}
\atop{i\neq k}
}}
\prod_{j \in L^c \atop{l\in L}}
\frac{[v_l-u_i+\frac{s}{2}]_r
[u_k-v_j+\frac{s}{2}]_r
[u_i-v_l+\frac{s}{2}-1]_r
[v_j-u_k+\frac{s}{2}-1]_r}{
[u_k-u_i]_r[v_l-v_j]_r[u_i-u_k-1]_r[v_j-v_l-1]_r}
\nonumber\\
&\times&
\frac{\widehat{\vartheta}_{\alpha,r}(
\sum_{j\in K^c\cup \{\alpha\}}u_j-\sum_{j\in L^c \cup \{\alpha\}}v_j)
\widehat{\vartheta}_{\beta,r}(\sum_{j\in K\cup\{\alpha\}}u_j-
\sum_{j \in L \cup \{\alpha\}}v_j)}
{\widehat{\vartheta}_{\gamma,r}(\sum_{j=1}^{n+m}(u_j-v_j))|_{u_\alpha=u_\beta}}=0.
\end{eqnarray}
Hence $u_\alpha=u_\beta, (\alpha \neq \beta)$ are not pole.
They are regular points.
By the same manner, we have 
\begin{eqnarray}
{\rm Res}_{u_\alpha=u_\beta}{\rm RHS}(n,m)=0,~~
{\rm Res}_{v_\alpha=v_\beta}{\rm LHS}(n,m)=0,~~
{\rm Res}_{v_\alpha=v_\beta}{\rm RHS}(n,m)=0.
\end{eqnarray}
Therefore points $u_\alpha=u_\beta$ and $v_\alpha=v_\beta$
are not poles. 
Therefore candidates of poles are restricted to
only
$u_\alpha=u_\beta+1$, $v_\alpha=v_\beta+1$ and
$\widehat{\vartheta}_{\gamma,r}(\{u_\alpha\}|\{u_\beta\})=0$.
We show the equation ${\rm LHS}(n,m)={\rm RHS}(n,m)$ 
by the induction of the number $n+m$.
We assume $n>m\geqq 1$ without loosing generality.
(The case $n=m$ is trivial.)
At first we show the starting point $n>m=1$.
We would like to show ${\rm LHS}(n,1)={\rm RHS}(n,1)$.
%\begin{eqnarray}
%&&\sum_{k,l=1}^{n+1}
%\frac{\vartheta_{\alpha} (\sum_{j=1\atop{j\neq k}}^{n+1}u_j-
%\sum_{j=1\atop{j\neq l}}^{n+1}v_j) \vartheta_{1,\beta}(
%u_k-v_l)}{\vartheta_{n+1,\gamma} (\sum_{j=1}^{n+1}(u_j-v_j))}\nonumber\\
%&\times&
%\prod_{j=1\atop{j \neq k}}^{n+1}
%\frac{[v_l-u_j+\frac{s}{2}-1][u_j-v_l+\frac{s}{2}]}
%{[u_k-u_j-1][u_j-u_k]}
%\prod_{j=1\atop{j \neq l}}^{n+1}
%\frac{[u_k-v_j+\frac{s}{2}-1][v_j-u_k+\frac{s}{2}]}
%{[v_l-v_j-1][v_j-v_l]}\nonumber\\
%&=&\sum_{k,l=1}^{n+1}
%\frac{\vartheta_{n,\alpha} (\sum_{j=1\atop{j\neq k}}^{n+1}u_j-
%\sum_{j=1\atop{j\neq l}}^{n+1}v_j) \vartheta_{1,\beta}(
%u_k-v_l)}{\vartheta_{n+1,\gamma} (\sum_{j=1}^{n+1}(u_j-v_j))}\nonumber\\
%&\times&
%\prod_{j=1\atop{j \neq k}}^{n+1}
%\frac{[u_j-v_l+\frac{s}{2}-1]_r[v_l-u_j+\frac{s}{2}]_r}
%{[u_j-u_k-1]_r[u_k-u_j]_r}
%\prod_{j=1\atop{j \neq l}}^{n+1}
%\frac{[v_j-u_k+\frac{s}{2}-1]_r[u_k-v_j+\frac{s}{2}]_r}
%{[v_j-v_l-1]_r[v_l-v_j]_r}.
%\end{eqnarray}
Let us take the residue at $u_1=u_2+1$ and $v_1=v_2+1$.
We have
\begin{eqnarray}
&&{\rm Res}_{u_1=u_2+1}
{\rm Res}_{v_1=v_2+1}{\rm LHS}(n,1)\nonumber\\
&=&
{\rm Res}_{u=0}\frac{[v_1-u+\frac{s}{2}]_r
[u-v_1+\frac{s}{2}-1]_r}{[u]_r[1]_r}
{\rm Res}_{u=0}\frac{[u-v_1+\frac{s}{2}]_r[v_1-u+\frac{s}{2}+1]_r}
{[u]_r[1]_r}\nonumber\\
&\times&
\prod_{j=3}^{n+1}\frac{
[v_1-u_j+\frac{s}{2}-1]_r
[u_j-v_1+\frac{s}{2}]_r
[u_1-v_j+\frac{s}{2}-1]_r[v_j-u_1+\frac{s}{2}]_r}
{[u_1-u_j-1]_r[u_j-u_1]_r
[v_1-v_j-1]_r[v_j-v_1]_r}\nonumber\\
&\times&\left.\frac{
\widehat{\vartheta}_{\alpha,r} (\sum_{j=1\atop{j\neq 1}}^{n+1}u_j-
\sum_{j=1\atop{j\neq 1}}^{n+1}v_j) 
\widehat{\vartheta}_{\beta,r}(u_1-v_1)}
{\widehat{\vartheta}_{\gamma,r} (\sum_{j=1}^{n+1}(u_j-v_j))}\right|_{u_1=u_2+1
\atop{v_1=v_2+1}}\nonumber\\
&=&{\rm Res}_{u_1=u_2+1}
{\rm Res}_{v_1=v_2+1}{\rm RHS}(n,1).
\end{eqnarray}
Because both LHS$(n,1)$ and RHS$(n,1)$ are symmetric with respect with
$u_1,u_2,\cdots,u_{n+1}$ and
$v_1,v_2,\cdots,v_{n+1}$,
we have 
\begin{eqnarray}
{\rm Res}_{u_\alpha=u_\beta+1}
{\rm Res}_{v_\gamma=v_\delta+1}{\rm LHS}(n,1)=
{\rm Res}_{u_\alpha=u_\beta+1}
{\rm Res}_{v_\gamma=v_\delta+1}{\rm RHS}(n,1).\label{eqn:NLNL1}
\end{eqnarray}
After taking the residues finitely many times,
every residue relation which comes from 
${\rm LHS}(n,1)={\rm RHS}(n,1)$,
is reduced to the above residue relation (\ref{eqn:NLNL1}) at
$u_\alpha=u_\beta+1, v_\gamma=v_\delta+1$.
Hence we have proved the starting point $n>m=1$.
For the second, we show the general $n>m\geqq 1$ case.
We assume the equation 
${\rm LHS}(n-1,m-1)={\rm RHS}(n-1,m-1)$
for some $(n,m)$.
Let us take the residue at
$u_1=u_2+1$ and $v_1=v_2+1$ of $L(n,m)-R(n,m)$.
We have
\begin{eqnarray}
&&{\rm Res}_{u_1=u_2+1}
{\rm Res}_{v_1=v_2+1}
\left({\rm LHS}(n,m)-{\rm RHS}(n,m)\right)\nonumber\\
&=&
{\rm Res}_{u=0}{\rm Res}_{v=0}
\frac{[u_1-v_1+\frac{s}{2}]_r
[v_1-u_1+\frac{s}{2}]_r
[u_1-v_1+\frac{s}{2}-1]_r
[v_1-u_1+\frac{s}{2}-1]_r}{[u]_r[v]_r[1]_r[1]_r}\nonumber\\
&\times&
\prod_{j=3}^{n+m}\frac{[v_1-u_j+\frac{s}{2}-1]_r
[u_j-v_1+\frac{s}{2}]_r[v_j-u_1+\frac{s}{2}]_r
[u_1-v_j+\frac{s}{2}-1]_r}{
[u_1-u_j-1]_r[u_j-u_1]_r[v_1-v_j-1]_r[v_j-v_1]_r}\nonumber\\
&\times&\left.
\sum_{K \cup K^c=\{3,\cdots,n+m\}
\atop{|K|=m-1,|K^c|=n-1}}
\sum_{L \cup L^c=\{3,\cdots,n+m\}
\atop{|L|=m-1,|L^c|=n-1}}
\frac{\displaystyle
\widehat{\vartheta}_{\alpha,r}
(\sum_{j\in K^c}u_j-\sum_{j\in L^c}v_j+u_1-v_1)
\widehat{\vartheta}_{\beta,r}(\sum_{j\in K}u_j-\sum_{j\in L}v_j+u_1-v_1)}
{\displaystyle
\widehat{\vartheta}_{\gamma,r}(\sum_{j=3}^{n+m}(u_j-v_j)
+2(u_1-v_1))}\right.\nonumber\\
&\times&\left(\frac{
\displaystyle \prod_{j\in L^c\atop{k \in K}}
[v_j-u_k+\frac{s}{2}]_r[u_k-v_j+\frac{s}{2}-1]_r
\prod_{i \in K^c \atop{l \in L}}
[u_i-v_l+\frac{s}{2}]_r[v_l-u_i+\frac{s}{2}-1]_r
}{\displaystyle \prod_{i\in K^c 
\atop{k \in K}}[u_i-u_k]_r[u_k-u_i-1]_r 
\prod_{j\in L^c \atop{l \in L}}[v_j-v_l]_r[v_l-v_j-1]_r}
\right.\nonumber\\
&-&\left.
\frac{\displaystyle
\prod_{l\in L\atop{i \in K^c}}
[v_l-u_i+\frac{s}{2}]_r[u_i-v_l+\frac{s}{2}-1]_r
\prod_{k \in K \atop{j \in L^c}}
[u_k-v_j+\frac{s}{2}]_r[v_j-u_k+\frac{s}{2}-1]_r
}{\displaystyle
\prod_{i\in K^c 
\atop{k \in K}}[u_k-u_i]_r[u_i-u_k-1]_r 
\prod_{j\in L^c \atop{l \in L}}
[v_l-v_j]_r[v_j-v_l-1]_r}
\right)=0.\nonumber\\
\end{eqnarray}
We have used the hypothesis of the induction:
${\rm LHS}(n-1,m-1)={\rm RHS}(n-1,m-1)$.
Because both LHS$(n,m)$ and RHS$(n,m)$ 
are symmetric with respect with
$u_1,u_2,\cdots,u_{m+n}$ and
$v_1,v_2,\cdots,v_{m+n}$,
we have 
\begin{eqnarray}
{\rm Res}_{u_\alpha=u_\beta+1}
{\rm Res}_{v_\gamma=v_\delta+1}{\rm LHS}(n,m)=
{\rm Res}_{u_\alpha=u_\beta+1}
{\rm Res}_{v_\gamma=v_\delta+1}{\rm RHS}(n,m),\label{eqn:NLNL2}
\end{eqnarray}
for arbitrary $1\leqq \alpha\neq \beta \leqq n+m$
and $1\leqq \gamma \neq \delta \leqq n+m$.
After taking the residues finitely many times,
every residue relation which comes from 
${\rm LHS}(n,m)={\rm RHS}(n,m)$,
is reduced to the above residue relation (\ref{eqn:NLNL2}). 
Hence we have shown ${\rm LHS}(n,m)={\rm RHS}(n,m)$
for general $n,m=1,2,\cdots$.~~~
Q.E.D.
%%%%%%%%%%%%%%%%%%%%%%%%%%%%%%%%%%%%%%%%
%%%%%%%%%%%%%%%%%%%%%%%%%%%%%%%%%%%%%%%%
~\\

Now let us show the commutation relation,
$[{\cal G}_n,{\cal G}_m]=0$.

~\\
{\it Proof~of~Theorem \ref{thm:Nonlocal-Com1}}~~~
We study the case : $0<{\rm Re}(s)<2$ and ${\rm Re}(r)>0$.
Other cases are similar.
Hence the integrand of the nonlocal integrals of motion
satisfies the $S_n$-invarince
\begin{eqnarray}
&&
F_1(z_{\sigma(1)})F_1(z_{\sigma(2)})
\cdots F_1(z_{\sigma(n)})
F_0(w_{\rho(1)})F_0(w_{\rho(2)})
\cdots
F_0(w_{\rho(n)})
\nonumber\\
&\times&
\frac{\prod_{1\leqq i<j \leqq n}
[u_{\sigma(i)}-u_{\sigma(j)}]_r
[u_{\sigma(j)}-u_{\sigma(i)}-1]_r
[v_{\rho(i)}-v_{\rho(j)}]_r[v_{\rho(j)}-v_{\rho(i)}-1]_r
}{
\prod_{i,j=1}^n
[u_{\sigma(i)}-v_{\rho(j)}+\frac{s}{2}]_r
[v_{\rho(j)}-u_{\sigma(i)}+\frac{s}{2}-1]_r}
\nonumber\\
&=&
\prod_{\rightarrow
\atop{1\leqq j \leqq n}}F_1(z_j)
\prod_{\rightarrow
\atop{1\leqq j \leqq n}}F_0(w_j)
\frac{\prod_{1\leqq i<j \leqq n}
[u_i-u_j]_r[u_j-u_i-1]_r[v_i-v_j]_r[v_j-v_i-1]_r
}{
\prod_{i,j=1}^n
[u_i-v_j+\frac{s}{2}]_r[v_j-u_i+\frac{s}{2}-1]_r},\nonumber\\
&&~~~~~~~~~~~~~~~~~~~~~~~~~~~~~~~~~~~~~~~~~~~~~
{\rm for}~~\sigma, \rho \in S_n.
\end{eqnarray}
Hence we have
\begin{eqnarray}
{\cal G}_n\cdot {\cal G}_m
&=&
\left[
\prod_{\rightarrow
\atop{1\leqq j \leqq n+m}}F_1(z_j)
\prod_{\rightarrow
\atop{1\leqq j \leqq n+m}}F_0(w_j)\right.\nonumber\\
&&\times
\frac{\prod_{1\leqq i<j \leqq n+m}
[u_i-u_j]_r[u_j-u_i-1]_r[v_i-v_j]_r[v_j-v_i-1]_r
}{
\prod_{i,j=1}^{n+m}
[u_i-v_j+\frac{s}{2}]_r[v_j-u_i+\frac{s}{2}-1]_r}
\nonumber\\
&&\times
\frac{1}{((n+m)!)^2}
\sum_{\sigma,\rho \in S_{n+m}}
\vartheta_{\alpha,r}\left(\sum_{j=1}^n(u_{\sigma(j)}-v_{\rho(j)})\right)
\vartheta_{\beta,r}\left(\sum_{j=n+1}^{n+m}(u_{\sigma(j)}-v_{\rho(j)})
\right)\nonumber\\
&&\times
\prod_{i=1}^n\prod_{j=n+1}^{n+m}
\frac{[v_{\rho(i)}-u_{\sigma(j)}+\frac{s}{2}]_r
[u_{\sigma(i)}-v_{\rho(j)}+\frac{s}{2}]_r}
{[u_{\sigma(i)}-u_{\sigma(j)}]_r
[v_{\rho(i)}-v_{\rho(j)}]_r}\nonumber\\
&&\times
\left.
\prod_{i=1}^n\prod_{j=n+1}^{n+m}
\frac{
[u_{\sigma(j)}-v_{\rho(i)}+\frac{s}{2}-1]_r
[v_{\rho(j)}-u_{\sigma(i)}+\frac{s}{2}-1]_r}
{
[u_{\sigma(j)}-u_{\sigma(i)}-1]_r
[v_{\rho(j)}-v_{\rho(i)}-1]_r}
\right]_{1, z_1 \cdots z_{n+m} w_1 \cdots w_{n+m}}.
\end{eqnarray}
Therefore we have the following theta function identity
as a sufficient condition of the commutation relation
${\cal G}_n \cdot {\cal G}_m={\cal G}_m \cdot {\cal G}_n$.
\begin{eqnarray}
&&\sum_{K \cup K^c=\{1,\cdots,n+m\}
\atop{|K|=n,~K \cap K^c=\varnothing}}
\sum_{L \cup L^c=\{1,\cdots,n+m\}
\atop{|L|=n,~ L \cap L^c=\varnothing}}
\vartheta_{\alpha,r}(\sum_{j\in K}u_j-\sum_{j\in L}v_j)
\vartheta_{\beta,r}(\sum_{j\in K^c}u_j-\sum_{j \in L^c}v_j)\nonumber
\\
&&\times\prod_{i\in K^c \atop{k\in K}}
\prod_{j \in L^c \atop{l\in L}}
\frac{[v_j-u_k+\frac{s}{2}]_r
[u_i-v_l+\frac{s}{2}]_r
[u_k-v_j+\frac{s}{2}-1]_r
[v_l-u_i+\frac{s}{2}-1]_r}{
[u_i-u_k]_r[v_j-v_l]_r[u_k-u_i-1]_r[v_l-v_j-1]_r}
\nonumber\\
&=&\sum_{K \cup K^c=\{1,\cdots,n+m\}
\atop{|K|=m,~K \cap K^c=\varnothing}}
\sum_{L \cup L^c=\{1,\cdots,n+m\}
\atop{|L|=m,~ L \cap L^c=\varnothing}}
\vartheta_{\alpha,r}(\sum_{j\in K^c}u_j-\sum_{j\in L^c}v_j)
\vartheta_{\beta,r}(\sum_{j\in K}u_j-\sum_{j \in L}v_j)
\nonumber\\
&&\times\prod_{i\in K^c \atop{k\in K}}
\prod_{j \in L^c \atop{l\in L}}
\frac{[v_l-u_i+\frac{s}{2}]_r
[u_k-v_j+\frac{s}{2}]_r
[u_i-v_l+\frac{s}{2}-1]_r
[v_j-u_k+\frac{s}{2}-1]_r}{
[u_k-u_i]_r[v_l-v_j]_r[u_i-u_k-1]_r[v_j-v_l-1]_r}.\nonumber\\
\end{eqnarray}
This is special case $\nu_\alpha=\nu_\beta=0$ 
of the equation
(\ref{eqn:Nonlocaltheta}).
(In order to use induction,
we have introduced additional parameters
$\nu_\alpha, \nu_\beta$ to 
(\ref{eqn:Nonlocaltheta}).)
We have shown the commutation reltion,
$[{\cal G}_m,{\cal G}_n]=0$.
~~~
Q.E.D.

\subsection{Proof of $[{\cal G}_m,{\cal G}_n^*]=0$}

In this section we show the commutation relation
$[{\cal G}_m,{\cal G}_n^*]=0$.
The screening currents $E_j(z)$ and $F_j(z)$ almost commute
\begin{eqnarray}
[E_1(z_1),F_1(z_2)]=\frac{1}{x-x^{-1}}
(\delta(xz_2/z_1)H(x^rz_2)-\delta(xz_1/z_2)H(x^{-r}z_2)).
\nonumber
\end{eqnarray}
Hence, in order to show the commutation relation, 
remaining task for us is to check whether
delta-function factors cancell out or not.

~\\
{\it Proof~of~Theorem \ref{thm:Nonlocal-Com2}}~~~
For a while we study the following parameter case:
$0<{\rm Re}(r)<1$ and 
$0<{\rm Re}(s)<2$.
For reader's convenience, we show
the simple case $[{\cal G}_1^*,{\cal G}_1]=0$ at first.
Using the commutation relations of the screening currents
$E_j(z), F_j(z)$, we have
\begin{eqnarray}
[{\cal G}_1^*,{\cal G}_1]
&=&\int \int \int_{C_1}
\frac{dz_1}{2\pi \sqrt{-1} z_1} 
\frac{dz_2}{2\pi \sqrt{-1} z_2} 
\frac{dw_2}{2\pi \sqrt{-1} w_2}
B_1(x^rz_1|z_1,z_2,w_2)
\nonumber\\
&-&
\int \int \int_{\widetilde{C}_1}
\frac{dz_1}{2\pi \sqrt{-1} z_1} 
\frac{dz_2}{2\pi \sqrt{-1} z_2} 
\frac{dw_2}{2\pi \sqrt{-1} w_2} 
B_1(x^{-r}z_1|z_1,z_2,w_2)\nonumber\\
&+&
\int \int \int_{C_2}
\frac{dz_1}{2\pi \sqrt{-1} z_1} 
\frac{dz_2}{2\pi \sqrt{-1} z_2} 
\frac{dw_1}{2\pi \sqrt{-1} w_1} 
B_{2}(x^{r}z_2|z_1,z_2,w_1)\nonumber\\
&-&
\int \int \int_{\widetilde{C}_2}
\frac{dz_1}{2\pi \sqrt{-1} z_1} 
\frac{dz_2}{2\pi \sqrt{-1} z_2} 
\frac{dw_1}{2\pi \sqrt{-1} w_1} 
B_{2}(x^{-r}z_2|z_1,z_2,w_1).\label{eqn:NLNL*1}
\end{eqnarray}
Here we have set
\begin{eqnarray}
B_{1}(z|z_1,z_2,w_2)&=&
\frac{1}{x-x^{-1}}H(z)E_0(w_2)F_0(z_2)\nonumber
\\
&&\times
\frac{\vartheta_{\alpha,r}(u_1-u_2)
\vartheta_{\beta,-r^*}(u-v_2-\frac{r^*}{2})
}{[u-\frac{r^*}{2}-v_2+\frac{s}{2}]_{-r^*}
[u-\frac{r^*}{2}-v_2-\frac{s}{2}+1]_{-r^*}
[u_1-u_2+\frac{s}{2}]_{r}
[u_1-u_2-\frac{s}{2}+1]_{r}},
\nonumber\\
\\
B_{2}(z|z_1,z_2,w_1)&=&
\frac{1}{x-x^{-1}}F_1(z_1)E_1(w_1)\tau(H(z))\nonumber
\\
&\times&
\frac{\vartheta_{\alpha,r}(u_1-u_2)
\vartheta_{\beta,-r^*}(v_1-u+\frac{r^*}{2})
}{[v_1-u+\frac{r^*}{2}+\frac{s}{2}]_{-r^*}
[v_1-u+\frac{r^*}{2}-\frac{s}{2}+1]_{-r^*}
[u_1-u_2+\frac{s}{2}]_{r}
[u_1-u_2-\frac{s}{2}+1]_{r}}.
\nonumber\\
\end{eqnarray}
where the integral contours $C_{1}, C_2,
\widetilde{C}_{1}, \widetilde{C}_2$
are given by
\begin{eqnarray}
C_{1}&:&
|x^sz_2|,|x^{2-s}z_2|<|z_1|<|x^{-2r+s-2}z_2|,|x^{-2r-s}z_2|,\nonumber\\
&&
|x^{s-1-2r^*}w_2|, |x^{-s+1-2r^*}w_2|<|z_1|<
|x^{-s-1}w_2|,|x^{s-3}w_2|,~~~
|z_2|<|xw_2|,
\\
\widetilde{C}_{1}&:&
|x^{s+2r}z_2|,|x^{2-s+2r}z_2|<|z_1|<|x^{s-2}z_2|,|x^{-s}z_2|,\nonumber\\
&&
|x^{s+1}w_2|, |x^{-s+3}w_2|<|z_1|<
|x^{-s-1+2r}w_2|,|x^{s-3+2r}w_2|,~~~
|z_2|<|xw_2|,
\end{eqnarray}
and
\begin{eqnarray}
C_{2}&:&|x^{s-1-2r^*}w_1|,|x^{-s+1+2r^*}w_1|<|z_2|<|x^{s-3}w_1|,|x^{-s-1}w_1|,
\nonumber
\\
&&
|x^{s}z_1|,|x^{2-s}z_1|<|z_2|<|x^{s-2-2r}z_1|,|x^{-s-2r}z_1|,
~~~|z_1|<|xw_1|,\\
\widetilde{C}_{2}&:&
|x^{s+1}w_1|,|x^{-s+3}w_1|<|z_2|<|x^{s+2r^*-1}w_1|,|x^{-s+2r^*}w_1|,
\nonumber\\
&&
|x^{s+2r}z_1|,|x^{2-s+2r}z_1|<|z_2|<|x^{s-2}z_1|,|x^{-s}z_1|,
~~~|z_1|<|xw_1|.
\end{eqnarray}
When we change the variablw $z_1 \to x^{-2r}z_1$ in
the first term in
RHS of the equation (\ref{eqn:NLNL*1}),
the integrand 
$B_{1}(x^{r}z_1|z_1,z_2,w_2)$, 
is deformed to
$B_1(x^{-r}z_1|z_1,z_2,w_2)=B_{1}(x^{-r}z_1|x^{-2r}z_1,z_2,w_2)$, 
and the contour $C_{1}$ is deformed to exactly the same as 
$\widetilde{C}_{1}$.
Therefore we have
\begin{eqnarray}
\int \int \int_{C_1}
\frac{dz_1}{2\pi \sqrt{-1} z_1} 
\frac{dz_2}{2\pi \sqrt{-1} z_2} 
\frac{dw_2}{2\pi \sqrt{-1} w_2} 
B_{1}(x^{r}z_1|z_1,z_2,w_2)\nonumber\\
=
\int \int \int_{\widetilde{C}_1}
\frac{dz_1}{2\pi \sqrt{-1} z_1} 
\frac{dz_2}{2\pi \sqrt{-1} z_2} 
\frac{dw_2}{2\pi \sqrt{-1} w_2} 
B_{1}(x^{-r}z_1|z_1,z_2,w_2).
\end{eqnarray}
As the same manner we have
\begin{eqnarray}
\int \int \int_{C_2}
\frac{dz_1}{2\pi \sqrt{-1} z_1} 
\frac{dz_2}{2\pi \sqrt{-1} z_2} 
\frac{dw_1}{2\pi \sqrt{-1} w_1} 
B_{2}(x^{r}z_1|z_1,z_2,w_1)\nonumber\\
=
\int \int \int_{\widetilde{C}_2}
\frac{dz_1}{2\pi \sqrt{-1} z_1} 
\frac{dz_2}{2\pi \sqrt{-1} z_2} 
\frac{dw_1}{2\pi \sqrt{-1} w_1} 
B_{2}(x^{-r}z_1|z_1,z_2,w_1).
\end{eqnarray}
Therefore we have the commutation relation $[{\cal G}_1,{\cal G}_1^*]=0$.
Generalization to generic parameter $s \in {\mathbb C},
0<{\rm Re}(r)<1$ should be
understood as analytic continuation.

For the second we show the commutation relation
$[{\cal G}_n,{\cal G}_m^*]=0$.
For a while we study the following parameter case:
$0<{\rm Re}(r)<1$ and 
$0<{\rm Re}(s)<2$.
Using the commutation relations of 
the screening currents, we have
\begin{eqnarray}
&&[{\cal G}_n^*,{\cal G}_m]\nonumber\\
&=&
\sum_{i=1}^n \sum_{j=1}^m
\int \cdots \int_{{C}_{ij}}
\prod_{k=1}^n\frac{dz_k}{2\pi \sqrt{-1} z_k} 
\prod_{k=1\atop{k\neq i}}^m
\frac{dw_k}{2\pi \sqrt{-1} w_k} 
B_{i,j}^{(n,m)}\left(x^{r}z_i,
\{z_k\}_{k=1}^{2n},
\{w_k\}_{k=1\atop{k\neq j}}^{2m}
\right)\nonumber\\
&-&
\sum_{i=1}^n \sum_{j=1}^m
\int \cdots \int_{\widetilde{C}_{ij}}
\prod_{k=1}^n\frac{dz_k}{2\pi \sqrt{-1} z_k} 
\prod_{k=1\atop{k\neq i}}^m
\frac{dw_k}{2\pi \sqrt{-1} w_k} 
B_{i,j}^{(n,m)}\left(x^{-r}z_j,
\{z_k\}_{k=1}^{2n},
\{w_k\}_{k=1\atop{k\neq i}}^{2m}
\right)
\nonumber\\
&+&
\sum_{i={n+1}}^{2n}
\sum_{j={m+1}}^{2m}
\int \cdots \int_{{C}_{ij}}
\prod_{k=1}^n\frac{dz_k}{2\pi \sqrt{-1} z_k} 
\prod_{k=1\atop{k\neq i}}^m
\frac{dw_k}{2\pi \sqrt{-1} w_k} 
B_{i,j}^{(n,m)}\left(x^{-r}z_j,
\{z_k\}_{k=1}^{2n},
\{w_k\}_{k=1\atop{k\neq i}}^{2m}
\right)
\nonumber\\
&-&
\sum_{i={n+1}}^{2n}
\sum_{j={m+1}}^{2m}
\int \cdots \int_{\widetilde{C}_{ij}}
\prod_{k=1}^n\frac{dz_k}{2\pi \sqrt{-1} z_k} 
\prod_{k=1\atop{k\neq i}}^m
\frac{dw_k}{2\pi \sqrt{-1} w_k} 
B_{i,j}^{(n,m)}\left(x^{-r}z_j,
\{z_k\}_{k=1}^{2n},
\{w_k\}_{k=1\atop{k\neq i}}^{2m}
\right).\nonumber\\
\label{eqn:NLNL*2}
\end{eqnarray}
Here we have set for $1\leqq i \leqq n$ and $1\leqq j \leqq m$,
\begin{eqnarray}
&&B_{i,j}^{(n,m)}\left(z,
\{z_k\}_{k=1}^{2n},
\{w_k\}_{k=1\atop{k\neq i}}^{2m}\right)\nonumber\\
&=&
\frac{1}{x-x^{-1}}
F_1(z_1)\cdots F_1(z_{j-1})
E_1(w_1)\cdots E_1(w_{i-1})
H(z)\nonumber\\
&\times&F_1(z_{j+1})\cdots F_1(z_m)E_1(w_{i+1})\cdots E_1(w_n)
E_0(w_{n+1})\cdots E_0(w_{2n})F_0(z_{m+1})\cdots F_0(z_{2m})\nonumber\\
&\times&
\vartheta_{\alpha,-r^*}\left(\sum_{k=1}^n
v_k-\sum_{l=n+1}^{2n}v_l\right)
\vartheta_{\beta,r}\left(\sum_{k=1}^m
u_k-\sum_{l=m+1}^{2m}u_l\right)
\nonumber
\\
&\times&
\frac{\displaystyle
\prod_{1 \leqq k<l \leqq  n}
[v_k-v_l]_{-r^*}[v_l-v_k-1]_{-r^*}
\prod_{n+1 \leqq k<l \leqq 2n}
[v_k-v_l]_{-r^*}[v_l-v_k-1]_{-r^*}
}{\displaystyle
\prod_{k=1}^n \prod_{l=n+1}^{2n}
\left[v_k-v_l+\frac{s}{2}\right]_{-r^*}
\left[v_k-v_l-\frac{s}{2}+1\right]_{-r^*}
}
\nonumber\\
&\times&
\left.
\frac{\displaystyle
\prod_{1 \leqq k<l \leqq  m}
[u_k-u_l]_{r}[u_l-u_k-1]_{r}
\prod_{m+1 \leqq k<l \leqq 2m}
[u_k-u_l]_{r}[u_l-u_k-1]_{r}
}{\displaystyle
\prod_{k=1}^m \prod_{l=m+1}^{2m}
\left[u_k-u_l+\frac{s}{2}\right]_{r}
\left[u_k-u_l-\frac{s}{2}+1\right]_{r}
}
\right|_{v_i=u-\frac{r^*}{2}},
\end{eqnarray}
and for $n+1\leqq i \leqq 2n$ and $m+1\leqq j \leqq 2m$,
\begin{eqnarray}
&&B_{i,j}^{(n,m)}\left(z,
\{z_k\}_{k=1}^{2m},
\{w_k\}_{k=1\atop{k\neq i}}^{2n}\right)\nonumber\\
&=&
\frac{1}{x-x^{-1}}
F_1(z_1)\cdots F_1(z_m)
E_1(w_1)\cdots E_1(w_n)\nonumber\\
&\times&
F_0(z_{m+1})\cdots F_0(z_{j-1})
E_0(w_{n+1})\cdots E_0(w_{i-1})
\tau(H(z))\nonumber\\
&\times&F_0(z_{j+1})\cdots F_0(z_{2m})
E_0(w_{i+1})\cdots E_0(w_{2n})\nonumber\\
&\times&
\vartheta_{\alpha,-r^*}\left(\sum_{k=1}^n
v_k-\sum_{l=n+1}^{2n}v_l\right)
\vartheta_{\beta,r}\left(\sum_{k=1}^m
u_k-\sum_{l=m+1}^{2m}u_l\right)
\nonumber
\\
&\times&
\frac{\displaystyle
\prod_{1 \leqq k<l \leqq  n}
[v_k-v_l]_{-r^*}[v_l-v_k-1]_{-r^*}
\prod_{n+1 \leqq k<l \leqq 2n}
[u_k-v_l]_{-r^*}[v_l-v_k-1]_{-r^*}
}{\displaystyle
\prod_{k=1}^n \prod_{l=n+1}^{2n}\left[v_k-v_l+\frac{s}{2}\right]_{-r^*}
\left[v_k-v_l-\frac{s}{2}+1\right]_{-r^*}
}
\nonumber\\
&\times&
\left.
\frac{\displaystyle
\prod_{1 \leqq k<l \leqq  m}
[u_k-u_l]_{r}[u_l-u_k-1]_{r}
\prod_{m+1 \leqq k<l \leqq 2m}
[u_k-u_l]_{r}[u_l-u_k-1]_{r}
}{\displaystyle
\prod_{k=1}^m \prod_{l=m+1}^{2m}
\left[u_k-u_l+\frac{s}{2}\right]_{r}
\left[u_k-u_l-\frac{s}{2}+1\right]_{r}}
\right|_{v_i=u-\frac{r^*}{2}},
\end{eqnarray}
where the integral contours $C_{i,j}, \widetilde{C}_{i,j}$
are given by as follows.
For $1\leqq i \leqq n, 1\leqq j \leqq m$ we set
\begin{eqnarray}
C_{i,j}&:&|x^{s-1-2r^*}w_k|,|x^{-s+1-2r^*}w_k|<|z_j|<
|x^{-s-1}w_k|,|x^{s-3}w_k|,~~(n+1\leqq k \neq i\leqq 2n),\nonumber
\\
&&|x^sz_k|,|x^{2-s}z_k|<|z_j|<|x^{-2r+s-2}z_k|,|x^{-2r-s}z_k|,
~~(m+1\leqq k \neq j\leqq 2m),\\
\widetilde{C}_{i,j}&:&
|x^{s+1}w_k|,|x^{-s+3}w_k|<|z_j|<
|x^{-s+1+2r^*}w_k|,|x^{s-1+2r^*}w_k|,~~(n+1\leqq k \neq i\leqq 2n),\nonumber
\\
&&|x^{s+2r}z_k|,|x^{2-s+2r}z_k|<|z_j|<|x^{s-2}z_k|,|x^{-s}z_k|,
~~(m+1\leqq k \neq j\leqq 2m),
\end{eqnarray}
and for $n+1\leqq i \leqq 2n, m+1\leqq j \leqq 2m$ we set
\begin{eqnarray}
C_{i,j}&:&|x^{s-1-2r^*}w_k|,|x^{-s+1+2r^*}w_k|<|z_j|
<|x^{s-3}w_k|,|x^{-s-1}w_k|,~~(n+1\leqq k \neq i \leqq 2n),
\nonumber
\\
&&
|x^{s}z_k|,|x^{2-s}z_k|<|z_j|<|x^{s-2-2r}z_k|,|x^{-s-2r}z_k|,~~
(m+1\leqq k \neq j\leqq 2m),\\
\widetilde{C}_{i,j}&:&
|x^{s+1}w_k|,|x^{-s+3}w_k|<|z_j|
<|x^{s+2r^*-1}w_k|,|x^{-s+2r^*}w_k|,~~(n+1\leqq k \neq i\leqq 2n),
\nonumber\\
&&
|x^{s+2r}z_k|,|x^{2-s+2r}z_k|<|z_j|<|x^{s-2}z_k|,|x^{-s}z_k|,~~
(m+1\leqq k\neq j \leqq 2m).
\end{eqnarray}
When we change the variable $z_j \to x^{-2r}z_j$ in
the first term in LHS of (\ref{eqn:NLNL*2}),
both integrand and
integral contour are deformed to the same as the second term of
(\ref{eqn:NLNL*2}).
Hence we have
\begin{eqnarray}
&&\int \cdots \int_{{C}_{ij}}
\prod_{k=1}^n\frac{dz_k}{2\pi \sqrt{-1} z_k} 
\prod_{k=1\atop{k\neq i}}^m
\frac{dw_k}{2\pi \sqrt{-1} w_k} 
B_{i,j}^{(n,m)}\left(x^{r}z_i,
\{z_k\}_{k=1}^{2n},
\{w_k\}_{k=1\atop{k\neq j}}^{2m}
\right)\nonumber\\
&=&
\int \cdots \int_{\widetilde{C}_{ij}}
\prod_{k=1}^n\frac{dz_k}{2\pi \sqrt{-1} z_k} 
\prod_{k=1\atop{k\neq i}}^m
\frac{dw_k}{2\pi \sqrt{-1} w_k} 
B_{i,j}^{(n,m)}\left(x^{-r}z_j,
\{z_k\}_{k=1}^{2n},
\{w_k\}_{k=1\atop{k\neq i}}^{2m}
\right).
\end{eqnarray}
By the same arguments as above we have
$[{\cal G}_n^*,{\cal G}_m]=0$ for $0<{\rm Re}(r)<1$ and $0<{\rm Re}(s)<2$.
Generalization to generic parameter $0<{\rm Re}(r)<1$ and $s \in {\mathbb C}$
should be understood as analytic continuation.~~~
Q.E.D.

\subsection{Proof of Dynkin Automorphism Invariance
$\eta({\cal G}_n)={\cal G}_n$}

In this section we show the invariance condition
$\eta({\cal G}_n)={\cal G}_n$ and 
$\eta({\cal G}_n^*)={\cal G}_n^*$.
For reader's convenience, we explain 
$\eta({\cal G}_1)={\cal G}_1$
at first.
We study the case : $0<{\rm Re}(r)$ and
$0<{\rm Re}(s)<2$.
From the definition of the Dynkin automorphism $\eta$,
we have
\begin{eqnarray}
\eta({\cal G}_1)=
\int \int_I \prod_{j=1}^2
\frac{dz_j}{2\pi \sqrt{-1}z_j}
F_0(z_1)F_1(z_2)\frac{\vartheta_{\alpha,r}(u_1-u_2)|_{\widehat{\pi} \to 
-\widehat{\pi}}}{
[u_1-u_2+\frac{s}{2}]_r[u_2-u_1+\frac{s}{2}-1]_r}.
\end{eqnarray}
Exchanging the ordering of $F_1(z_1)$ and $F_0(z_2)$,
and changing the variables $u_1 \to u_2$ and $u_2 \to u_1$,
we have
\begin{eqnarray}
\eta({\cal G}_1)&=&\int \int_I \prod_{j=1}^2
\frac{dz_j}{2\pi \sqrt{-1}z_j}
F_1(z_1)F_0(z_2)\frac{\vartheta_{-\alpha,r}(u_1-u_2)}{
[u_1-u_2+\frac{s}{2}]_r[u_2-u_1+\frac{s}{2}-1]_r}.
\end{eqnarray}
The relation $\vartheta_{\alpha,r}(u)=\vartheta_{-\alpha,r}(u)$
implies $\eta({\cal G}_1)={\cal G}_1$.
Generalization to generic parameter
$0<{\rm Re}(r)$ and $s \in {\mathbb C}$
should be understood as analytic continuation.\\
~\\
{\it Proof.}~~
Let us show
$\tau({\cal G}_n)={\cal G}_n$.
We study the case : $0<{\rm Re}(r)$ and
$0<{\rm Re}(s)<2$.
From the definition of the Dynkin automorphism $\tau$,
we have
\begin{eqnarray}
\eta({\cal G}_n)&=&\int \cdots \int_{I}
\prod_{j=1}^n\frac{dz_j}{2\pi\sqrt{-1}z_j}
\prod_{j=1}^n\frac{dw_j}{2\pi\sqrt{-1}w_j}
\prod_{\longrightarrow
\atop{1\leqq j \leqq n}}F_0(z_j)
\prod_{\longrightarrow
\atop{1\leqq j \leqq n}}F_1(w_j)\nonumber\\
&\times&
\frac{\prod_{1\leqq i<j \leqq n}
[u_i-u_j]_r[u_j-u_i-1]_r[v_i-v_j]_r[v_j-v_i-1]_r
}{
\prod_{i,j=1}^n
[u_i-v_j+\frac{s}{2}]_r[v_j-u_i+\frac{s}{2}-1]_r}
\left.\vartheta_{\alpha,r}\left(
\sum_{j=1}^n(u_j-v_j)\right)
\right|_{\hat{\pi}\to -\hat{\pi}}.\nonumber\\
\end{eqnarray}
Exchanging the ordering of $F_1(z_j)$ and $F_0(w_k)$,
and changing the variables $u_j \to v_j$ and $v_j \to u_j$,
we have
\begin{eqnarray}
\eta({\cal G}_n)&=&\int \cdots \int_{I}
\prod_{j=1}^n\frac{dz_j}{2\pi\sqrt{-1}z_j}
\prod_{j=1}^n\frac{dw_j}{2\pi\sqrt{-1}w_j}
\prod_{\longrightarrow
\atop{1\leqq j \leqq n}}F_1(w_j)
\prod_{\longrightarrow
\atop{1\leqq j \leqq n}}F_0(z_j)\nonumber\\
&\times&
\frac{\prod_{1\leqq i<j \leqq n}
[u_i-u_j]_r[u_j-u_i-1]_r[v_i-v_j]_r[v_j-v_i-1]_r
}{
\prod_{i,j=1}^n
[u_i-v_j+\frac{s}{2}]_r[v_j-u_i+\frac{s}{2}-1]_r}
\left.\vartheta_{\alpha,r}\left(
\sum_{j=1}^n(v_j-u_j)\right)
\right|_{\hat{\pi}\to -\hat{\pi}}.\nonumber\\
\end{eqnarray}
Using the relation 
$\vartheta_{\alpha,r}(u)=\vartheta_{\alpha,r}(-u)
|_{\widehat{\pi}\to -\widehat{\pi}}$,
we have $\eta({\cal G}_n)={\cal G}_n$.
Generalization to generic parameter
$0<{\rm Re}(r)$ and $s \in {\mathbb C}$
should be understood as analytic continuation.
The proof of the invariance
$\eta({\cal G}_n^*)={\cal G}_n^*$ is given 
by the same manner.
~~~~
Q.E.D.

\subsection{Proof of $[{\cal I}_m,{\cal G}_n]=0$}

In this section we show the commutation relations
$[{\cal I}_m,{\cal G}_n]=0$.
The operators $\Lambda_j(z)$ and $F_j(z)$
commute almost everywhere.
$$
[\Lambda_1(z_1),F_1(z_2)]=(x^{-r^*}-x^{r^*})
\delta(x^rz_1/z_2){\cal A}(x^{-r}z_2).
$$
Hence,
in order to show the commutation relation,
remaining task for us is
to check whether delta-function factors cancell out or not.
At first we summarize simple case $[{\cal I}_1,{\cal G}_n]=0$,
for reader's convenience.\\\\
{\it Proof of $[{\cal I}_1,{\cal G}_n]=0$}~~~
For a while we restrict
our interest to the case 
$0<{\rm Re}(s)<2$, $0<{\rm Re}(r)<1$, and 
${\rm Re}(2r)<{\rm Re}(s)$.
We have
\begin{eqnarray}
&&[{\cal I}_1,{\cal G}_n]\nonumber\\
&=&(x^{-r^*}-x^{r^*})\sum_{j=1}^n \int \cdots \int_{\widetilde{{I}}(j)}
\prod_{k=1}^n \frac{dz_k}{2\pi \sqrt{-1}z_k}
\frac{dw_k}{2\pi \sqrt{-1} w_k}
F_1(z_1)\cdots F_1(z_{j-1})\nonumber\\
&\times&{\cal A}(x^{-r}z_j)F_1(z_{j+1})\cdots F_1(z_n)
F_0(w_1)\cdots F_0(w_n)\nonumber\\
&\times&
\frac{\prod_{1\leqq k<l \leqq n}
[u_k-u_l]_r[u_k-u_l+1]_r[v_k-v_l]_r[v_k-v_l+1]_r
}{
\prod_{k,l=1}^n[u_k-v_l+\frac{s}{2}]_r[v_l-u_k+\frac{s}{2}-1]_r
}
\vartheta_{\alpha,r}
\left(\sum_{j=1}^n (u_j-v_j)\right)\nonumber\\
&-&(x^{-r^*}-x^{r^*})\sum_{j=1}^n \int \cdots \int_{\widetilde{{I}}'(j)}
\prod_{k=1}^n \frac{dz_k}{2\pi \sqrt{-1}z_k}
\frac{dw_k}{2\pi \sqrt{-1} w_k}
F_1(z_1)\cdots F_1(z_{j-1})\nonumber\\
&\times&{\cal A}(x^{r}z_j)F_1(z_{j+1})\cdots F_1(z_n)
F_0(w_1)\cdots F_0(w_n)\nonumber\\
&\times&
\frac{\prod_{1\leqq k<l \leqq n}
[u_k-u_l]_r[u_k-u_l+1]_r[v_k-v_l]_r[v_k-v_l+1]_r
}{
\prod_{k,l=1}^n[u_k-v_l+\frac{s}{2}]_r[v_l-u_k+\frac{s}{2}-1]_r
}
\vartheta_{\alpha,r}
\left(\sum_{j=1}^n (u_j-v_j)\right)\nonumber\\
&+&
\eta
\left(
\sum_{j=1}^n
\int_C \frac{d\zeta}{2\pi \sqrt{-1}\zeta}
\int \cdots \int_I
\prod_{k=1}^n \frac{dz_k}{2\pi \sqrt{-1}z_k}
\frac{dw_k}{2\pi \sqrt{-1} w_k}
F_0(z_1)\cdots F_0(z_{n})\right.\nonumber\\
&\times&
F_1(w_{1})\cdots F_1(w_{j-1})
[T_1(\zeta),F_1(w_j)]
F_1(w_{j+1})\cdots F_1(w_n)\nonumber\\
&\times&
\left.\frac{\prod_{1\leqq k<l \leqq n}
[u_k-u_l]_r[u_k-u_l+1]_r[v_k-v_l]_r[v_k-v_l+1]_r
}{
\prod_{k,l=1}^n[u_k-v_l+\frac{s}{2}]_r[v_l-u_k+\frac{s}{2}-1]_r
}
\vartheta_{\alpha,r}
\left(\sum_{j=1}^n (u_j-v_j)\right)
\right).\nonumber\\
\label{eqn:IG1}
\end{eqnarray}
Here the contours 
$\widetilde{{I}}(j)$ and $\widetilde{{I}}'(j)$ are given by
\begin{eqnarray}
\widetilde{{I}}(j)&:& |x^{4r-2}z_{j+1}|,\cdots,|x^{4r-2}z_n|
<|z_j|<|x^{-2r+2}z_1|,\cdots,|x^{-2r+2}z_{j-1}|,\nonumber\\
&&|x^sw_k|,|x^{2r-s+2}w_k|<|z_j|<|x^{s-2}w_k|,|x^{-s+2r}w_k|,
~~(1\leqq k \leqq n),
\nonumber\\
\widetilde{{I}}'(j)&:&
|x^{2r-2}z_{j+1}|,\cdots,|x^{2r-2}z_n|
<|z_j|<|x^{-4r+2}z_1|,\cdots,|x^{-4r+2}z_{j-1}|,\nonumber\\
&&|x^{s-2r}w_k|,|x^{-s+2}w_k|<|z_j|<|x^{s-2-2r}w_k|,|x^{-s}w_k|,
~~(1\leqq k \leqq n).
\end{eqnarray}
Let us change the variable $z_j \to x^{2r}z_j$ of the first term
of (\ref{eqn:IG1}), upon
condition 
$0<{\rm Re}(s)<2$, $0<{\rm Re}(r)<1$, ${\rm Re}(2r)<{\rm Re}(s)$.
Using periodicity of the integrand, we have
\begin{eqnarray}
&&\int \cdots \int_{\widetilde{{I}}(j)}
\prod_{k=1}^n \frac{dz_k}{2\pi \sqrt{-1}z_k}
\frac{dw_k}{2\pi \sqrt{-1} w_k}
F_1(z_1)\cdots F_1(z_{j-1})
\nonumber\\
&\times&{\cal A}(x^{-r}z_j)F_1(z_{j+1})\cdots F_1(z_n)
F_0(w_1)\cdots F_0(w_n)\nonumber\\
&\times&
\frac{\prod_{1\leqq k<l \leqq n}
[u_k-u_l]_r[u_k-u_l+1]_r[v_k-v_l]_r[v_k-v_l+1]_r
}{
\prod_{k,l=1}^n[u_k-v_l+\frac{s}{2}]_r[v_l-u_k+\frac{s}{2}-1]_r}
\vartheta_{\alpha,r}
\left(\sum_{j=1}^n (u_j-v_j)\right)\nonumber\\
&=&\int \cdots \int_{\widetilde{{I}}'(j)}
\prod_{k=1}^n \frac{dz_k}{2\pi \sqrt{-1}z_k}
\frac{dw_k}{2\pi \sqrt{-1} w_k}
F_1(z_1)\cdots F_1(z_{j-1})\nonumber\\
&\times&{\cal A}(x^{r}z_j)F_1(z_{j+1})\cdots F_1(z_n)
F_0(w_1)\cdots F_0(w_n)\nonumber\\
&\times&
\frac{\prod_{1\leqq k<l \leqq n}
[u_k-u_l]_r[u_k-u_l+1]_r[v_k-v_l]_r[v_k-v_l+1]_r
}{
\prod_{k,l=1}^n[u_k-v_l+\frac{s}{2}]_r[v_l-u_k+\frac{s}{2}-1]_r
}
\vartheta_{\alpha,r}
\left(\sum_{j=1}^n (u_j-v_j)\right).\nonumber
\end{eqnarray}
By the same manner as above, we have
\begin{eqnarray}
&&\int_C \frac{d\zeta}{2\pi \sqrt{-1}\zeta}
\int \cdots \int_I
\prod_{k=1}^n \frac{dz_k}{2\pi \sqrt{-1}z_k}
\frac{dw_k}{2\pi \sqrt{-1} w_k}
F_0(z_1)\cdots F_0(z_{n})\nonumber\\
&\times&
F_1(w_{1})\cdots F_1(w_{j-1})
[T_1(\zeta),F_1(w_j)]
F_1(w_{j+1})\cdots F_1(w_n)\\
&\times&\frac{\prod_{1\leqq k<l \leqq n}
[u_k-u_l]_r[u_k-u_l+1]_r[v_k-v_l]_r[v_k-v_l+1]_r
}{
\prod_{k,l=1}^n[u_k-v_l+\frac{s}{2}]_r[v_l-u_k+\frac{s}{2}-1]_r
}
\eta \left(\vartheta_{\alpha,r}
\left(\sum_{j=1}^n (u_j-v_j)\right)\right)=0.\nonumber
\end{eqnarray}
Therefore we have $[{\cal I}_1,{\cal G}_n]=0$.
Generalization to generic ${\rm Re}(s)>0$ and ${\rm Re}(r)>0$
case, should be understood as analytic continuation.
~~
Q.E.D.
\\\\
Proof of general case 
$[{\cal I}_m,{\cal G}_n]=0$
is given essentially the same manner as above.
Because the integral contour of 
the local integral motion ${\cal I}_m$
is not cylinder, we are not allowed to
use the symbol $\delta(z)$ directly.
Before starting general proof of
$[{\cal I}_m,{\cal G}_n]=0$, 
we review elementary fact about $\delta(z)$.
Let us set the operators ${\cal F}_{1}$ and ${\cal F}_{2}$ by
\begin{eqnarray}
&&{\cal F}_{1}=\int \cdots \int_{C_1} 
\prod_{j=1}^m \frac{dw_j}{2\pi \sqrt{-1}w_j}
f_1(w_1,\cdots,w_m)\Lambda_1(w_1),
\\
&&{\cal F}_{2}=\int \cdots \int_{C_2} 
\prod_{j=1}^n \frac{dz_j}{2\pi \sqrt{-1}z_j}f_2(z_1,\cdots,z_n)F_1(z_1),
\end{eqnarray}
where $f_1(w_1,\cdots,w_m)$ and 
$f_2(z_1,\cdots,z_n)$ are meromorphic functions. 
Let us set the region $D_1$ in which 
Laurent-series of $f_1(w_1,\cdots,w_m)$ exists,
and the region $D_2$ in which Laurent-series of
$f_2(z_1,\cdots,z_n)$ has Laurant-expansion exists.
Let us set 
$\widetilde{C}_1(z_1)=\{(x^{-r}z_1,w_2,\cdots,w_m)\in {\mathbb C}^m|
(w_1,\cdots,w_m)\in C_1\}$.
If the contours $C_1 \subset D_1$, $C_2 \subset D_2$,
and $\widetilde{C}_1(z_1) \subset D_1$ for
any $(z_1,z_2,\cdots,z_n)\in C_2$, we have
the following formulae, which we have already used above.
\begin{eqnarray}
[{\cal F}_{1},{\cal F}_{2}]
=(x^{-r^*}-x^{r^*})\left[
\delta(x^{-r}z_1/w_1){\cal A}(x^{-r}z_1)
f_2(z_1,\cdots,z_n)f_1(x^{-r}z_1,w_2,\cdots, w_m)
\right]_{1,z_1 \cdots z_n w_1 \cdots w_m}.\nonumber
\end{eqnarray}
Even if $\widetilde{C}_1(z_1)$ is not included in $D_1$ for some
$(z_1,z_2,\cdots,z_n)\in C_2$, we have a similar formulae.
\begin{prop}~~
\begin{eqnarray}
\left[{\cal F}_{1}, {\cal F}_{2}\right]&=&
(x^{-r^*}-x^{r^*})\int \cdots \int_{C_2}
\prod_{j=1}^n
\frac{dz_j}{2\pi \sqrt{-1}z_j}f_2(z_1,\cdots,z_n){\cal A}(x^{-r}z_1)\\
&\times&\left(
\int \cdots \int_{C_1 \cap \left\{\left|\frac{x^{-r}z_1}{w_1}\right|<1
\right\}}\prod_{j=1}^m \frac{dw_j}{2\pi \sqrt{-1}w_j}
\frac{f_1(w_1,w_2,\cdots,w_m)
}{(1-x^{-r}z_1/w_1)}\right.\nonumber\\
&+&\left.
\int \cdots
\int_{C_1 \cap \left\{\left|\frac{x^{-r}z_1}{w_1}\right|>1
\right\}}
\prod_{j=1}^m
\frac{dw_j}{2\pi \sqrt{-1}w_j}
\frac{\frac{x^rw_1}{z_1}
f_1(w_1,w_2,\cdots,w_m)
}{(1-x^{r}w_1/z_1)}\right).\nonumber
\end{eqnarray}
\end{prop}
Now let us start to show theorem for general case.\\\\
{\it Proof of $[{\cal I}_m,{\cal G}_n]=0$}~~~
For a while we restrict our interest to
the regime $0<{\rm Re}(s)<2$, $0<{\rm Re}(r)<1$ 
and ${\rm Re}(2r)<{\rm Re}(s)$.
We have
\begin{eqnarray}
&&[{\cal I}_m,{\cal G}_n]\nonumber\\
&=&(x^{-r^*}-x^{r^*})\sum_{i=1}^m
\sum_{j=1}^n
\int \cdots \int_{\widetilde{{I}}(j)}
\prod_{k=1}^n \frac{dz_k}{2\pi \sqrt{-1}z_k}
\prod_{k=1}^n \frac{dw_k}{2\pi \sqrt{-1} w_k}\nonumber\\
&\times&\frac{\prod_{1\leqq k<l \leqq n}
[u_k-u_l]_r[u_k-u_l+1]_r[v_k-v_l]_r[v_k-v_l+1]_r
}{
\prod_{k,l=1}^n[u_k-v_l+\frac{s}{2}]_r[v_l-u_k+\frac{s}{2}-1]_r
}\vartheta_{\alpha,r}
\left(\sum_{j=1}^n (u_j-v_j)\right)\nonumber\\
&\times&
\left(\int_{I \cap \{|\frac{x^{-r}z_j}{z_i'}|<1
\}}
\prod_{k=1}^m \frac{dz_k'}{2\pi \sqrt{-1}z_k'}
\frac{1}{(1-x^{-r}z_j/z_i')}+
\int_{I \cap \{|\frac{x^{-r}z_j}{z_i'}|>1
\}}
\prod_{k=1}^m \frac{dz_k'}{2\pi \sqrt{-1}z_k'}
\frac{x^{r}z_i'/z_j}{(1-x^{r}z_i'/z_j)}
\right)\nonumber\\
&\times&
\prod_{1\leqq k<l \leqq m}h(u_k'-u_l')
T_1(z_1')\cdots T_1(z_{i-1}')
F_1(z_1)\cdots F_1(z_{j-1})\nonumber\\
&\times&{\cal A}(x^{-r}z_j)
F_1(z_{j+1})\cdots F_1(z_n)F_0(w_1)\cdots F_0(w_n)T_1(z_{i+1}')
\cdots T_1(z_m')\nonumber\\
&-&(x^{-r^*}-x^{r^*})\sum_{i,j=1}^n
\int \cdots \int_{\widetilde{{I}}'(j)}
\prod_{k=1}^n \frac{dz_k}{2\pi \sqrt{-1}z_k}
\prod_{k=1}^n \frac{dw_k}{2\pi \sqrt{-1} w_k}\nonumber\\
&\times&\frac{\prod_{1\leqq k<l \leqq n}
[u_k-u_l]_r[u_k-u_l+1]_r[v_k-v_l]_r[v_k-v_l+1]_r
}{
\prod_{k,l=1}^n[u_k-v_l+\frac{s}{2}]_r[v_l-u_k+\frac{s}{2}-1]_r
}\vartheta_{\alpha,r}
\left(\sum_{j=1}^n (u_j-v_j)\right)\nonumber\\
&\times&
\left(\int_{I \cap \{|\frac{x^{r}z_j}{z_i'}|<1
\}}
\prod_{k=1}^m \frac{dz_k'}{2\pi \sqrt{-1}z_k'}
\frac{1}{(1-x^{r}z_j/z_i')}+
\int_{I \cap \{|\frac{x^{-r}z_j}{z_i'}|>1
\}}
\prod_{k=1}^m \frac{dz_k'}{2\pi \sqrt{-1}z_k'}
\frac{x^{r}z_i'/z_j}{(1-x^{r}z_i'/z_j)}
\right)\nonumber\\
&\times&
\prod_{1\leqq k<l \leqq m}h(u_k'-u_l')
T_1(z_1')\cdots T_1(z_{i-1}')
F_1(z_1)\cdots F_1(z_{j-1})\nonumber\\
&\times&{\cal A}(x^{r}z_j)
F_1(z_{j+1})\cdots F_1(z_n)F_0(w_1)\cdots F_0(w_n)T_1(z_{i+1}')
\cdots T_1(z_m')\nonumber\\
&+&\eta\left(\sum_{i,j=1}^n
\int \cdots \int_{I}
\prod_{k=1}^n \frac{dz_k}{2\pi \sqrt{-1}z_k}
\prod_{k=1}^n \frac{dw_k}{2\pi \sqrt{-1} w_k}
\int \cdots \int_C
\prod_{k=1}^n \frac{dz_k'}{2\pi \sqrt{-1}z_k'}
\right.\nonumber\\
&\times&\frac{\prod_{1\leqq k<l \leqq n}
[u_k-u_l]_r[u_k-u_l+1]_r[v_k-v_l]_r[v_k-v_l+1]_r
}{
\prod_{k,l=1}^n
[u_k-v_l+\frac{s}{2}]_r[v_l-u_k+\frac{s}{2}-1]_r}\nonumber\\
&\times&
\vartheta_{\alpha,r}
\left(\sum_{j=1}^n (u_j-v_j)\right)
\prod_{1\leqq k<l \leqq n}h(u_k'-u_l')\nonumber\\
&\times&
T_1(z_1')\cdots T_1(z_{i-1}')
F_0(z_1)\cdots F_0(z_n)
F_1(w_1)\cdots F_1(w_{j-1})\nonumber\\
&\times&
\left.[T_1(z_i'),F_1(w_j)]
F_1(w_{j+1})\cdots F_1(w_n)T_1(z_{i+1}')
\cdots T_1(z_n')\right).
\end{eqnarray}
Here $\widetilde{I}(j)$ and 
$\widetilde{I}'(j)$ are the same as given in proof of 
$[{\cal I}_1,{\cal G}_n]=0$.
Let us change the variable $z_j\to x^{2r}z_j$ of the first term,
upon condition $0<{\rm Re}(s)<2, 0<{\rm Re}(r)<1$
and ${\rm Re}(2r)<{\rm Re}(s)$.
Using periodicity of integrand, we have
\begin{eqnarray}
&&\int \cdots \int_{\widetilde{{I}}(j)}
\prod_{k=1}^n \frac{dz_k}{2\pi \sqrt{-1}z_k}
\prod_{k=1}^n \frac{dw_k}{2\pi \sqrt{-1} w_k}
\int_{I \cap \{|\frac{x^{-r}z_j}{z_i'}|<1\}}
\prod_{k=1}^m \frac{dz_k'}{2\pi \sqrt{-1}z_k'}
\frac{1}{(1-x^{-r}z_j/z_i')}
\nonumber\\
&\times&\frac{\prod_{1\leqq k<l \leqq n}
[u_k-u_l]_r[u_k-u_l+1]_r[v_k-v_l]_r[v_k-v_l+1]_r
}{
\prod_{k,l=1}^n[u_k-v_l+\frac{s}{2}]_r[v_l-u_k+\frac{s}{2}-1]_r}
\nonumber\\
&\times&
\vartheta_{\alpha,r}
\left(\sum_{j=1}^n (u_j-v_j)\right)
\prod_{1\leqq k<l \leqq m}h(u_k'-u_l')
\nonumber\\
&\times&
T_1(z_1')\cdots T_1(z_{i-1}')
F_1(z_1)\cdots F_1(z_{j-1}){\cal A}(x^{-r}z_j)
F_1(z_{j+1})\cdots F_1(z_n)\nonumber\\
&\times&F_0(w_1)\cdots F_0(w_n)T_1(z_{i+1}')
\cdots T_1(z_m')\nonumber\\
&=&
\int \cdots \int_{\widetilde{{I}}'(j)}
\prod_{k=1}^n \frac{dz_k}{2\pi \sqrt{-1}z_k}
\prod_{k=1}^n \frac{dw_k}{2\pi \sqrt{-1} w_k}
\int_{I \cap \{|\frac{x^{r}z_j}{z_i'}|<1\}}
\prod_{k=1}^m \frac{dz_k'}{2\pi \sqrt{-1}z_k'}
\frac{1}{(1-x^{r}z_j/z_i')}\nonumber
\\
&\times&\frac{\prod_{1\leqq k<l \leqq n}
[u_k-u_l]_r[u_k-u_l+1]_r[v_k-v_l]_r[v_k-v_l+1]_r
}{
\prod_{k,l=1}^n[u_k-v_l+\frac{s}{2}]_r[v_l-u_k+\frac{s}{2}-1]_r
}\nonumber\\
&\times&
\vartheta_{\alpha,r}
\left(\sum_{j=1}^n (u_j-v_j)\right)
\prod_{1\leqq k<l \leqq m}h(u_k'-u_l')
\nonumber\\
&\times&
T_1(z_1')\cdots T_1(z_{i-1}')
F_1(z_1)\cdots F_1(z_{j-1}){\cal A}(x^{r}z_j)
F_1(z_{j+1})\cdots F_1(z_n)\nonumber\\
&\times&F_0(w_1)\cdots F_0(w_n)T_1(z_{i+1}')
\cdots T_1(z_m').
\end{eqnarray}
By the similar way, we have
\begin{eqnarray}
&&\eta\left(\sum_{i=1}^m
\sum_{j=1}^n
\int \cdots \int_{I}
\prod_{k=1}^n \frac{dz_k}{2\pi \sqrt{-1}z_k}
\prod_{k=1}^n \frac{dw_k}{2\pi \sqrt{-1} w_k}
\int \cdots \int_C
\prod_{k=1}^m \frac{dz_k'}{2\pi \sqrt{-1}z_k'}
\right.\nonumber\\
&\times&\frac{\prod_{1\leqq k<l \leqq n}
[u_k-u_l]_r[u_k-u_l+1]_r[v_k-v_l]_r[v_k-v_l+1]_r
}{
\prod_{k,l=1}^n[u_k-v_l+\frac{s}{2}]_r[v_l-u_k+\frac{s}{2}-1]_r
}\nonumber\\
&\times&
\vartheta_{\alpha,r}
\left(\sum_{j=1}^n (u_j-v_j)\right)
\prod_{1\leqq k<l \leqq m}
h(u_k'-u_l')\nonumber\\
&\times&
T_1(z_1')\cdots T_1(z_{i-1}')
F_0(z_1)\cdots F_0(z_n)
F_1(w_1)\cdots F_1(w_{j-1})\nonumber\\
&\times&
\left.[T_1(z_i'),F_1(w_j)]
F_1(z_{j+1})\cdots F_1(z_n)T_1(z_{i+1}')
\cdots T_1(z_m')\right)=0.
\end{eqnarray}
Therefore we have $[{\cal I}_m,{\cal G}_n]=0$.
Generalization to generic ${\rm Re}(s)>0$ and ${\rm Re}(r)>0$
case, should be understood as analytic continuation.~~~ 
Q.E.D.

\section{Specialization upon $s=2$}

In this section we study
the specialization to $s=2$,
and discuss relation to 
the Poisson-Virasoro algebra \cite{FR}.

\subsection{Local Integrals of Motion}

In what follows we restrict our interst to the case $1<{\rm Re}(s)\leqq 2$.
\begin{df}~~~We set the local integrals of motion 
${\cal I}_n^{DV}$
for the deformed Virasoro algebra $(s=2)$ by
\begin{eqnarray}
{\cal I}_n^{DV}=
\prod_{j=1}^n
\int_{|z_j|=1}
\frac{dz_j}{2\pi \sqrt{-1}z_j} 
\prod_{1\leqq j<k \leqq n}h(u_k-u_j)|_{s=2}
T^{DV}(z_1)\cdots T^{DV}(z_n).
\end{eqnarray}
\end{df}

\begin{conj}~~~
Upon specialization $s=2$ we have
\begin{eqnarray}
~[{\cal I}_n^{DV},{\cal I}_m^{DV}]=0,~~~\eta(
{\cal I}_n^{DV})={\cal I}_n^{DV},
~~(m,n=1,2,\cdots).
\end{eqnarray}
\end{conj}
In what follows we give a supporting argument of
this conjecture.
\begin{df}~~~Let us set the auxiliary 
operators ${\cal I}_{m,l}$ by
\begin{eqnarray}
{\cal I}_{m,l}&=& 
\prod_{j=1}^{m+l}
\int_{|z_j|=1}
\frac{dz_j}{2\pi \sqrt{-1}z_j}
T_1(z_1)\cdots T_1(z_m)T_2(x^{-1}z_{m+1})\cdots T_2(x^{-1}z_{m+l})\nonumber\\
&\times&
\prod_{1\leqq i<j\leqq m}h(u_j-u_i)
~\prod_{i=1}^m \prod_{j=m+1}^{m+l} h_{12}(u_j-u_i)
\prod_{m+1\leqq i<j\leqq m+l}h_{22}(u_j-u_i).
\end{eqnarray}
Here we have set
\begin{eqnarray}
h_{12}(u)=h(u-1)h(u+1),~~~h_{22}(u)=h_{12}(u-1)h_{12}(u+1).
\end{eqnarray}
\end{df}

\begin{prop}~~~
The local integrals of motion ${\cal I}_n$ are written by linear
combination of the auxiliary operators ${\cal I}_{m,l}$
\begin{eqnarray}
{\cal I}_n={\cal I}_{n,0}+\sum_{\alpha=1}^{[\frac{n}{2}]}
(-c)^\alpha s(x^{-2})^{\alpha}\frac{n!}{\alpha! (n-2\alpha)!}
{\cal I}_{n-2\alpha,\alpha}.
\end{eqnarray}
\end{prop}
{\bf Example}
\begin{eqnarray}
&&{\cal I}_2={\cal I}_{2,0}-2c s(x^{-2}) {\cal I}_{0,1},~~
{\cal I}_3={\cal I}_{3,0}-6c s(x^{-2}) {\cal I}_{1,1},\\
&&{\cal I}_4={\cal I}_{4,0}-12c s(x^{-2}){\cal I}_{2,1}+12c^2 s(x^{-2})^2
{\cal I}_{0,2},\\
&&{\cal I}_5={\cal I}_{5,0}-20c s(x^{-2}) {\cal I}_{3,1}+
60c^2 s(x^{-2})^2 {\cal I}_{1,2},\\
&&{\cal I}_6={\cal I}_{6,0}-30c s(x^{-2}) {\cal I}_{4,1}+
180 c^2 s(x^{-2})^2 {\cal I}_{2,2}-120
c^3 s(x^{-2})^3 {\cal I}_{0,3}.
\end{eqnarray}

\begin{df}~~~
Let us set ``renormalized'' local integrals of motion ${\cal I}_n(s)$ by
\begin{eqnarray}
{\cal I}_n(s)=
{\cal I}_n+\sum_{\alpha=1}^{[\frac{n}{2}]}
\sum_{\beta=\alpha}^{[\frac{n}{2}]}
(-1)^{\alpha+\beta}c^\beta \cdot s(x^{-2})^\beta 
\frac{n!}{\beta! (n-2\beta)!}
\left(
\sum_{k_1,k_2,\cdots,k_{\alpha}\geqq 1
\atop{k_1+\cdots+k_\alpha=\beta}}
\frac{\beta!}{k_1! k_2! \cdots k_\alpha!}
\right){\cal I}_{n-2\beta}.\nonumber\\
\end{eqnarray}
\end{df}

\begin{prop}~~~
We have the commutation relation
\begin{eqnarray}
[{\cal I}_n(s),{\cal I}_m(s)]=0~~~(m,n=1,2,\cdots).
\end{eqnarray}
\end{prop}
{\bf Example}
\begin{eqnarray}
&&{\cal I}_2(s)={\cal I}_2+2c s(x^{-2}) {\cal I}_0,~
{\cal I}_3(s)={\cal I}_3+6c s(x^{-2}) {\cal I}_1,\\
&&{\cal I}_4(s)={\cal I}_4+12c s(x^{-2}) {\cal I}_2+
12 c^2 s(x^{-2})^2{\cal I}_0,\\
&&{\cal I}_5(s)={\cal I}_5+20 cs(x^{-2}){\cal I}_3
+60 c^2 s(x^{-2})^2 {\cal I}_1
\\
&&{\cal I}_6(s)={\cal I}_6+30 c s(x^{-2}){\cal I}_4+180 c^2 s(x^{-2})^2
{\cal I}_2+120 c^3 s(x^{-2})^3 {\cal I}_0.
\end{eqnarray}

\begin{prop}~~~
``Renormalized'' 
local integrals of motion ${\cal I}_n(s)$ is written by
linear combination of difference $({\cal I}_{m,l}-{\cal I}_{m,0})$.
\begin{eqnarray}
{\cal I}_n(s)&=&
{\cal I}_{n,0}+
\sum_{\delta=1}^{[\frac{n}{2}]}
(-c)^\delta
\frac{n!}{\delta!(n-2\delta)!}
({\cal I}_{n-2\delta,\delta}-{\cal I}_{n-2\delta,0})
s(x^{-2})^\delta\nonumber\\
&-&\sum_{\alpha=1}^{[\frac{n}{2}]-1}
\sum_{\delta=\alpha+1}^{[\frac{n}{2}]}
\sum_{\beta=1}^{\delta-\alpha}
(-1)^\alpha (-c)^\delta \frac{n!}{(n-2\delta)!\beta!}
\nonumber\\
&\times&
\left(
\sum_{k_1,k_2,\cdots,k_{\alpha}\geqq 1
\atop{k_1+\cdots+k_\alpha=\delta-\beta}}
\frac{\delta!}{k_1! k_2! \cdots k_\alpha!}
\right)
s(x^{-2})^\delta
( {\cal I}_{n-2\delta, \beta}-{\cal I}_{n-2\delta, 0}).
\end{eqnarray}
\end{prop}
{\bf Example}
\begin{eqnarray}
{\cal I}_2(s)&=&{\cal I}_{2,0}-2cs(x^{-2})({\cal I}_{0,1}-{\cal I}_{0,0}),\\
{\cal I}_3(s)&=&{\cal I}_{3,0}-6c s(x^{-2})({\cal I}_{1,1}-{\cal I}_{1,0}),\\
{\cal I}_4(s)&=&{\cal I}_{4,0}-12cs(x^{-2})({\cal I}_{2,1}-{\cal I}_{2,0})
\nonumber\\
&+&12c^2s(x^{-2})^2({\cal I}_{0,2}-{\cal I}_{0,0})
-24c^2s(x^{-2})^2({\cal I}_{0,1}-{\cal I}_{0,0}),\\
{\cal I}_5(s)&=&{\cal I}_{5,0}-20cs(x^{-2})({\cal I}_{3,1}-{\cal I}_{3,0})\nonumber\\
&+&60c^2s(x^{2})^2({\cal I}_{1,2}-{\cal I}_{1,0})-120c^2s(x^{-2})^2
({\cal I}_{1,1}-{\cal I}_{1,0}),\\
{\cal I}_6(s)&=&
{\cal I}_{6,0}-30cs(x^{-2})({\cal I}_{4,1}-{\cal I}_{4,0})
+180c^2s(x^{-2})^2({\cal I}_{2,2}-{\cal I}_{2,0})\nonumber\\
&-&360c^2s(x^{-2})^2({\cal I}_{2,1}-{\cal I}_{2,0})
-120c^3 s(x^{-2})^3({\cal I}_{0,3}-{\cal I}_{0,0})\nonumber\\
&+&360c^3 s(x^{-2})^3({\cal I}_{0,2}-{\cal I}_{0,0})
-360c^3 s(x^{-2})^3({\cal I}_{0,1}-{\cal I}_{0,0}).
\end{eqnarray}
When we take the limit $s\to2$, we have
\begin{eqnarray}
&&\frac{1}{(s-2)^\delta}
({\cal I}_{n-2\delta,\beta}-{\cal I}_{n-2\delta,0})\nonumber\\
&\to& 
\prod_{j=1}^{
n-2\delta}
\int_{|z_j|=1}
\frac{dz_j}{2\pi \sqrt{-1}z_j}
\prod_{1\leqq j<k \leqq n-2\delta}h(u_k-u_j)|_{s=2}
T^{DV}(z_1)\cdots T^{DV}(z_{n-2\delta})\nonumber\\
&\times&
\prod_{j=n-2\delta+1}^{
n-2\delta+\beta}
\int_{|z_j|=1}\frac{dz_j}{2\pi \sqrt{-1}z_j}
\left.\left(\frac{\partial}{\partial s}\right)^\delta
\right|_{s=2}
T_2(x^{-1}z_{n-2\delta+1})\cdots T_2(x^{-1}z_{n-2\delta+\beta})\nonumber\\
&\times&
\prod_{j=1}^{n-2\delta}
\prod_{k=n-2\delta+1}^{n-2\delta+\beta}
h_{12}(u_k-u_j)
\prod_{n-2\delta+1\leqq j<k \leqq n-2\delta+\beta}h_{22}(u_k-u_j).
\end{eqnarray}
Simplifications occure for $s=2$.
\begin{eqnarray}
h_{12}(u)|_{s=2}=1,~~h_{22}(u)|_{s=2}=1,~~T_2(z)|_{s=2}=1,~~
\left.\frac{\partial}{\partial s}\right|_{s=2}T_2(z)=0,
\end{eqnarray}
We conjecture the following. 
\begin{conj}~~~
When we take the limit $s\to2$, we have
\begin{eqnarray}
{\cal I}_n(s) \to {\cal I}_n^{DV},~~~~~(s\to 2).\label{eqn:DVLocal}
\end{eqnarray}
\end{conj}
We have checked relation (\ref{eqn:DVLocal}) for 
small $n$, {\it i.e.},
$1\leqq n \leqq 9$.
As a consequence of (\ref{eqn:DVLocal}), we have the commutation relation
$[{\cal I}^{DV}_m, {\cal I}_n^{DV}]=0$
and invariance
$\eta\left({\cal I}^{DV}_n\right)={\cal I}_n^{DV}$.
\\
\\
{\bf Example}~~({\bf The Virasoro-Poisson algebra})
\\
Let us set two parameters
$(q,\beta)$ by $q=x^{2r}, \beta=\frac{r-1}{r}$.
When we take the limit $\beta \to 0$ with $q$ fixed,
we have the Virasoro-Poisson algebra \cite{FR}.
\begin{eqnarray}
\{T_m,T_n\}_{P.B.}=\sum_{l\in {\mathbb Z}}
\frac{1-q^l}{1+q^l}T_{n-l}T_{m+l}+(q^n-q^{-n})\delta_{n+m,0}.
\end{eqnarray}
In this limit
the integral of motion ${\cal I}_1^{DV}$
degenerates to the conservation law $H_1$
in \cite{F}.
\\\\
{\bf Example}~~({\bf CFT-limit})
\\
Let us set two parameters $(h,\beta)$ by
$e^h=x^{2r}, \beta=\frac{r-1}{r}$.
We take the limit $h\to 0$ with $\beta$ fixed 
under the following $h$-expansion
\begin{eqnarray}
T_n=2\delta_{n,0}+\beta\left(L_n+\frac{(1-\beta)^2}{4\beta}\delta_{n,0}\right)h^2
+O(h^4),
\end{eqnarray}
we have the defining relation of the Virasoro algebra
\begin{eqnarray}
[L_m,L_n]=(n-m)L_{m+n}+\frac{c_{CFT}}{12}n(n^2-1)\delta_{n+m,0},~~
c_{CFT}=1-\frac{6(1-\beta)^2}{\beta}.
\end{eqnarray}
We call this limit the CFT-limit.
In this limit the local integrals of motion
${\cal I}_1^{DV}, {\cal I}_2^{DV}$ reproduce
those of CFT $I_1, I_3$ by \cite{BLZ}.

\subsection{Nonlocal Integrals of Motion}

In what follows 
we restrict our interst to the case $1<{\rm Re}(s)\leqq 2$.

\begin{df}~~~We set the 
nonlocal integrals of motion ${\cal G}_n^{DV}$ 
for the deformed Virasoro algebra $(s=2)$ by
\begin{eqnarray}
{\cal G}_n^{DV}&=&\int \cdots \int_{I_{Arg}}
\prod_{j=1}^n \frac{dz_j}{2\pi \sqrt{-1}z_j}
\prod_{j=1}^n \frac{dw_j}{2\pi \sqrt{-1}w_j}\nonumber\\
&\times&
\frac{F_1^{DV}(z_1)F_0^{DV}(w_1)}{[u_1-v_1+1]_r[v_1-u_1]_r}
\cdots 
\frac{F_1^{DV}(z_n)F_0^{DV}(w_n)}{[u_n-v_n+1]_r[v_n-u_n]_r}\\
&\times&
\prod_{1\leqq i<j \leqq n}\frac{[u_i-u_j]_r[u_i-u_j+1]_r
[v_i-v_j]_r[v_i-v\j+1]_r}{
[u_i-v_j+1]_r[v_j-u_i]_r[v_i-u_j+1]_r
[u_j-v_i]_r}\left.\vartheta_{\alpha,r}\left(
\sum_{j=1}^n (u_j-v_j)\right)\right|_{s=2},\nonumber
\end{eqnarray}
where the contour $I_{Arg}$ are given by
\begin{eqnarray}
|x^2w_n|,|x^{2r}w_n|<|z_1|<|w_1|<|z_2|<|w_2|<\cdots <|z_n|<|w_n|.
\end{eqnarray}
\end{df}

\begin{conj}~~~
Upon specialization $s=2$ we have
\begin{eqnarray}
~[{\cal G}_n^{DV},{\cal G}_m^{DV}]=0,~~~\eta(
{\cal G}_n^{DV})={\cal G}_n^{DV},
~~(m,n=1,2,\cdots).
\end{eqnarray}
\end{conj}

\begin{conj}~~~
Upon specialization $s=2$ we have
\begin{eqnarray}
~[{\cal I}_n^{DV},{\cal G}_m^{DV}]=0,~~~(m,n=1,2,\cdots),
\end{eqnarray}
\end{conj}
In what follows we give a supporting argument of
the above conjecture.
\begin{df}~~~
Let us set the the auxiliary operators ${\cal G}_{m,l}$ by
\begin{eqnarray}
{\cal G}_{m,l}&=&\int \cdots \int_{\widetilde{I}(m,l)}
\prod_{j=1}^{m}
\frac{dz_j}{2\pi \sqrt{-1}z_j}
\prod_{j=1}^{m+l}
\frac{dw_j}{2\pi \sqrt{-1}w_j}
\prod_{1\leqq i<j\leqq m}j_{1,1}(u_i,v_i|u_j,v_j)
\nonumber\\
&\times&\prod_{i=1}^m
\prod_{j=m+1}^{m+l}j_{1,2}(u_i,v_i|v_j)
\prod_{m+1\leqq i<j \leqq m+l}j_{2,2}(v_i|v_j)\times
\vartheta_{\alpha,r}\left(\sum_{j=1}^m(u_j-v_j)+l(\frac{s}{2}-1)\right)
\nonumber\\
&\times&\frac{F_1(z_1)F_0(w_1)}{[u_1-v_1+\frac{s}{2}]_r
[v_1-u_1+\frac{s}{2}-1]_r}\cdots
\frac{F_1(z_m)F_0(w_m)}{[u_m-v_m+\frac{s}{2}]_r
[v_m-u_m+\frac{s}{2}-1]_r}\nonumber\\
&\times&
:F_1(x^{2-s}w_{m+1})F_0(w_{m+1}):\cdots
:F_1(x^{2-s}w_{m+l})F_0(w_{m+l}):
\end{eqnarray}
Here we have set
\begin{eqnarray}
j_{1,1}(u_1,v_1|u_2,v_2)&=&\frac{[u_1-u_2]_r[u_1-u_2+1]_r
[v_1-v_2]_r[v_1-v_2+1]_r}{
[u_1-v_2+\frac{s}{2}]_r
[v_2-u_1+\frac{s}{2}-1]_r
[v_1-u_2+\frac{s}{2}]_r
[u_2-v_1+\frac{s}{2}-1]_r},\nonumber\\
j_{1,2}(u_1,v_1|v_2)&=&
j_{1,1}\left(u_1,v_1\left|v_2+1-\frac{s}{2},v_2\right.\right),
\nonumber\\
j_{2,2}(v_1|v_2)&=&
j_{1,1}\left(\left.v_1+1-\frac{s}{2},v_1\right|
v_2+1-\frac{s}{2},v_2\right).
\end{eqnarray}
Here the contour $\widetilde{I}(m,l)$ is given by
\begin{eqnarray}
\widetilde{I}(m,l)&:&|x^sw_m|,|x^{2r}w_m|<|z_1|<|w_1|<|z_2|<|w_2|<\cdots
<|z_m|<|w_m|,\nonumber
\\
&&|x^{2s-2}w_j|, |x^{4-2s}w_j|<|w_i|<|x^{2-2s}w_j|,
|x^{2s-4}w_j|,~~~(m+1\leqq i<j \leqq m+l),\\
&&
|x^sw_j|<|z_i|<|x^{s-2}w_j|,~
|x^{4-2s}w_j|<|w_i|<|x^{2-2s}w_j|,~~
(1\leqq i \leqq m,~m+1\leqq j \leqq m+l).\nonumber
\end{eqnarray}
\end{df}

\begin{prop}~
The local integrals of motion ${\cal G}_n$ are written by
linear combination of ${\cal G}_{m,l}$.
\begin{eqnarray}
{\cal G}_n={\cal G}_{n,0}+\sum_{\alpha=1}^n
\frac{\{n(n-1)\cdots (n-\alpha+1)\}^2}{\alpha!}
\left(\frac{t(s)}{[s-2]_r}\right)^\alpha
{\cal G}_{n-\alpha,\alpha},
\end{eqnarray}
where we have set
\begin{eqnarray}
t(s)=\frac{x^{-\frac{2r^*}{r}}}{[1]_r}
\frac{(x^{2r+2s-4};x^{2r})_\infty 
(x^{2r-2};x^{2r})_\infty}{
(x^{2s-2};x^{2r})_\infty 
(x^{2r};x^{2r})_\infty}.
\end{eqnarray}
\end{prop}
{\bf Example}
\begin{eqnarray}
{\cal G}_1&=&{\cal G}_{1,0}+\frac{t(s)}{[s-2]_r}{\cal G}_{0,1},\\
{\cal G}_2&=&{\cal G}_{2,0}+
4\frac{t(s)}{[s-2]_r}{\cal G}_{1,1}+
2\left(\frac{t(s)}{[s-2]_r}\right)^2{\cal G}_{0,2},\\
{\cal G}_3&=&{\cal G}_{3,0}+9\frac{t(s)}{[s-2]_r}{\cal G}_{2,1}+
18 \left(\frac{t(s)}{[s-2]_r}\right)^2{\cal G}_{1,2}+
6\left(\frac{t(s)}{[s-2]_r}\right)^3{\cal G}_{0,3}.
\end{eqnarray}
\begin{df}~~~Let us set 
``renormalized'' nonlocal integrals of motion,
${\cal G}_n(s)$ by
\begin{eqnarray}
{\cal G}_n(s)&=&{\cal G}_{n}+
\sum_{\alpha=1}^n
\sum_{\beta=\alpha}^n
(-1)^\alpha
\left(\frac{t(s)}{[s-2]_r}\right)^\beta
\left(\frac{n!}{(n-\beta)!}\right)^2
\left(\sum_{k_1,k_2,\cdots,k_\alpha \geqq 1
\atop{k_1+k_2+\cdots+k_\alpha=\beta}}
\frac{1}{k_1!k_2!\cdots k_\alpha!}\right)
{\cal G}_{n-\beta}.\nonumber\\
\end{eqnarray}
\end{df}
\begin{prop}~~~
We have the commutation relation.
\begin{eqnarray}
[{\cal G}_n(s),{\cal G}_m(s)]=0~~(m,n=1,2,\cdots).
\end{eqnarray}
\end{prop}
{\bf Example}
\begin{eqnarray}
{\cal G}_1(s)&=&{\cal G}_1
-\frac{t(s)}{[s-2]_r}{\cal G}_0,\\
{\cal G}_2(s)&=&{\cal G}_2-4\frac{t(s)}{[s-2]_r}{\cal G}_1+
2 \left(\frac{t(s)}{[s-2]_r}\right)^2{\cal G}_0,\\
{\cal G}_3(s)&=&{\cal G}_3-9\frac{t(s)}{[s-2]_r}{\cal G}_2
+18\left(\frac{t(s)}{[s-2]_r}\right)^2{\cal G}_1
-6\left(\frac{t(s)}{[s-2]_r}\right)^3{\cal G}_0.
\end{eqnarray}

\begin{prop}~~~``Renormalized'' nonlocal 
integrals of motion 
${\cal G}_n(s)$ is written by linear combination of
difference $({\cal G}_{m,l}-{\cal G}_{m,0})$.
\begin{eqnarray}
{\cal G}_n(s)&=&{\cal G}_{n,0}+
\sum_{\delta=1}^n
\left(\frac{t(s)}{[s-2]_r}\right)^\delta
\left(\frac{n!}{(n-\delta)!}\right)^2
\frac{1}{\delta!}({\cal G}_{n-\delta,\delta}-{\cal G}_{n-\delta,0})
\nonumber\\
&+&
\sum_{\alpha=1}^{n-1}
\sum_{\delta=\alpha+1}^n
\sum_{\beta=1}^{\delta-\alpha}
(-1)^\alpha
\left(\frac{t(s)}{[s-2]_r}\right)^{\delta}
\left(\frac{n!}{(n-\delta)!}\right)^2
\frac{1}{\beta!}\nonumber\\
&\times&
\left(\sum_{k_1,k_2,\cdots,k_\alpha \geqq 1
\atop{k_1+k_2+\cdots+k_\alpha=\delta-\beta}}
\frac{1}{k_1!k_2!\cdots k_\alpha!}\right)(
{\cal G}_{n-\delta,\beta}-{\cal G}_{n-\delta,0}).
\end{eqnarray}
\end{prop}
{\bf Example}
\begin{eqnarray}
{\cal G}_1(s)&=&{\cal G}_{1,0}+
\frac{t(s)}{[s-2]_r}({\cal G}_{0,1}-{\cal G}_{0,0}),\\
{\cal G}_2(s)&=&{\cal G}_{2,0}+
4\frac{t(s)}{[s-2]_r}({\cal G}_{1,1}-{\cal G}_{1,0})\nonumber\\
&-&4\left(\frac{t(s)}{[s-2]_r}\right)^2({\cal G}_{0,1}-{\cal G}_{0,0})
+2\left(\frac{t(s)}{[s-2]_r}\right)^2({\cal G}_{0,2}-{\cal G}_{0,0}),\\
{\cal G}_3(s)&=&{\cal G}_{3,0}+9\frac{t(s)}{[s-2]_r}(
{\cal G}_{2,1}-{\cal G}_{2,0})\nonumber\\
&+&18\left(\frac{t(s)}{[s-2]_r}\right)^2
({\cal G}_{1,2}-{\cal G}_{1,0})
-36\left(\frac{t(s)}{[s-2]_r}\right)^2
({\cal G}_{1,1}-{\cal G}_{1,0})\\
&+&6\left(\frac{t(s)}{[s-2]_r}\right)^3
({\cal G}_{0,3}-{\cal G}_{0,0})
-18\left(\frac{t(s)}{[s-2]_r}\right)^3
({\cal G}_{0,2}-{\cal G}_{0,0})\nonumber\\
&+&18\left(\frac{t(s)}{[s-2]_r}\right)^3
({\cal G}_{0,1}-{\cal G}_{0,0}).\nonumber
\end{eqnarray}
When we take the limit $s\to 2$, we have
\begin{eqnarray}
&&\frac{1}{(s-2)^\delta}({\cal G}_{n-\delta,\beta}-{\cal G}_{n-\delta,0})
\nonumber\\
&\to&
\int \cdots \int_{\widetilde{I}(n-\delta,\beta)}
\prod_{j=1}^{n-\delta}
\frac{dz_j}{2\pi \sqrt{-1}z_j}
\prod_{j=1}^{n+\delta-\beta}
\frac{dw_j}{2\pi \sqrt{-1}w_j}
\prod_{1\leqq i<j\leqq n-\delta}j_{1,1}(u_i,v_i|u_j,v_j)\nonumber\\
&\times&
\frac{F_1(z_1)F_0(w_1)}{[u_1-v_1+\frac{s}{2}]_r[v_1-u_1+\frac{s}{2}-1]_r}
\cdots \frac{F_1(z_{n-\delta})F_0(w_{n-\delta})}{
[u_{n-\delta}-v_{n-\delta}+\frac{s}{2}]_r
[v_{n-\delta}-u_{n-\delta}+\frac{s}{2}-1]_r}
\nonumber\\
&\times&
\left.\left(\frac{\partial}{\partial s}\right)^\delta\right|_{s=2}
\prod_{i=1}^{n-\delta} \prod_{j=n-\delta+1}^{n-\delta+\beta}
j_{1,2}(u_i,v_i|v_j)\prod_{n-\delta+1
\leqq i<j\leqq n-\delta+\beta}
j_{2,2}(v_i|v_j)\nonumber\\
&\times&
\vartheta_{\alpha,r}\left(\sum_{j=1}^{n-\delta
}(u_j-v_j)+\beta(\frac{s}{2}-1)\right)
\nonumber\\
&\times&
:F_1(x^{2-s}w_{n-\delta+1})F_0(w_{n-\delta+1}):\cdots 
:F_1(x^{2-s}w_{n-\delta+\beta})F_0(w_{m-\delta+\beta}):.\end{eqnarray}
Simplifications occure for $s=2$.
\begin{eqnarray}
j_{1,2}(u_1,v_1|v_2)|_{s=2}=1,~~j_{2,2}(u|v)|_{s=2}=1,
~~\left.:F_1(x^{2-s}w)F_0(w):\right|_{s=2}=id,
\end{eqnarray}
We conjecture the following. 
\begin{conj}~~~
When we take the limit $s\to2$, we have
\begin{eqnarray}
{\cal G}_n(s) \to {\cal G}_n^{DV},~~~(s\to 2).\label{eqn:DVNonlocal}
\end{eqnarray}
\end{conj}
%We have checked the (\ref{eqn:DVNonlocal}) for small $n$.
As a consequence of (\ref{eqn:DVNonlocal}), 
we have the commutation relation
$[{\cal G}_m^{DV},{\cal G}_n^{DV}]=0$,
$[{\cal I}_m^{DV},{\cal G}_n^{DV}]=0$
and the invariance $\eta({\cal G}_n^{DV})={\cal G}_n^{DV}$.
\\\\
{\bf Example}~~({\bf CFT-limit})
\\
Let us set two parameters $(h,\beta)$ by
$e^h=x^{2r}, \beta=\frac{r-1}{r}$.
We call the limit $h\to 0$ with $\beta$ fixed, the CFT-limit.
In this limit the nonlocal integrals of motion,
${\cal G}_k^{DV} (k=1,2,\cdots)$ reproduce
those of CFT, $Q_k (k=1,2,\cdots)$ by \cite{BLZ}.

~\\
{\bf Acknowledgements.}~~We would like to thank 
Professor M.Jimbo
for his useful discussions and
interest in this work.
We would like to thank 
Professor V.Bazhanov,
Professor A.Belavin, Professor S.Duzhin,
Professor E.Frenkel, Professor P.Kulish,
and Professor M.Wadati
for their interests in this work.
B.F.is partly supported by Grant RFBR (05-01-01007),
SS (2044.2003.2), and RFBR-JSPS (05-01-02934).
T.K. is partly supported by Grant-in Aid for 
Young Scientist {\bf B} (18740092) from JSPS,
and Grant for Research in Abroad {\bf B}
from Nihon University. 
J.S. is partly supported by Grant-in Aid for 
Scientific Research {\bf C} (16540183) from JSPS.
We would like to dedicate this paper to Professor 
Masaki Kashiwara on the occasion on the 60th birathday.

\begin{appendix}

\section{Normal Ordering}
We summarize the normal orderings of the
currents.
\begin{eqnarray}
\Lambda_i(z_1)\Lambda_i(z_2)&=&::(1-z_2/z_1)
\frac{(x^2z_2/z_1;x^{2s})_\infty 
(x^{2r+2s-2}z_2/z_1;x^{2s})_\infty 
(x^{2s-2r}z_2/z_1;x^{2s})_\infty}{
(x^{2s-2}z_2/z_1;x^{2s})_\infty
(x^{2r}z_2/z_1;x^{2s})_\infty
(x^{2-2r}z_2/z_1;x^{2s})_\infty},
\nonumber\\
&&~~~~~~~~~~~~~~~~~~~~~~~~~~~~~~~~
~~~~~~~~~~~~~~~~~~~~~~~~~~~~~~~~~~(i=1,2),
\\
\Lambda_1(z_1)\Lambda_2(z_2)&=&::
\frac{(x^2z_2/z_1;x^{2s})_\infty 
(x^{-2r}z_2/z_1;x^{2s})_\infty 
(x^{2r-2}z_2/z_1;x^{2s})_\infty}{
(x^{-2}z_2/z_1;x^{2s})_\infty
(x^{2r}z_2/z_1;x^{2s})_\infty
(x^{2-2r}z_2/z_1;x^{2s})_\infty},\\
\Lambda_2(z_1)\Lambda_1(z_2)&=&::
\frac{(x^{2+2s}z_2/z_1;x^{2s})_\infty 
(x^{2s-2r}z_2/z_1;x^{2s})_\infty 
(x^{2s+2r-2}z_2/z_1;x^{2s})_\infty}{
(x^{2s-2}z_2/z_1;x^{2s})_\infty
(x^{2s+2r}z_2/z_1;x^{2s})_\infty
(x^{2s+2-2r}z_2/z_1;x^{2s})_\infty}.
\end{eqnarray}

\begin{eqnarray}
\Lambda_1(z_1)F_1(z_2)&=&
:\Lambda_1(z_1)F_1(z_2):x^{-2(r-1)}\frac{(1-x^{r-2}z_2/z_1)}{
(1-x^{-r}z_2/z_1)},\\
F_1(z_1)\Lambda_1(z_2)&=&
:F_1(z_1)\Lambda_1(z_2):\frac{(1-x^{2-r}z_2/z_1)}{(1-x^rz_2/z_1)},\\
\Lambda_2(z_1)F_1(z_2)&=&:
\Lambda_2(z_1)F_1(z_2):x^{2(r-1)}\frac{(1-x^{2-r}z_2/z_1)}{(1-x^rz_2/z_1)},
\\
F_1(z_1)\Lambda_2(z_2)&=&:
F_1(z_1)\Lambda_2(z_2):
\frac{(1-x^{r-2}z_2/z_1)}{(1-x^{-r}z_2/z_1)},\\
%%%%%%%%%%%%%%%%%%%%%%%%%%%%%%%%%%%%%%%%%%%%%%%%%%%%%%%%%
\Lambda_1(z_1)F_0(z_2)&=&:
\Lambda_1(z_1)F_0(z_2)
:x^{2(r-1)}\frac{(1-x^{2-r-s}z_2/z_1)}{
(1-x^{r-s}z_2/z_1)},\\
F_0(z_1)\Lambda_1(z_2)&=&:
F_0(z_1)\Lambda_1(z_2)
:\frac{(1-x^{r+s-2}z_2/z_1)}{(1-x^{-r+s}z_2/z_1)},\\
\Lambda_2(z_1)F_0(z_2)&=&:
\Lambda_2(z_1)F_0(z_2):
x^{-2(r-1)}\frac{(1-x^{r+s-2}z_2/z_1)}{(1-x^{-r+s}z_2/z_1)},
\\
F_0(z_1)\Lambda_2(z_2)&=&:
F_0(z_1)\Lambda_2(z_2):
\frac{(1-x^{2-r-s}z_2/z_1)}{(1-x^{r-s}z_2/z_1)}.
\end{eqnarray}

\begin{eqnarray}
\Lambda_1(z_1)E_1(z_2)&=&
:\Lambda_1(z_1)E_1(z_2):x^{2r}\frac{(1-x^{-r-1}z_2/z_1)}{
(1-x^{r-1}z_2/z_1)},\\
E_1(z_1)\Lambda_1(z_2)&=&
:E_1(z_1)\Lambda_1(z_2):\frac{(1-x^{r+1}z_2/z_1)}{(1-x^{-r+1}z_2/z_1)},\\
\Lambda_2(z_1)E_1(z_2)&=&:
\Lambda_2(z_1)E_1(z_2):x^{-2r}\frac{(1-x^{r+1}z_2/z_1)}{(1-x^{-r+1}z_2/z_1)},
\\
E_1(z_1)\Lambda_2(z_2)&=&:
E_1(z_1)\Lambda_2(z_2):
\frac{(1-x^{-r-1}z_2/z_1)}{(1-x^{r-1}z_2/z_1)},\\
%%%%%%%%%%%%%%%%%%%%%%%%%%%%%%%%%%%%%%%%%%%%%%%%%%%%%%%%%%%%%%%%%
\Lambda_1(z_1)E_0(z_2)&=&
:\Lambda_1(z_1)E_0(z_2):x^{-2r}\frac{(1-x^{r+1-s}z_2/z_1)}{
(1-x^{-r+1-s}z_2/z_1)},\\
E_0(z_1)\Lambda_1(z_2)&=&
:E_0(z_1)\Lambda_1(z_2):\frac{(1-x^{-r-1+s}z_2/z_1)}{
(1-x^{r-1+s}z_2/z_1)},\\
\Lambda_2(z_1)E_0(z_2)&=&:
\Lambda_2(z_1)E_0(z_2):x^{2r}\frac{(1-x^{-r-1+s}z_2/z_1)}{
(1-x^{r-1+s}z_2/z_1)},
\\
E_0(z_1)\Lambda_2(z_2)&=&:
E_0(z_1)\Lambda_2(z_2):
\frac{(1-x^{r+1-s}z_2/z_1)}{(1-x^{-r+1-s}z_2/z_1)},
\end{eqnarray}

\begin{eqnarray}
E_j(z_1)E_j(z_2)&=&::
z_1^{\frac{2r}{r^*}}(1-z_2/z_1)\frac{(x^{-2}z_2/z_1;x^{2r^*})_\infty}{
(x^{2r^*+2}z_2/z_1;x^{2r^*})_\infty}
,\\
E_1(z_1)E_0(z_2)&=&::z_1^{-\frac{2r}{r^*}}
\frac{(x^{2r^*+s}z_2/z_1;x^{2r^*})_\infty 
(x^{2r^*+2-s}z_2/z_1;x^{2r^*})_\infty}
{(x^{-s}z_2/z_1;x^{2r^*})_\infty 
(x^{s-2}z_2/z_1;x^{2r^*})_\infty},\\
E_0(z_1)E_1(z_2)&=&::
z_1^{-\frac{2r}{r^*}}
\frac{(x^{2r^*+s}z_2/z_1;x^{2r^*})_\infty 
(x^{2r^*+2-s}z_2/z_1;x^{2r^*})_\infty}
{(x^{-s}z_2/z_1;x^{2r^*})_\infty 
(x^{s-2}z_2/z_1;x^{2r^*})_\infty},\\
F_j(z_1)F_j(z_2)&=&::x^{\frac{2r^*}{r}}
(1-z_2/z_1)\frac{(x^2z_2/z_1;x^{2r})_\infty}{
(x^{2r-2}z_2/z_1;x^{2r})_\infty},\\
F_1(z_1)F_0(z_2)&=&::x^{-\frac{2r^*}{r}}
\frac{(x^{2r-2+s}z_2/z_1;x^{2r})_\infty 
(x^{2r-s}z_2/z_1;x^{2r})_\infty}{
(x^{s}z_2/z_1;x^{2r})_\infty 
(x^{2-s}z_2/z_1;x^{2r})_\infty},\\
F_0(z_1)F_1(z_2)&=&::x^{-\frac{2r^*}{r}}
\frac{(x^{2r-2+s}z_2/z_1;x^{2r})_\infty 
(x^{2r-s}z_2/z_1;x^{2r})_\infty}{
(x^{s}z_2/z_1;x^{2r})_\infty 
(x^{2-s}z_2/z_1;x^{2r})_\infty},
\end{eqnarray}
\begin{eqnarray}
F_1(z_1)E_1(z_2)&=&::\frac{1}{z_1^2(1-xz_2/z_1)(1-x^{-1}z_2/z_1)},\\
E_1(z_1)F_1(z_2)&=&::\frac{1}{z_1^2(1-xz_2/z_1)(1-x^{-1}z_2/z_1)},\\
F_0(z_1)E_0(z_2)&=&::\frac{1}{z_1^2(1-xz_2/z_1)(1-x^{-1}z_2/z_1)},\\
E_0(z_1)F_0(z_2)&=&::\frac{1}{z_1^2(1-xz_2/z_1)(1-x^{-1}z_2/z_1)},\\
F_1(z_1)E_0(z_2)&=&::z_1^2(1-x^{s-1}z_2/z_1)(1-x^{1-s}z_2/z_1),\\
E_0(z_1)F_1(z_2)&=&::z_1^2(1-x^{s-1}z_2/z_1)(1-x^{1-s}z_2/z_1),\\
F_0(z_1)E_1(z_2)&=&::z_1^2(1-x^{s-1}z_2/z_1)(1-x^{1-s}z_2/z_1),\\
E_1(z_1)F_0(z_2)&=&::z_1^2(1-x^{s-1}z_2/z_1)(1-x^{1-s}z_2/z_1).
\end{eqnarray}

\begin{eqnarray}
{\cal A}(z_1)F_0(z_2)&=&::z_1^{-2}
\frac{(x^{3r-s}z_2/z_1;x^{2r})_\infty
(x^{r+s-2}z_2/z_1;x^{2r})_\infty}{
(x^{r-s+2}z_2/z_1;x^{2r})_\infty
(x^{-r+s}z_2/z_1;x^{2r})_\infty},\\
F_0(z_1){\cal A}(z_2)&=&::
z_1^{-2}
\frac{(x^{3r-s}z_2/z_1;x^{2r})_\infty
(x^{r+s-2}z_2/z_1;x^{2r})_\infty}{
(x^{r-s+2}z_2/z_1;x^{2r})_\infty
(x^{-r+s}z_2/z_1;x^{2r})_\infty},\\
{\cal A}(z_1)F_1(z_2)&=&::z_1^{\frac{2r^*}{r}}
\frac{(x^{-r+2}z_2/z_1;x^{2r})_\infty}{
(x^{3r-2}z_2/z_1;x^{2r})_\infty},\\
F_1(z_1){\cal A}(z_2)&=&::
z_1^{\frac{2r^*}{r}}
\frac{(x^{-r+2}z_2/z_1;x^{2r})_\infty}{
(x^{3r-2}z_2/z_1;x^{2r})_\infty},\\
%%%%%%%%%%%%%%%%%%%%%%%%%%%%%%%%%%%%%%%%%%%%%%%%
{\cal A}(z_1)E_0(z_2)&=&::z_1^2(1-x^{r+1-s}z_2/z_1)
(1-x^{-r+s-1}z_2/z_1),\\
E_0(z_1){\cal A}(z_2)&=&::z_1^2(1-x^{r+1-s}z_2/z_1)
(1-x^{-r+s-1}z_2/z_1),\\
{\cal A}(z_1)E_1(z_2)&=&::\frac{1}{z_1^2(1-x^{r-1}z_2/z_1)
(1-x^{1-r}z_2/z_1)},\\
E_1(z_1){\cal A}(z_2)&=&::\frac{1}{z_1^2(1-x^{r-1}z_2/z_1)
(1-x^{1-r}z_2/z_1)}.
\end{eqnarray}

\begin{eqnarray}
{\cal B}(z_1)F_0(z_2)&=&::z_1^2(1-x^{s+r^*-1}z_2/z_1)
(1-xs^{-r^*+1-s}z_2/z_1),\\
F_0(z_1){\cal B}(z_2)&=&::
z_1^2
(1-x^{s+r^*-1}z_2/z_1)
(1-xs^{-r^*+1-s}z_2/z_1),\\
{\cal B}(z_1)F_1(z_2)&=&::
\frac{1}{z_1^2(1-x^{r^*+1}z_2/z_1)(1-x^{-r^*-1}z_2/z_1)},\\
F_1(z_1){\cal B}(z_2)&=&::
\frac{1}{z_1^2(1-x^{r^*+1}z_2/z_1)(1-x^{-r^*-1}z_2/z_1)},\\
%%%%%%%%%%%%%%%%%%%%%%%%%%%%%%%%%%%%%%%%%%%%%%%%
{\cal B}(z_1)E_0(z_2)&=&::
z_1^{-2}
\frac{(x^{3r^*}z_2/z_1;x^{2r^*})_\infty 
(x^{r^*+2}z_2/z_1;x^{2r^*})_\infty}{
(x^{r^*-2}z_2/z_1;x^{2r^*};x^{2r^*})_\infty 
(x^{-r^*}z_2/z_1;x^{2r^*})_\infty},
\\
E_0(z_1){\cal B}(z_2)&=&::
z_1^{-2}
\frac{(x^{3r^*}z_2/z_1;x^{2r^*})_\infty 
(x^{r^*+2}z_2/z_1;x^{2r^*})_\infty}{
(x^{r^*-2}z_2/z_1;x^{2r^*};x^{2r^*})_\infty 
(x^{-r^*}z_2/z_1;x^{2r^*})_\infty},
\\
{\cal B}(z_1)E_1(z_2)&=&::z_1^{\frac{2r}{r^*}}
\frac{(x^{-r^*-2}z_2/z_1;x^{2r^*})_\infty}{
(x^{3r^*+2}z_2/z_1;x^{2r^*})_\infty},\\
E_1(z_1){\cal B}(z_2)&=&::
z_1^{\frac{2r}{r^*}}
\frac{(x^{-r^*-2}z_2/z_1;x^{2r^*})_\infty}{
(x^{3r^*+2}z_2/z_1;x^{2r^*})_\infty}.
\end{eqnarray}

For ${\rm Re}(r)>0$ we have
\begin{eqnarray}
F_0(z_1)H(z_2)&=&::z_1^{\frac{2}{r}}
\frac{(x^{r+s}z_2/z_1;x^{2r})_\infty 
(x^{r+2-s}z_2/z_1;x^{2r})_\infty}{
(x^{r+s-2}z_2/z_1;x^{2r})_\infty 
(x^{r-s}z_2/z_1;x^{2r})_\infty},\\
F_1(z_1)H(z_2)&=&::z_1^{-\frac{2}{r}}
\frac{(x^{r+2}z_2/z_1;x^{2r})_\infty}{
(x^{r-2}z_2/z_1;x^{2r})_\infty},\\
H(z_1)F_0(z_2)&=&::z_1^{\frac{2}{r}}
\frac{(x^{r+s}z_2/z_1;x^{2r})_\infty 
(x^{r+2-s}z_2/z_1;x^{2r})_\infty}{
(x^{r+s-2}z_2/z_1;x^{2r})_\infty 
(x^{r-s}z_2/z_1;x^{2r})_\infty},\\
H(z_1)F_1(z_2)&=&::z_1^{-\frac{2}{r}}
\frac{(x^{r+2}z_2/z_1;x^{2r})_\infty}{
(x^{r-2}z_2/z_1;x^{2r})_\infty}.
\end{eqnarray}

For ${\rm Re}(r^*)<0$ we have
\begin{eqnarray}
E_0(z_1)H(z_2)&=&::z_1^{-\frac{2}{r^*}}
\frac{(x^{-r^*+s}z_2/z_1;x^{-2r^*})_\infty 
(x^{-r^*+2-s}z_2/z_1;x^{-2r^*})_\infty}{
(x^{-r^*+s-2}z_2/z_1;x^{-2r^*})_\infty 
(x^{-r^*-s}z_2/z_1;x^{-2r^*})_\infty},\\
E_1(z_1)H(z_2)&=&::z_1^{\frac{2}{r^*}}
\frac{(x^{-r^*+2}z_2/z_1;x^{-2r^*})_\infty}{
(x^{-r^*-2}z_2/z_1;x^{-2r^*})_\infty},\\
H(z_1)E_0(z_2)&=&::z_1^{-\frac{2}{r^*}}
\frac{(x^{-r^*+s}z_2/z_1;x^{-2r^*})_\infty 
(x^{-r^*+2-s}z_2/z_1;x^{-2r^*})_\infty}{
(x^{-r^*+s-2}z_2/z_1;x^{-2r^*})_\infty 
(x^{-r^*-s}z_2/z_1;x^{-2r^*})_\infty},\\
H(z_1)E_1(z_2)&=&::z_1^{\frac{2}{r^*}}
\frac{(x^{-r^*+2}z_2/z_1;x^{-2r^*})_\infty}{
(x^{-r^*-2}z_2/z_1;x^{-2r^*})_\infty}.
\end{eqnarray}

\end{appendix}

\end{document}